\documentclass[useAMS,usenatbib]{mn2e}

\usepackage{graphicx}
\usepackage{scalefnt}

\newif\ifAMStwofonts
\bibpunct{(}{)}{;}{a}{}{,}
\title[Simulations of minor mergers. I.]{Simulations of 
minor mergers. I. General properties of thick disks}
\author[\'A. Villalobos and A.  Helmi]{\'Alvaro 
Villalobos\thanks{E-mail: villalobos@astro.rug.nl} and Amina 
Helmi\thanks{E-mail: ahelmi@astro.rug.nl}\\ Kapteyn Astronomical Institute,
University of Groningen, P.O. Box 800, 9700 AV Groningen, The
Netherlands}
\begin{document}

\date{}

\pagerange{\pageref{firstpage}--\pageref{lastpage}} \pubyear{}

\maketitle

\label{firstpage}

\begin{abstract}
We present simulations of the formation of thick disks via the
accretion of two-component satellites onto a pre-existing thin
disk. Our goal is to establish the detailed characteristics of the
thick disks obtained in this way, as well as their dependence on the
initial orbital and internal properties of the accreted objects. We
find that mergers with 10--20\% mass of the mass of the host lead to
the formation of thick disks whose characteristics are similar, both
in morphology as in kinematics, to those observed.  Despite the
relatively large mass ratios, the host disks are not fully destroyed
by the infalling satellites: a remaining kinematically cold and thin
component containing $\sim 15-25\%$ of the mass can be identified,
which is embedded in a hotter and thicker disk. This may for example,
explain the existence of a very old thin disk stars in the Milky Way.
The final scale-heights of the disks depend both on the initial
inclination and properties of the merger, but the fraction of
satellite stellar particles at $\sim 4$ scale-heights directly
measures the mass ratio between the satellite and host galaxy. Our
thick disks typically show boxy isophotes at very low surface
brightness levels ($> 6$ magnitudes below their peak value).
Kinematically, the velocity ellipsoids of the simulated thick disks
are similar to that of the Galactic thick disk at the solar radius.
The trend of $\sigma_Z/\sigma_R$ with radius is found to be a very
good discriminant of the initial inclination of the accreted
satellite.  In the Milky Way, the possible existence of a vertical
gradient in the rotational velocity of the thick disk as well as the
observed value of $\sigma_Z/\sigma_R$ at the solar vicinity appear to
favour the formation of the thick disk by a merger with either low or
intermediate orbital inclination.
\end{abstract}


\section{Introduction}

\defcitealias{quinn1986}{QG86}
\defcitealias{quinn1993}{QHF93}
\defcitealias{walker1996}{WMH96}
\defcitealias{huang1997}{HC97}
\defcitealias{velazquez1999}{VW99}
\defcitealias{dalcanton2002}{DB02}

The different components of disk galaxies contain key information
about various stages of the formation history of these systems.  In
this sense, one of the most significant components for studying
signatures of galaxy formation are the thick disks because they
contain imprints of the state of early disks and their interaction
with the galactic environment \citep{freeman2002}.

Until now thick-disks have been detected in S0 galaxies
\citep{burstein1979,tsikoudi1979}, in the Milky Way
\citep{gilmore1983} and in many other spirals \citep[][and references
therein]{vdkruit1981a,vdkruit1981b,jensen1982,
vdokkum1994,morrison1997,pohlen2000,dalcanton2002,yoachim2006}.  As
such, thick disks appear to be a rather ubiquitous structural
component of galaxies.

Historically, there have been two types of scenarii proposed to
explain the formation of thick disks: (i) ``dissipational'', and (ii)
``predominantly dissipationless'' models. In the first case,
thick-disk stars are formed during the dissipational collapse of gas
with a large scale-height, after the halo has formed and before the
thin disk has completely developed.  A variant of this model, more in
line with modern cosmology, is given by \citet{brook2004,brook2005}
and consists of the formation of the thick disk in an epoch of
multiple mergers of gas-rich building blocks. In this class of models,
one may expect a smooth transition between properties of the thin and
thick disks
\citep{eggen1962,gilmore1986,norris1991,burkert1992,pardi1995,
fuhrmann2004}.  Evidence of such a process at work may be the chain
and clumpy galaxies observed at high-redshift by
\citet{elmegreen2006}, which have also been suggested to be the
progenitors of thick disks \citep[e.g.][]{bournaud2007a}.  In the
``predominantly dissipationless'' models the thick disk stars were
either: (1) vertically ``heated'' from a pre-existent thin disk during
a (significant) minor merger \citep{quinn1993,mihos1995,
walker1996,robin1996,velazquez1999,aguerri2001,chen2001}; (2) directly
deposited at large scale-heights as tidally stripped debris during the
accretion of smaller satellite galaxies
\citep{bekki2001,gilmore2002,abadi2003,martin2004,
navarro2004,helmi2006}; or (3) the product of the
dissolution of massive thin-disk clusters with small radii and large
velocity dispersions \citep{kroupa2002}. The first and third models
have in common the presence of a pre-existing disk component.  In this
second class of models, the thick disk may be characterised as a
completely foreign component.

Models involving either one or more minor mergers --including the case
of gas-rich accretion proposed by Brook et al.-- are natural in the
context of current theories of hierarchical structure formation
\citep[e.g.,][]{kazantzidis2007}. Such models are also supported by
modern studies of both kinematical and chemical properties of the
thick disks in the Milky Way and in external galaxies
\citep{yoachim2005,yoachim2006,seth2005, mould2005}. The
observed lack of vertical colour and chemical gradients, as well as
the presence of counter-rotating disks, favours scenarios in which
thin and thick disks formed as separate entities. However, it is
unclear to what extent such models reproduce the detailed properties
of observed thick disks. For example, in the dissipationless minor
merger scenario, the thick disk should contain stars from both the
heated thin disk and the accreted satellite. What fraction of the
stars come from each constituent? If they are mostly satellite debris,
one may expect a significantly metal-poor thick disk, which appears to
be inconsistent with the observations reported in, e.g.,
\citet{mould2005}.

It is the lack of very detailed predictions that motivates us to
revisit the problem of thick disk formation. In this paper, we explore
the hypothesis of the heating of a pre-existing thin disk by a single
minor merger. In the scenario we envisage a new thin disk component
would form from the accretion of cold gas after the merger has taken
place. However, here we only study the global properties of the
final merger product. In a follow-up paper we will focus on its
phase-space structure with the aim of making a detailed comparison to
the thick disk in the Solar neighbourhood, and eventually to uncover
the debris from the object that may have triggered its formation
(Villalobos \& Helmi, in prep). It is important to note that we shall
neglect the effect of gas on the properties of the remnant thick disk
for the moment. This simplification should be borne in mind, in
particular since the structure of minor merger remnants may be
different, as suggested by \citet{naab-jesseit-burkert06} and
\citet{younger2008}.

Significant work has been carried out in the past to model the disk
heating process by a minor merger from the numerical point of view,
starting from \citet{quinn1986} \citepalias{quinn1986};
\citet*{quinn1993} \citepalias{quinn1993}; \citet{mihos1995};
\citet*{walker1996} \citepalias{walker1996}; \citet{huang1997}
\citepalias{huang1997}; \citet{sellwood1998}; \citet{velazquez1999}
\citepalias{velazquez1999}, and more recently \citet{kazantzidis2007},
and simultaneously with this work \citet{read2008}.  These papers 
have essentially shown that it is relatively easy to produce thick
disks whose general properties are consistent with those
observed. However, we believe there is still room for improvement,
both in the initial conditions used to model this process, as well as
in the degree of detail necessary for comparisons to the latest
observations of thick disks.

Besides the work mentioned above, in recent years several authors have
studied minor mergers with a number of different goals (other than the
formation of thick disks). These include for example, studies of the
merger remnants produced by encounters between disk galaxies \citep[as
a way of producing a bulge dominated
system, e.g.][]{naab-burkert2003,jesseit05,naab-trujillo2006}. More recently, also
the effects of gas, star formation and feedback have been taken
into account \citep{bournaud05,naab-jesseit-burkert06,dimatteo2007,cox08}. The
effects of repeated minor mergers have been considered in, e.g.
\citet{bournaud2007b,kazantzidis2007}.

The space of initial conditions to tackle the formation of the thick
disk via minor mergers is very large, and it is not desirable to probe
it randomly. We have therefore made the following physically motivated
choices. (i) We model the formation of thick disks at two different
redshifts, by scaling the properties of the host galaxy and accreted
satellite according to cosmological models, as in \citet*{mo1998} (ii)
We consider the accretion of relatively massive satellites (10\% or
20\% mass ratios), embedded in dark-matter halos, and with stellar
distributions that are initially either spherical (and on the
fundamental plane of dE+dSphs galaxies) or disky. (iii) The satellites
are released much farther away from the host disk compared to previous
studies, and their orbits are consistent with those of infalling
substructures in cosmological simulations (e.g.,
\citealt{benson2005}).

The outline of this paper is the following: In \S 2 we describe in
detail the numerical procedure used to build self-consistently the
different components of both the host galaxy and satellites including
our choices for the numerical parameters and the orbital parameters of
the satellites.  \S 3 describes the outcome of the simulations,
focusing on the final properties of thick disks and the evolution of
satellites. In \S 4 we present a discussion and limitations of our
approach.  In \S 5 we summarise the main results.

\section{Setting up the simulations}

In this section, we describe in broad terms the procedure adopted to
model the host disk galaxy and the (to be accreted) satellite. We
refer the interested reader to the Appendix for more details.  We
consider two configurations: a merger with a host whose properties
resemble the Milky Way today (our ``z=0'' experiment), and a merger
with a smaller host disk galaxy (this is our ``z=1'' experiment). The
``z=0'' configuration has been often used in the past in the same
context (\citetalias{quinn1986}; \citetalias{quinn1993};
\citetalias{walker1996}; \citetalias{huang1997};
\citetalias{velazquez1999}; \citealt{aguerri2001}; \citealt{font2001};
\citealt{ardi2003}; \citealt{hayashi2006}).  In the ``z=1''
configuration both the host system and the satellite's properties are
scaled to those expected at z=1 according to the model of
\citet{mo1998}. In this case, the aim is to simulate the merger event
that might have given rise to the Milky Way thick disk.  In this
configuration the mass of the present-day thick disk of the Milky Way
is roughly equal to the combined mass of the host disk galaxy and the
stellar component of the satellite.

In this section we also describe the orbital parameters of the various
experiments. Furthermore, we explain the choices made for the
numerical parameters employed in our simulations, and describe the
global stability of the system. 

\subsection{Main disk galaxy}

We model the main disk galaxy as a self-consistent two-component
system, containing a NFW dark matter halo \citep{navarro1997}
and a stellar disk.  The dark halo is adiabatically contracted in
response to the formation of a stellar disk in its central part
\citep{blumenthal1986,mo1998}. The disk component is constructed
following the procedure outlined by \citetalias{quinn1993} and
\citet{hernquist1993}, and follows a density profile of the form:
\begin{equation} \label{diskprof}
\rho_d(R,z) = \frac{M_d}{8\pi R_D^2 z_0}
\exp\left(-\frac{R}{R_D}\right)\ \textrm{sech}^2\left(\frac{z}{2
z_0}\right)
\end{equation}
where $M_d$ is the disk mass, $R_D$ is the exponential scale-length,
and $z_0$ is the exponential\footnote{Note that
$\textrm{sech}^2(z/2z_0) \approx \exp(-|z|/z_0)$ for $|z| \gg z_0$.}
scale-height. 

Following \citet{mo1998}, the ratios $M_{halo}/M_{disk}$ and
$R_{vir}/R_D$ have been kept nearly constant for all redshifts.  The
scale-height $z_0$ has been kept as 0.1$R_D$ according to observations
of external galaxies \citep{kregel2002}.  This has also been assumed
in our ``z=1'' experiments.  By keeping this ratios constant, the
way in which the mass and dimensions of the disk component scale with
redshift is simplified, since it follows the same scaling with
redshift as the halo within which it is embedded.  The chosen values for
the parameters of the main disk galaxy are listed in Table 
\ref{halo-disk-bulge-param} for our ``z=0'' and ``z=1'' experiments.

\begin{table} \scalefont{1.0}
\tabcolsep 2.8pt
 \caption{Properties of host disk galaxies.}
 \label{halo-disk-bulge-param}
 \begin{tabular}{@{}llll}
  \hline
                                    & ``z=0''                             & ``z=1''                         & \\
  \hline
  \hline
  NFW Halo                          &                                  &                               & \\
  \hline
  Virial mass, $M_{vir}$            & $10^{12}$                       & $5.07 \times 10^{11}$          & [$M_{\sun}$]\\
  Virial radius, $R_{vir}$          & 258.91                          & 122.22                         & [kpc]\\
  Concentration, $c$                & 13.12                           & 6.56                           & \\
  Circular velocity, $V_c(R_{vir})$ & 129.17                          & 133.87                         & [km/s]\\
  $N_H$                             & 500000                          & 500000                         & \\
  Softening, $\epsilon_{halo}$      & 0.35                            & 0.41                           & [kpc]\\
  \hline
  Disk                              &                                  &                                & \\
  \hline
  Disk mass, $M_{disk}$             & $2.8 \times 10^{10}$            & $1.2 \times 10^{10}$           & [$M_{\sun}$]\\
  Scale-length, $R_D$               & 3.5                             & 1.65                           & [kpc]\\
  Scale-height, $z_0$               & 0.35                            & 0.165                          & [kpc]\\
  Toomre Q($R=2.4R_D$)               & 2.0                              & 2.0                            & \\
  $N_D$                             & 100000                           & 100000                         & \\
  Softening, $\epsilon_{disk}$      & 0.05                            & 0.012                         & [kpc]\\
  \hline
  \hline
 \end{tabular}

\end{table}


\begin{table} \scalefont{1.0}
\tabcolsep 2.8pt
 \caption{Properties of satellite galaxies.}
 \label{halo-disk-sat-param}
 \begin{tabular}{@{}llll}
  \hline
                                    & ``z=0''                            & ``z=1''                         & \\
  \hline
  \hline
  NFW Halo                          &                                  &                               & \\
  \hline
  Virial mass, $M_{vir}$            & $2 \times 10^{11}$            & $1.01 \times 10^{11}$          & [$M_{\sun}$]\\
  Virial radius, $R_{vir}$          & 151.40                           & 71.35                          & [kpc]\\
  Concentration, $c$                & 16.18                            & 8.09                           & \\
  Circular velocity, $V_c(R_{vir})$ & 75.50                           & 78.10                        & [km/s]\\
  $N_H$                             & 100000                           & 100000                         & \\
  Softening, $\epsilon_{halo}$      & 0.14                            & 0.07/0.14$^1$                          & [kpc]\\
  \hline
  Disk                              &                                  &                                & \\
  \hline
  Disk mass, $M_{disk}$             & $5.6 \times 10^{9}$             & $2.4 \times 10^{9}$            & [$M_{\sun}$]\\
  Scale-length, $R_D$               & 1.69                            & 0.96                           & [kpc]\\
  Scale-height, $z_0$               & 0.17                            & 0.095                          & [kpc]\\
  Toomre Q($R=2.4R_D$)             & 2.0                              & 2.0                            & \\
  $N_D$                             & 100000                           & 100000                         & \\
  Softening, $\epsilon_{disk}$      & 0.024                           & 0.007                         & [kpc]\\
  \hline
  Bulge                             & & & \\
  \hline
  Bulge mass, $M_{bulge}$           & $5.6 \times 10^{9}$            &  $2.4 \times 10^{9}$          & [$M_{\sun}$]\\
  Scale radius, $a_b$               & 0.9                           &  0.709                      & [kpc]\\
  Velocity dispersion, $\sigma_0$ & 96.29                           &  69.06                        & [km/s]\\
  $N_B$                             & 100000                           &  100000                        & \\
  Softening, $\epsilon_{bulge}$     & 0.07                          &  0.07                         & [kpc]\\
  \hline
  \hline
 \end{tabular}

$^{1}$: \scriptsize{Softenings used in disky and spherical satellites respectively.}
\end{table}

\subsection{Satellite Galaxies}

The satellite galaxies are designed self-consistently with both dark
matter and stellar components.  The dark halo follows a NFW density
profile, and the initialisation procedure is identical to that for the
host system.

We consider two possible stellar distributions: an exponential disk
and a spherical Hernquist bulge. In this case, the density is given by
\citep{hernquist1990}:
\begin{equation}
\rho_b(r) = \frac{M_b}{2\pi} \frac{a_b}{r(r+a_b)^3}
\label{bulgeprof}
\end{equation}
where $M_b$ is the bulge mass and $a_b$ is the scale radius. 

Both types of satellites satisfy $M_{total,sat}=0.2 M_{total,host}$
for the ``z=0'' and ``z=1'' experiments. In particular, the mass ratio
between dark matter and luminous matter in the satellites is set to be
the same as in the host galaxy. This implies that the mass of the
stellar components of our satellites is 20\% of $M_{disk,host}$,
similar to those adopted in previous work on disk heating by satellite
accretion \citepalias{quinn1993,walker1996,huang1997,velazquez1999}.
Note however that the initial total mass of the satellite is
comparable to that of the host disk.  For completeness, we also
include the case of a satellite with a stellar mass of 10\% of
$M_{disk,host}$.  In this case the total mass of the satellite is also
10\% of that of the host galaxy.

The spherical satellites lie on the observed fundamental plane of
dE+dSphs galaxies \citep{derijcke2005}:
\begin{equation}
\log L_B \sim 4.39 + 2.55 \log \sigma_0
\label{l-sig0}
\end{equation}
\begin{equation}
\log L_B \sim 8.65 + 3.55 \log R_e
\label{l-re}
\end{equation}
where $L_B$ is the blue-band luminosity, and $\sigma_0$ is the central
velocity dispersion. The effective radius is related to the scale
radius by \mbox{$R_e \approx 1.82 a_b$} for a Hernquist density
profile.  Finally, we fix the mass-to-light ratio $\Upsilon_B = 2
\Upsilon_{B_\odot}$ to derive the stellar mass of our spherical
satellites.

The structural parameters of the disky satellites, $R_D$ and $z_0$ are
set in the same way as for the host. 

The above relations fully specify the properties of our satellites,
both for the ``z=0'' and ``z=1'' experiments.  Note that the
variation with redshift is always linked to that of the host halo.
Table \ref{halo-disk-sat-param} lists the properties of both 
spherical and disky satellites for ``z=0'' and ``z=1''.

\subsection{Orbital Parameters}

We release our satellites from significantly larger distances than
previous works (e.g., \citetalias{quinn1986};
\citetalias{velazquez1999}; \citealt{kazantzidis2007}).  For the
``z=0'' case, the satellite is launched from the virial radius of the
host galaxy computed at z=1 ($R_{vir}=122.22$ kpc $\approx$ $35R_D$).
For the ``z=1'' experiment, it is launched from a distance of $83.9$
kpc ($\approx$ 50$R_D$) which corresponds to the virial radius of the
host galaxy computed at z=1.6.

To initialise the orbital velocities of the satellites, we follow
\citet{benson2005}. Benson determined the orbital parameters of DM
substructures at the time they crossed the virial radius 
of their host halos 
\citep[see also][]{tormen1997,khochfar2006}.
We choose for the velocities of our satellites the most probable values
of the distributions as given for z=0 (since little variation is
visible as a function of redshift). The radial and tangential velocity
distributions peak, respectively, at 0.9 and 0.6 in units of $V_c(R_{vir})$.

We consider for our satellites initial orbital inclinations of
0$\degr$, 30$\degr$ and 60$\degr$ (with respect to the plane of the
host disk), in both prograde and retrograde directions with respect
the rotation of the host disk. In the case of disky satellites, their
midplanes are parallel to that of the primary disk, implying that they
are, respectively, inclined by 0$\degr$, 60$\degr$ and 30$\degr$ with
respect to their orbital planes.  Table \ref{orb-param} summarises the
orbital parameters adopted for both spherical and disky satellites,
for the configurations at ``z=0'', and ``z=1''.

Note that within each configuration, all the satellites are initially
released from the same distance with the same velocity modulus, i.e.,
they all have the same total energy and modulus of the angular
momentum, but the latter varying its orientation. This implies that the
initial apocentre, pericentre (and hence orbital eccentricity) are the
same for all experiments. However, as we shall see in
\S \ref{sec:orbit-evol} the orbits evolve due to dynamical friction,
with some dependence on the internal properties of the
satellite. Therefore, at the time the satellite merges with the disk,
the orbits of all our experiments are different.

With these initial conditions, we find that our satellites have completely 
merged with the host disks by z$\approx$0.4 and by z$\approx$1, 
respectively for the ``z=0'' and ``z=1'' cases.

\begin{table}
 \caption{Initial orbital parameters of satellite galaxies.}
 \label{orb-param}
 \begin{tabular}{@{}cccccccc}
  \hline
  $i$  & $x$ & $z$ & $v_x$ & $v_z$ 
               & L$_z$ \\
  \hline
  ``z=0'' & & & & & \\
  \hline
  $0\degr$  & 122.2 & 0.0 & -137.7 & 0.0 & 11219.8 \\
  $30\degr$ & 105.8 & 61.1 & -119.2 & -68.8 & 9716.6 \\
  $60\degr$ & 61.1 & 105.8 & -68.8 & -119.2 & 5609.9 \\
  \hline
  ``z=1'' & & & & & \\
  \hline
  $0\degr$  & 83.9 & 0.0 & -118.2 & 0.0 & 6615.9 \\
  $30\degr$ & 72.7 & 42.0 & -102.4 & -59.1 & 5729.5 \\
  $60\degr$ & 42.0 & 72.7 & -59.1 & -102.4 & 3307.9 \\
  \hline
  \hline
 \end{tabular}

 \medskip
 NOTES:\\
 - Distances in kpc, velocities in km/s, angular momentum in kpc$\times$km/s.\\
 - In all cases, $y =0$ kpc, and $v_y = 91.8$ km/s for ``z=0'' 
and $v_y = 78.8$ km/s for ``z=1''.\\
 - Listed $v_y$ and L$_z$ are for prograde orbits. Retrograde orbits have the opposite sign.\\
 - Initially $r_{apo} = 77$ kpc and 49 kpc, $r_{peri} \sim 10$ kpc and 5 kpc, respectively for
 the ``z=0'' and ``z=1'' configurations (as measured from the first apocentre and the subsequent pericentre). 
The corresponding eccentricities $e = (r_{apo} - r_{peri})/(r_{apo} + r_{peri})$ are 0.77 and 0.82.\\

\end{table}

\subsection{Numerical Parameters}

The $N$-body systems are evolved using \mbox{Gadget-2.0}
\citep{springel2005} a well documented massively parallel TreeSPH
code.  Our choices of the numerical parameters (number of particles
$N$ in each component in the system; softening $\epsilon$ and timestep
$\Delta t$) are described in detail in \S \ref{app:numerics}.  Tables
\ref{halo-disk-bulge-param} and \ref{halo-disk-sat-param} list the
values used for each component in our simulations. The maximum
timestep (not listed in the Tables) is $0.25$ Myr. Typically the
energy and angular momentum are conserved to better than 0.1\% over 9
Gyr of evolution for our main disk galaxy configured at ``z=0''.

\subsection{Evolution of Isolated Host Galaxy}

Before including the satellite, the host galaxy is simulated in
isolation to test its stability in the absence of any external
perturbation. As we show in \S \ref{app:isolation} our host galaxies
are very stable in their properties \citetext{cf.,
\citetalias{velazquez1999}; \citealt{gauthier2006}} for the amount of
time needed to complete the experiments.  Therefore, we are now ready
to focus on how these systems evolve when they suffer a minor merger.

\section{Results}

In total, 25 simulations have been carried out to study the formation
and global properties of thick disks as a result of the merger between
a host disk galaxy and a satellite.  The simulations explore
combinations of the following elements: two configurations for the
progenitors (``z=0'' and ``z=1''); two morphologies for the stellar
component of the satellite (spherical and disky); two total mass
ratios between the satellite and the host galaxy (10\% and 20\%); and
three initial orbital inclinations for the satellite with respect to
the midplane of the host disk (0$\degr$, 30$\degr$ and 60$\degr$), in
both prograde and retrograde directions.

\begin{figure}
\begin{center}
\includegraphics[width=41.7mm]{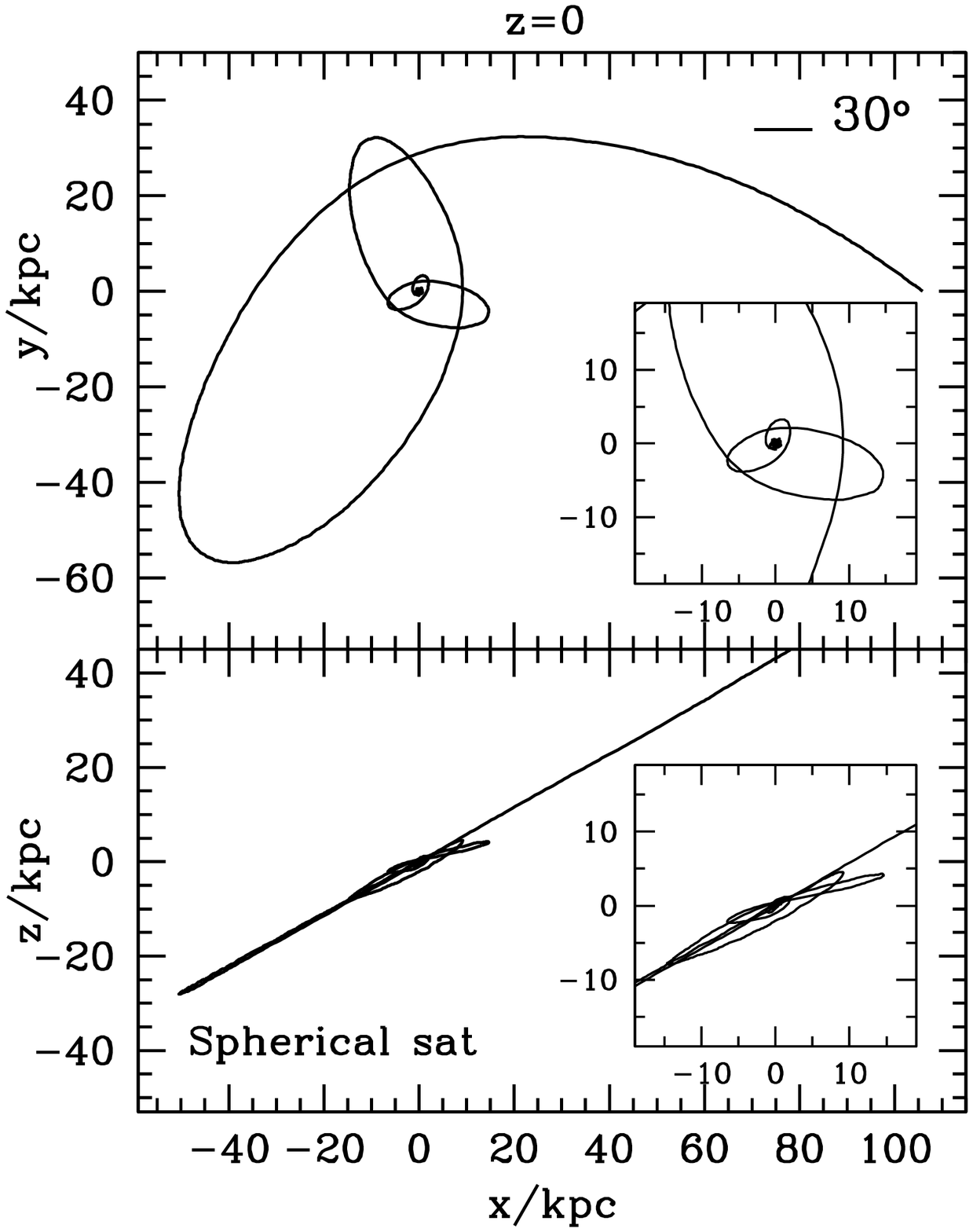}
\includegraphics[width=41.7mm]{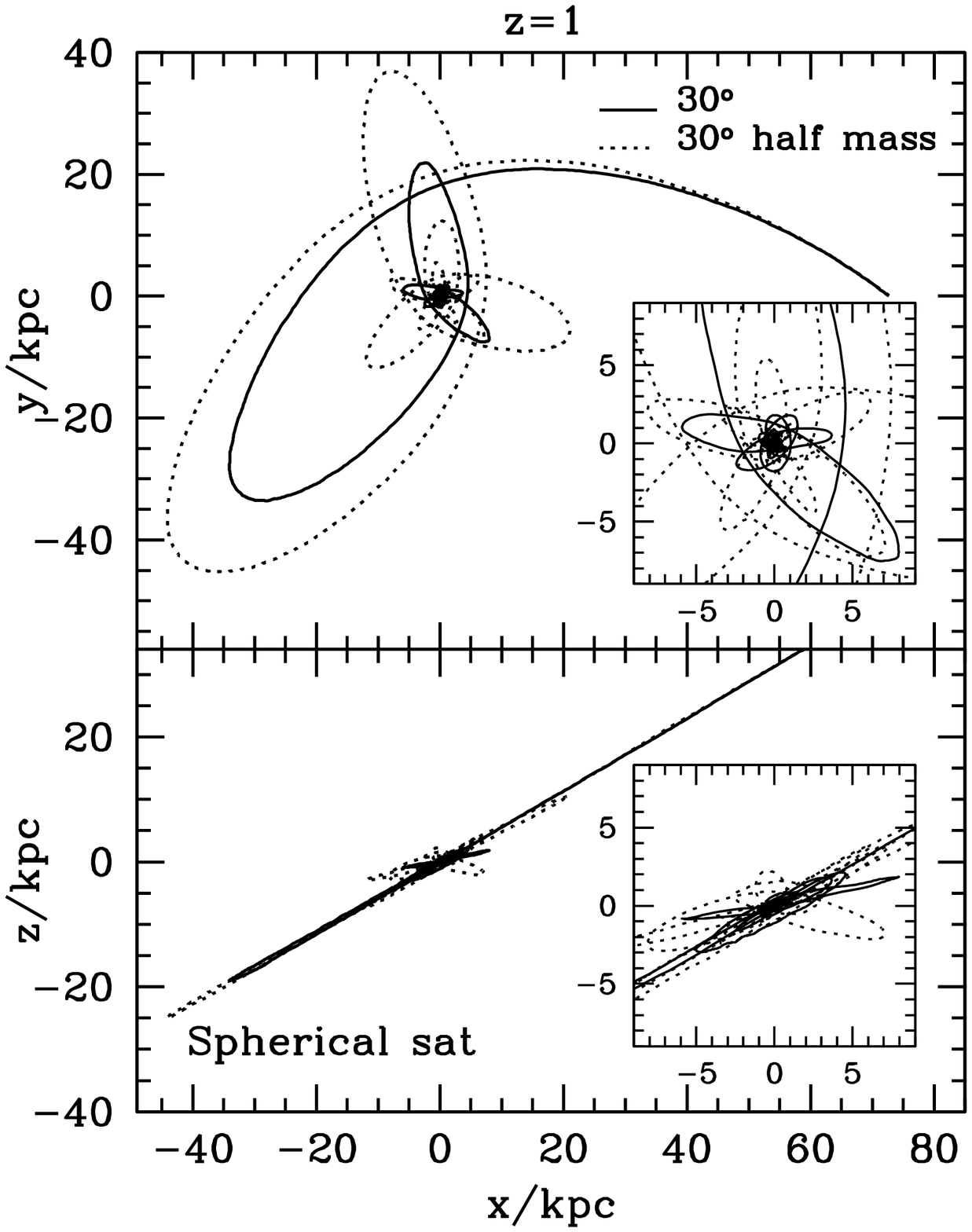}
\end{center}
\caption{XY and XZ projections of the trajectory of the centre of mass
of a spherical satellite as it decays towards the centre of the host
galaxy in our ``z=0'' and ``z=1'' experiments with initial inclination 30$\degr$.  
The coordinate system is centred on the centre of mass of the host disk.  
The inset panels show in more detail the trajectory of the satellite at late times
before it is fully disrupted.}
\label{orbit.sat}
\end{figure}

\begin{figure}
\includegraphics[width=85mm]{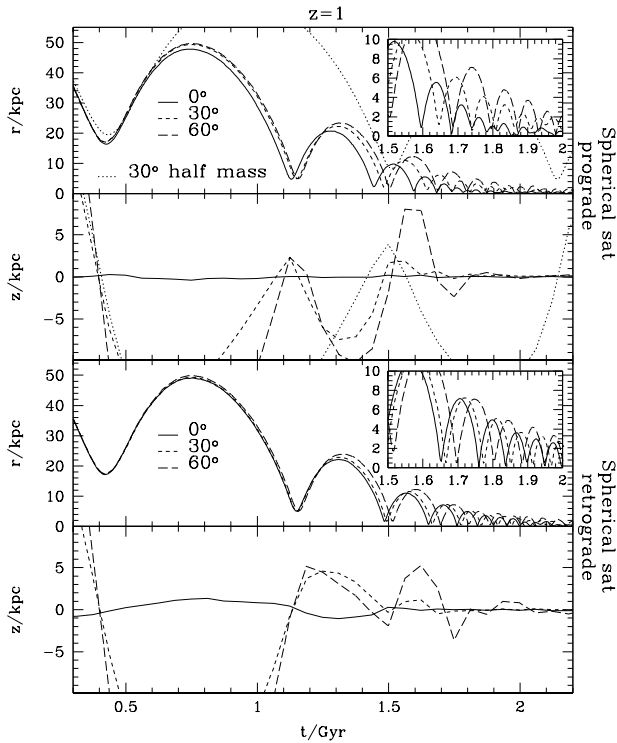}\\
\caption{Evolution of the radial separation between the centres of
mass of the host disk and the satellite, and of the $z$-distance of
the satellite's centre of mass with respect to the host disk plane,
for spherical satellites on prograde (top panels) and retrograde
(bottom panels) encounters.  The various curves correspond to different
orbital inclinations for the ``z=1'' configuration, and the 10\%
experiment is shown for the inclination of 30$\degr$.  The insets show
in more detail the trajectory of the satellites at late times.  The
lack of a clear trend in the amplitude of the vertical oscillations is
likely due to the difficulty in determining the exact orientation of
the disk plane at the $\sim 1$ kpc level. Similar behaviours are
observable for the ``z=0'' case.}
\label{orbit.r.sat}
\end{figure}

\subsection{Orbital Evolution of the Satellites}
\label{sec:orbit-evol}

To study the evolution of the orbits in our experiments we follow the
location of the centre of mass of the satellite (CM) with respect to
the host galaxy. The CM is defined as the mean position of bound
stellar and DM particles of the satellite.  Fig. \ref{orbit.sat} shows
the trajectories of the CMs for the experiments configured at ``z=0''
and ``z=1'' with a spherical satellite on a prograde orbit with
initial inclination $i = 30\degr$.  The XY projections clearly
illustrate how radial the orbits are and how rapidly they decrease in
amplitude due to dynamical friction between the satellite and the host
system.  Note that these orbits are quite different from the circular
ones usually used in earlier studies
\citepalias[e.g.,][]{quinn1986,quinn1993,walker1996}.  The XY and XZ
projections show that the satellites stay on their original orbital
planes as they decay, until they enter into the zone dominated by the
host disk, when they are drawn onto the disk plane and spiral in
towards its centre.  This is particularly clear for the lighter
satellite in our ``z=1'' experiment that spends $\sim$1 Gyr on the
disk plane before sinking in further.

Note that most of the angular momentum of the ``satellite + host
system'' is in the orbital motion of the satellites.  This is because
the satellites are relatively massive and have a very large initial
distance from the host. As a result, when the satellite decays in an
inclined orbit the disk is strongly tilted in both the prograde and
retrograde cases (see \S 3.3).

As expected, the trajectory of the lighter satellite is more extended  
because dynamical friction is less efficient in this case 
\citep{binney1987}.
Our 20\% satellites decay completely after $\sim$3 Gyr in the case of
``z=0'' experiments, and after $\sim$2 Gyr for ``z=1''
experiments. In comparison, the satellite with half of the mass takes
the double of time to sink starting from the same initial orbital
parameters. 

Initially, the orbital decay is due to the dynamical friction against
the host halo. This implies that the decay rate does not depend on
inclination, orbital direction or stellar mass distribution (see
Fig. \ref{orbit.r.sat}).  On the other hand, when the satellites
approach the centre of the system (where the disk is a significant
contributor to the global force field), the orbit decays by dynamical
friction also against the host disk.  Thus at later times, prograde low
inclination orbits decay faster than retrograde or high inclination
orbits \citetext{\citetalias{quinn1986}; \citetalias{walker1996}; 
see also \citetalias{huang1997}}.

\subsection{Satellite Mass Loss}

\begin{figure}
\begin{center}
\includegraphics[width=84mm]{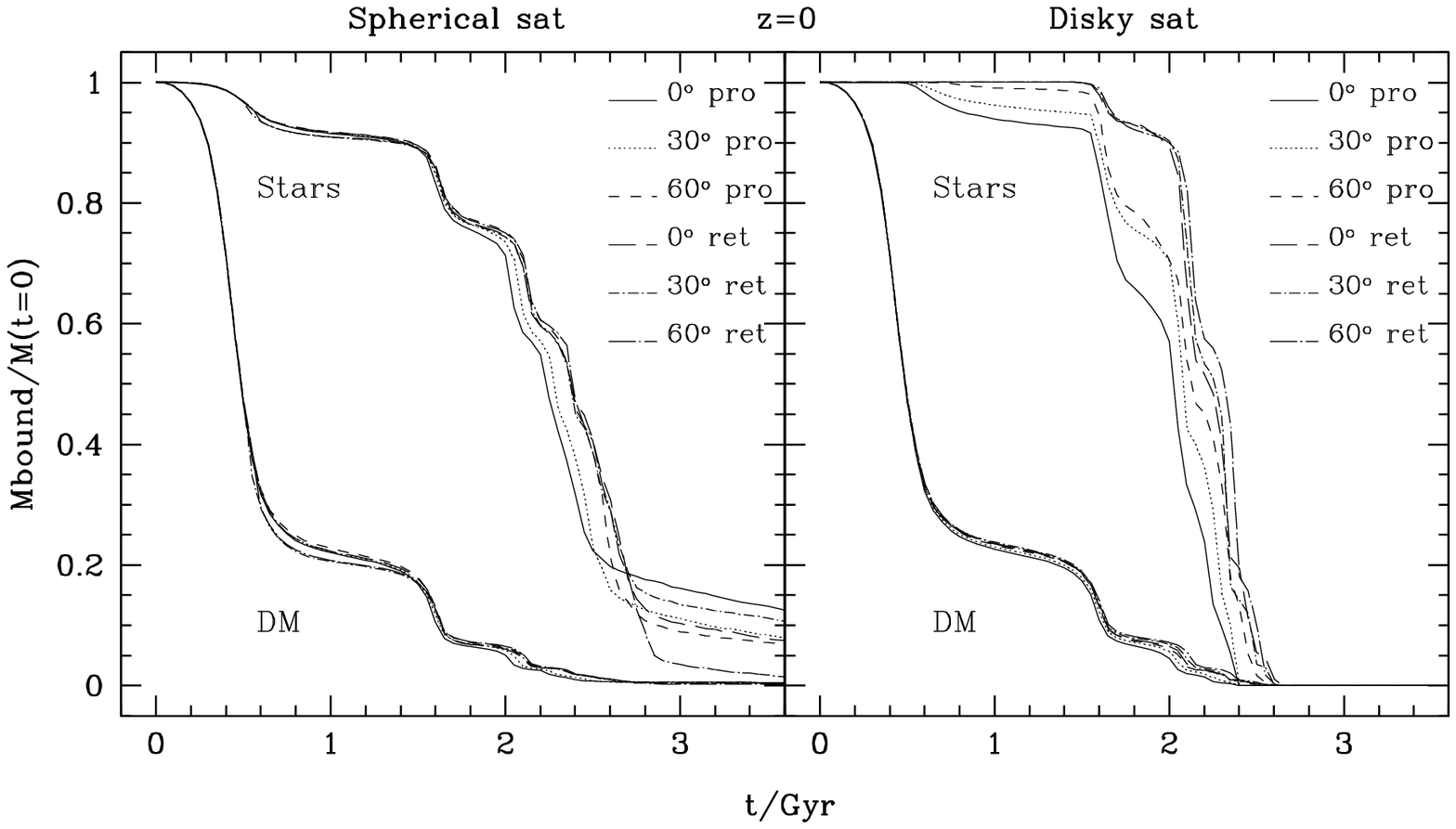}
\includegraphics[width=84mm]{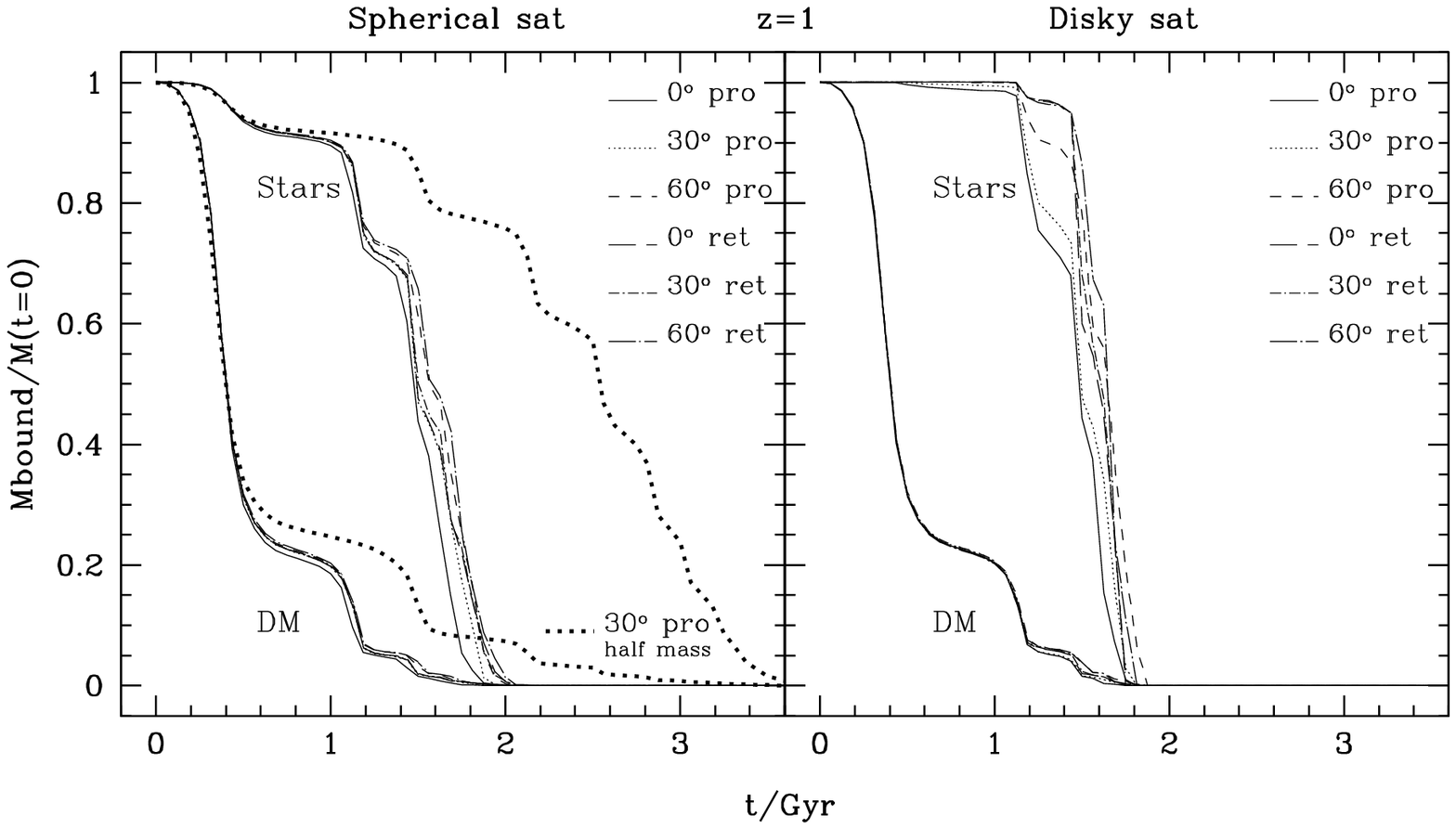}
\end{center}
\caption{ Mass that remains bound for both the dark matter and stellar
components of the satellites in our experiments at ``z=0'' (upper
panels) and at ``z=1'' (bottom panels).  The bound mass is in units of
the initial mass of the corresponding component. }
\label{massloss.spher.disky.z0}
\end{figure}

Fig. \ref{massloss.spher.disky.z0} shows the mass loss evolution of
both DM and stellar components of spherical and disky satellites in
the ``z=0'' and ``z=1'' experiments.

In order to calculate how much mass remains bound to the satellite at
a given time we implemented in Gadget-2.0 the following procedure
\citep{benson2004}:
\begin{enumerate}
\item
Start by considering all the satellite particles that were bound to
the satellite at the previous timestep (or simply all satellite
particles for the first timestep).
\item
Compute the mass of the satellite from these particles along with
the position and velocity of the CM.
\item
For each particle in this set, determine whether it is
gravitationally bound to the other particles in the set.
\item
Retain only those particles that are bound and go back to step (ii).
Repeat until the mass of the satellite has converged.
\end{enumerate}

Since the satellites were initialised in the absence of an external
potential, as soon as they are placed within the host potential a
large fraction (70\%) of the more extended DM component rapidly
becomes unbound before the first pericentric passage.  After that, the
mass loss rate of the DM component mostly depends on its initial mass.

The mass loss rate of the stellar components depends strongly on
initial mass, and orbital parameters \citepalias{walker1996,huang1997,
velazquez1999} but also on the stellar mass distribution.  As
satellites decay, prograde orbits with lower inclinations lose mass
faster than retrograde orbits with higher inclinations, due to the
stronger tidal interaction with the host galaxy.  This trend is more
notorious for disky satellites.  Spherical satellites are also
characterised by a more extended ``knee'' in the mass loss in
comparison to disky satellites.  The lighter satellite experiences a
slower mass loss compared to heavier satellites, because it suffers
less dynamical friction and hence is on a less bound orbit.

Once a satellite has sunk onto the plane of the host disk, its fate
will depend on its instantaneous mean density compared to the mean
density of the host at a given location.  If the mean density of the
satellite is larger than that of the host system then the most bound
particles of the satellite will reach the galactic centre as a
distinctive core causing more damage to the host disk.  Otherwise, the
satellite will receive most of the damage, being heated and torn apart
by the host disk.  In our simulations only the mergers with spherical
satellites in the ``z=0'' experiments deposit in the galactic centre
final cores of up to 20\% the initial stellar mass.  These final cores
are on the lighter side in comparison with previous studies using
high density satellites (\citetalias{quinn1993}, 20\%;
\citetalias{walker1996}, 45\%).  It is interesting to note that
spherical satellites with higher inclinations give rise to the
formation of less massive cores.  This can be explained by the fact
that satellites on higher inclinations experience more disk crossings
through the host disk as they decay, compared to ones on lower
inclinations.  In this case disk shocks perturb the structure of the
satellite and cause additional mass loss \citetext{see
\citealt{binney1987}; \citetalias{quinn1993}}.

Figure \ref{massloss.spher.disky.z0} shows that in general most of the
dark matter is stripped off early, and deposited at very large radii
(see Fig.~\ref{orbit.r.sat}). Therefore the fraction of dark matter
accreted from the satellite and deposited in a disk-like structure is
very small in our simulations, in comparison to what \citet{read2008}
find. This may be explained
by the fact that our satellites, although of comparable mass, are
launched from much larger distances.

\begin{figure*}
\begin{center}
\includegraphics[width=88mm]{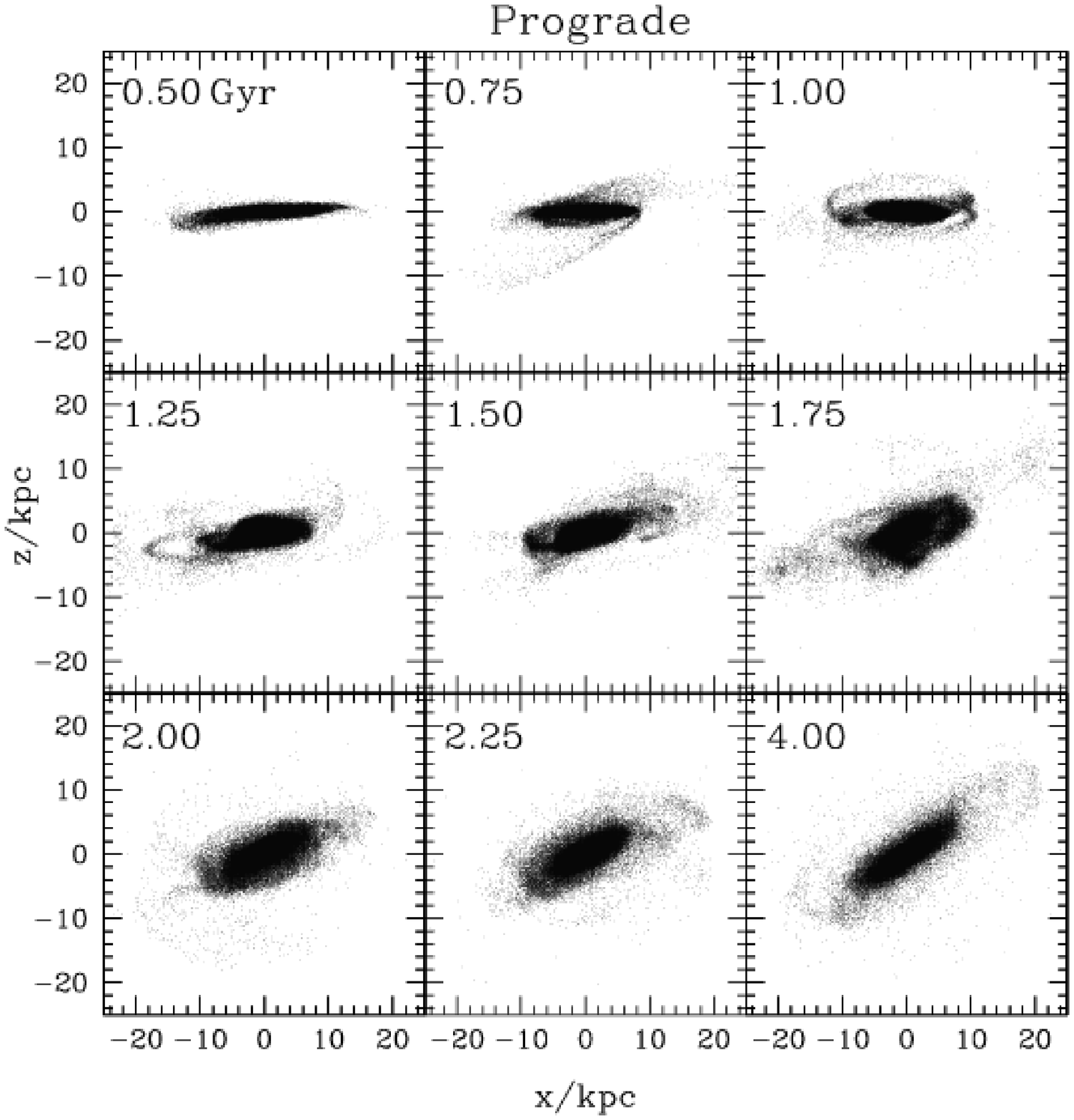}
\includegraphics[width=88mm]{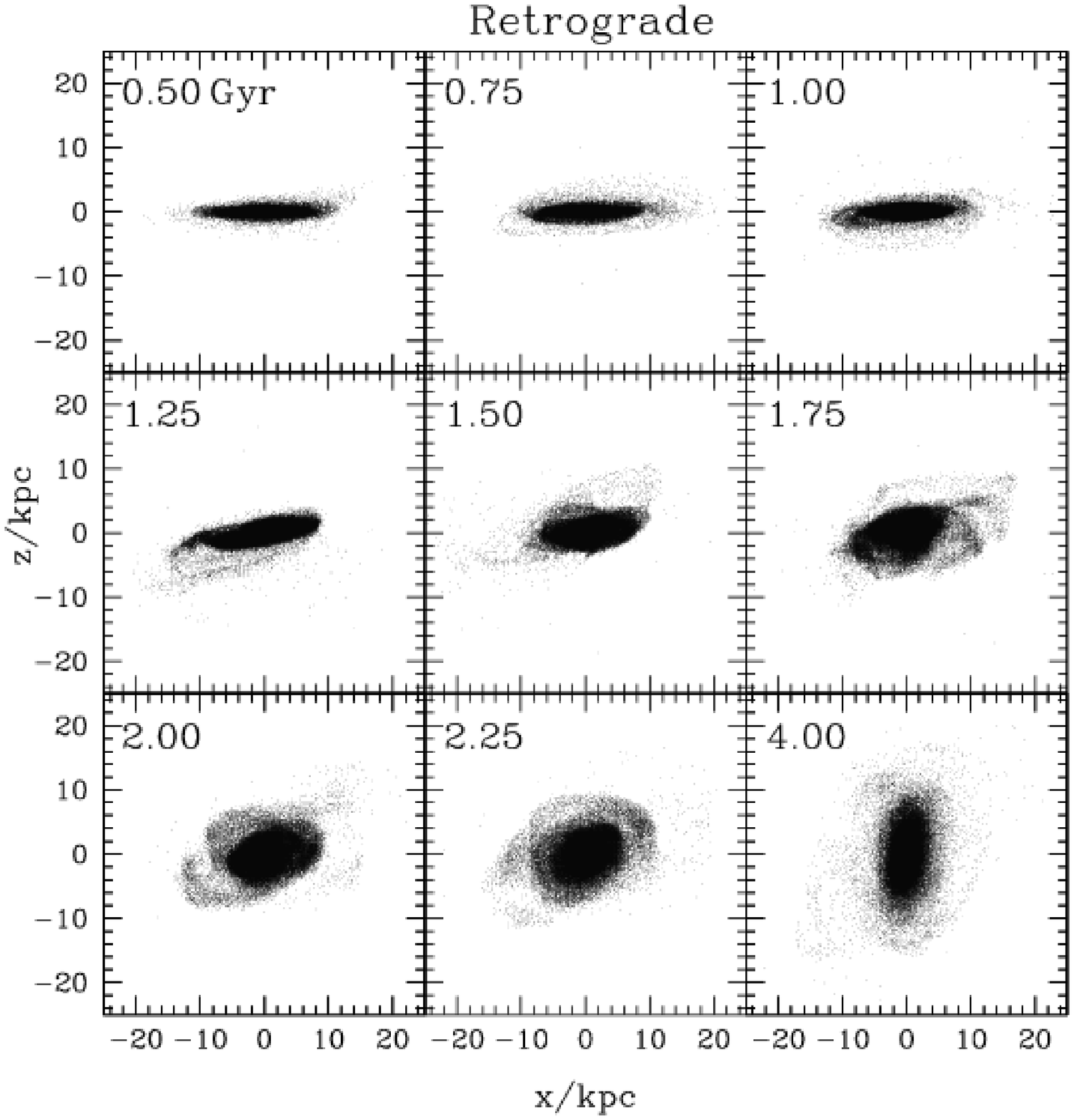}\\
\includegraphics[width=88mm]{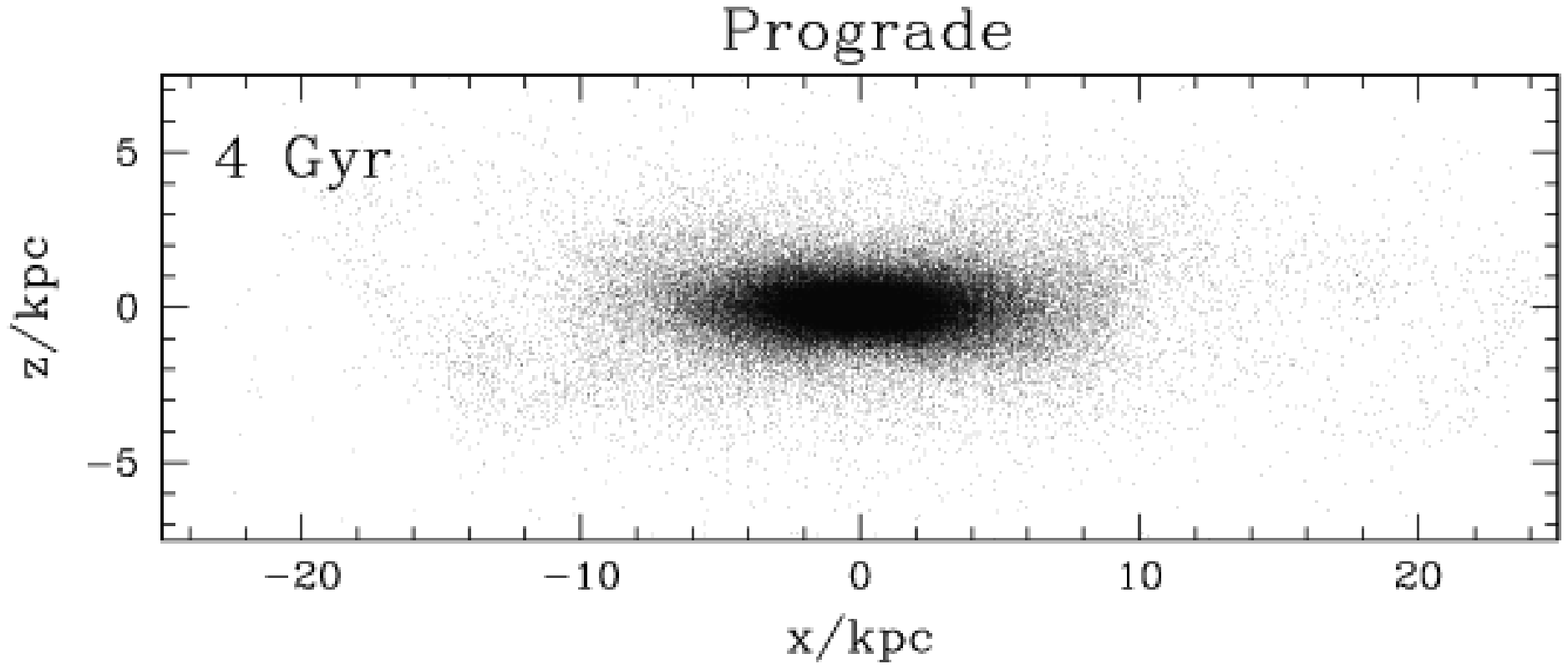}
\includegraphics[width=88mm]{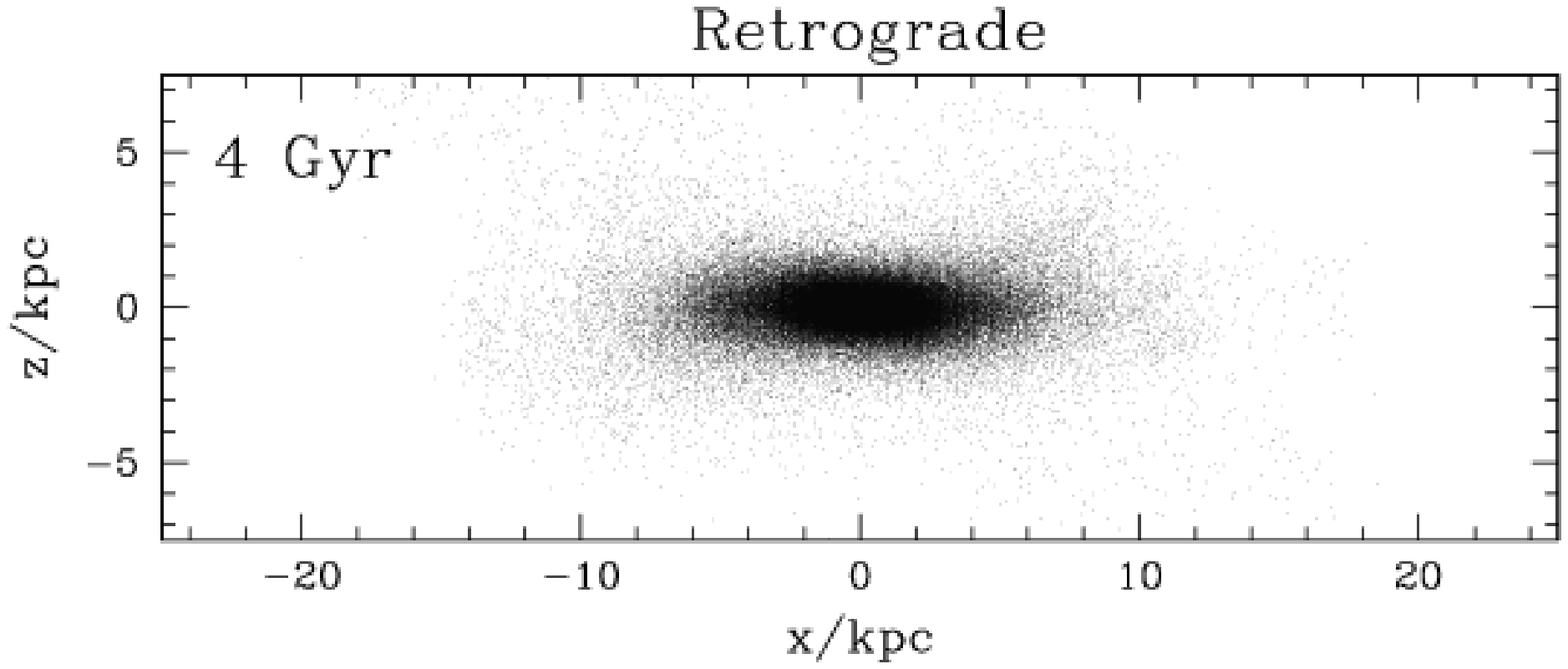}
\caption{ Evolution of the host disk (initially shown edge-on) during
the merger with a spherical satellite with inclination of 30$\degr$
for prograde and retrograde orbits in the ``z=1'' experiment.  Only
host disk particles are shown for clarity.  Note that the disk appears
over thickened or distorted because of projection effects, especially
in the case of the retrograde orbit. The bottom panels show edge-on
views of the final system.  }
\label{snaps.edgeon.proret.diski30}
\end{center}
\end{figure*}

\begin{figure*}
\begin{center}
\includegraphics[width=88mm]{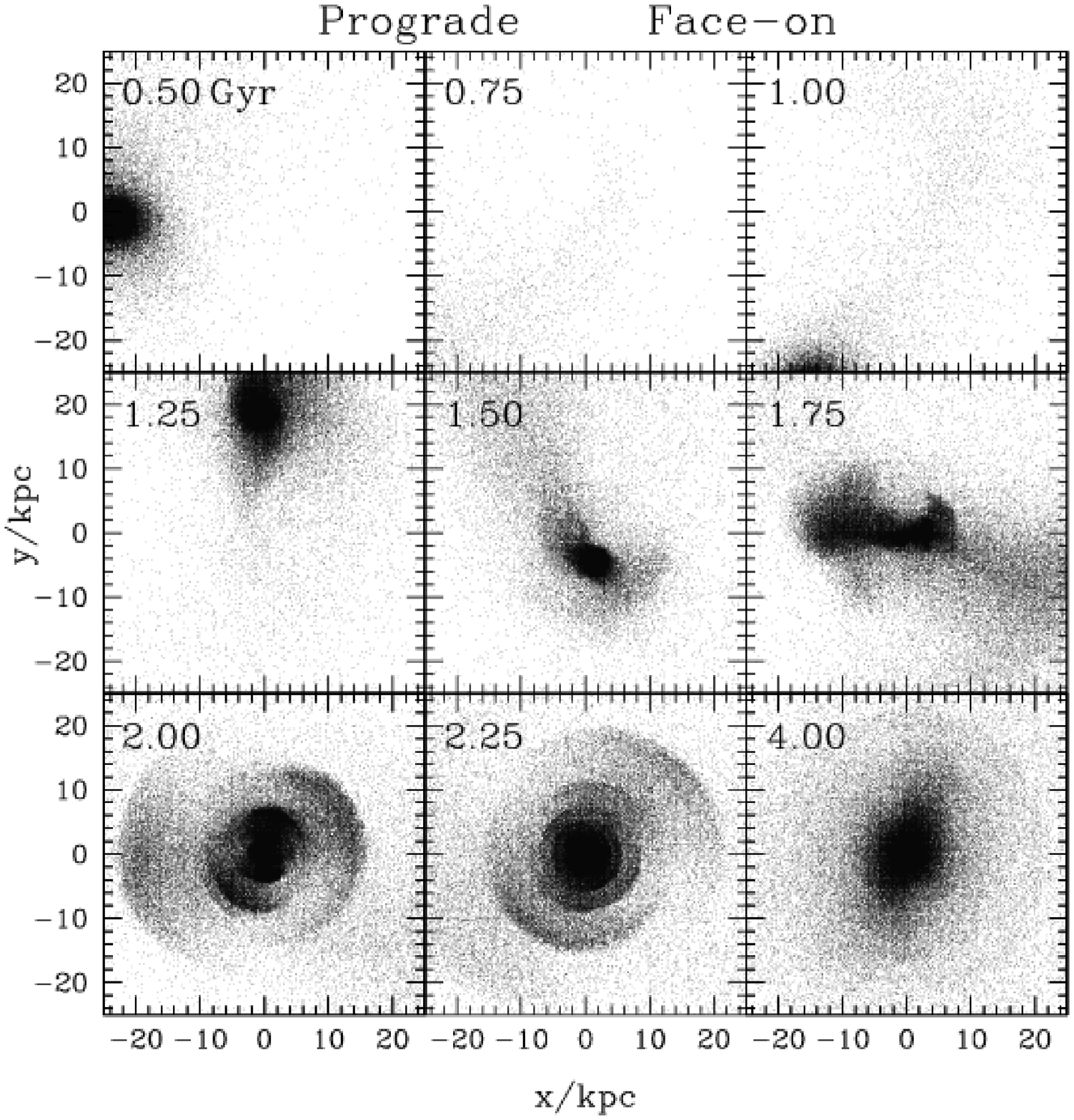}
\includegraphics[width=88mm]{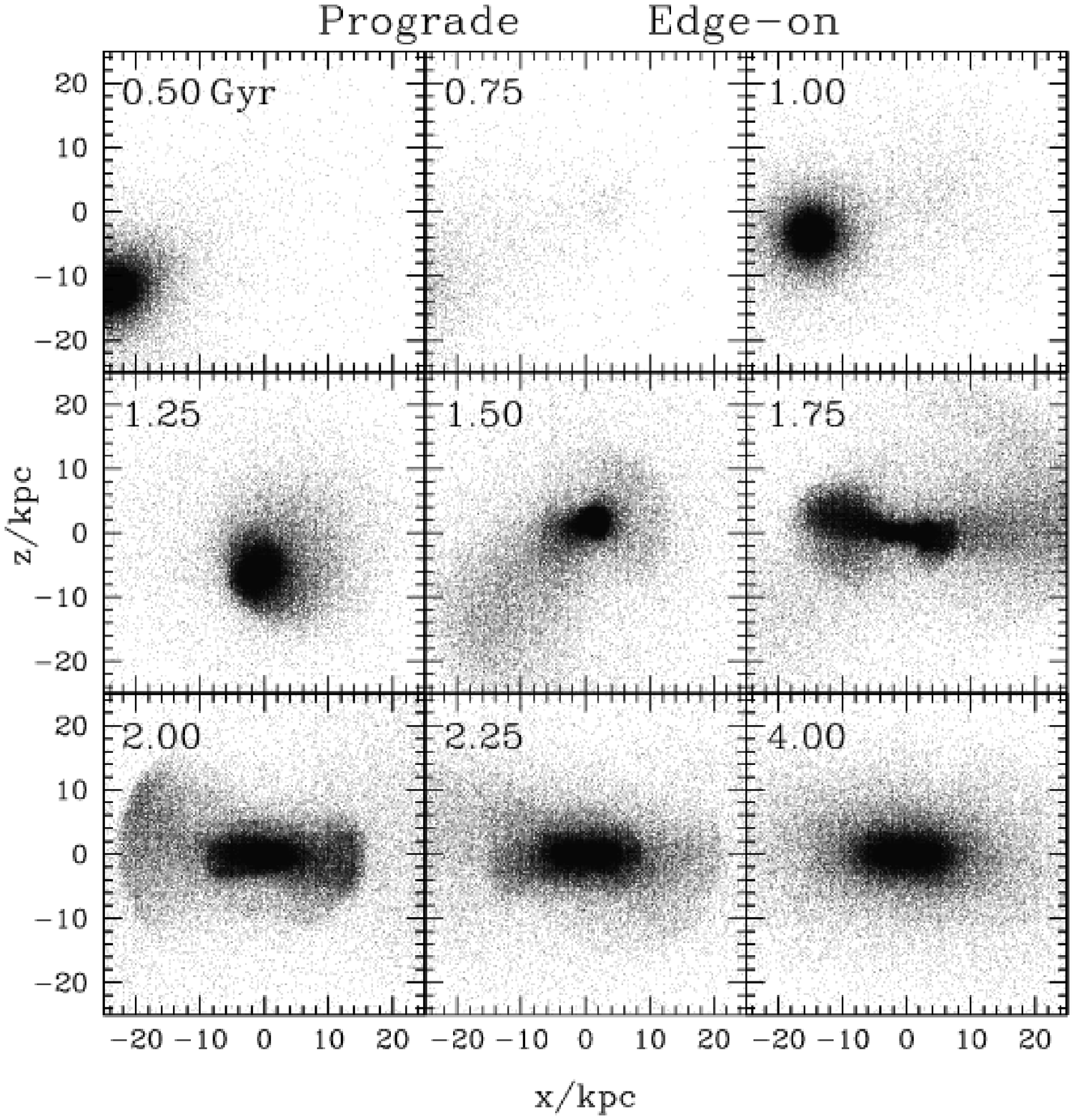}
\caption{ Evolution of the spherical satellite with inclination
30$\degr$ during the prograde merger with the host galaxy in the
``z=1'' experiment. In each snapshot the reference frame has been
centred on the centre of mass of the host disk and has also been
rotated to eliminate the tilting of the whole system with respect to
the original frame. The labels \emph{face-on} and \emph{edge-on} are
relative to the host disk (not shown here for clarity). 
}
\label{snaps.faceedgeon.proret.sati30}
\end{center}
\end{figure*}
\begin{figure*}
\begin{center}
\includegraphics[width=88mm]{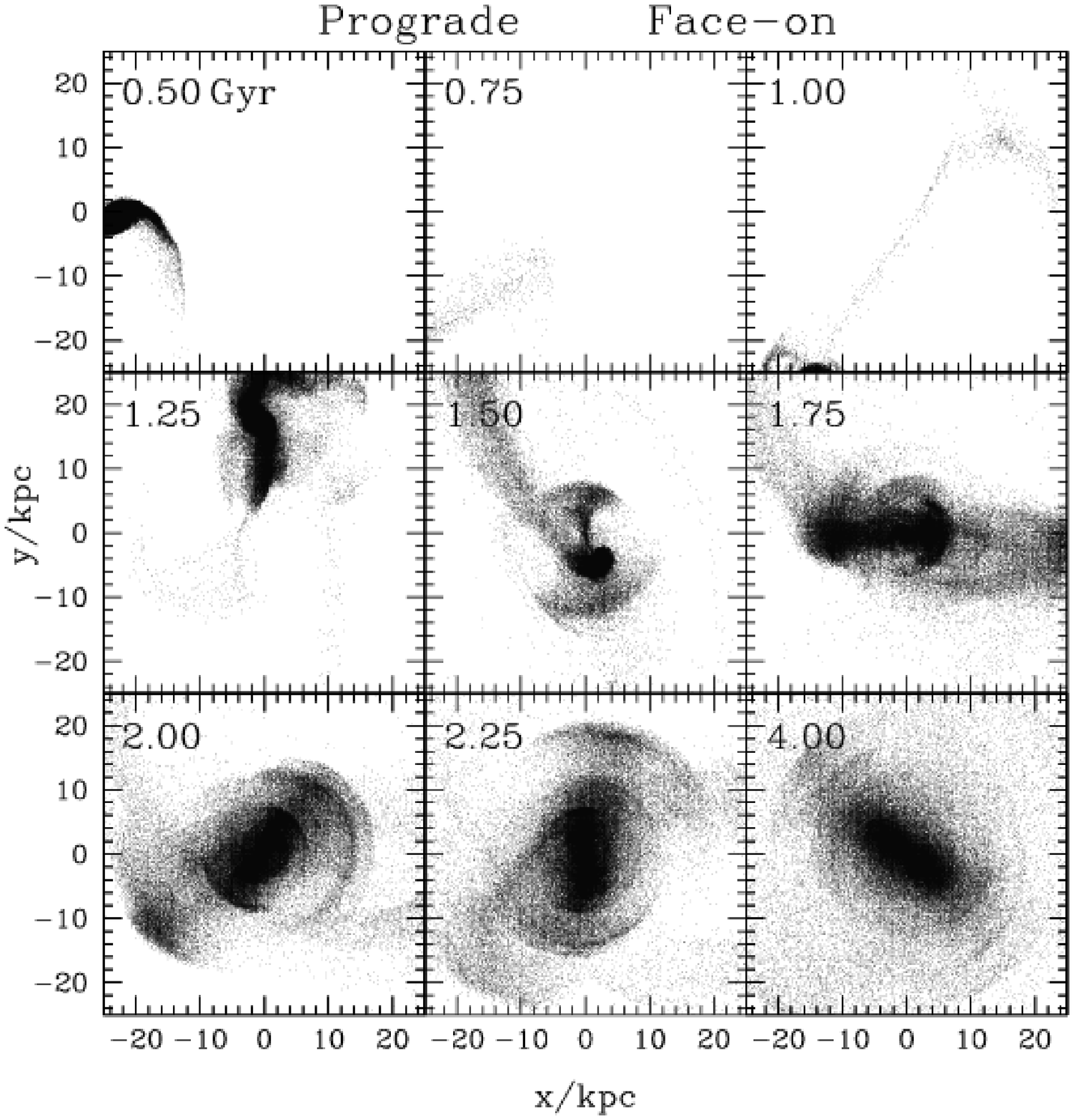}
\includegraphics[width=88mm]{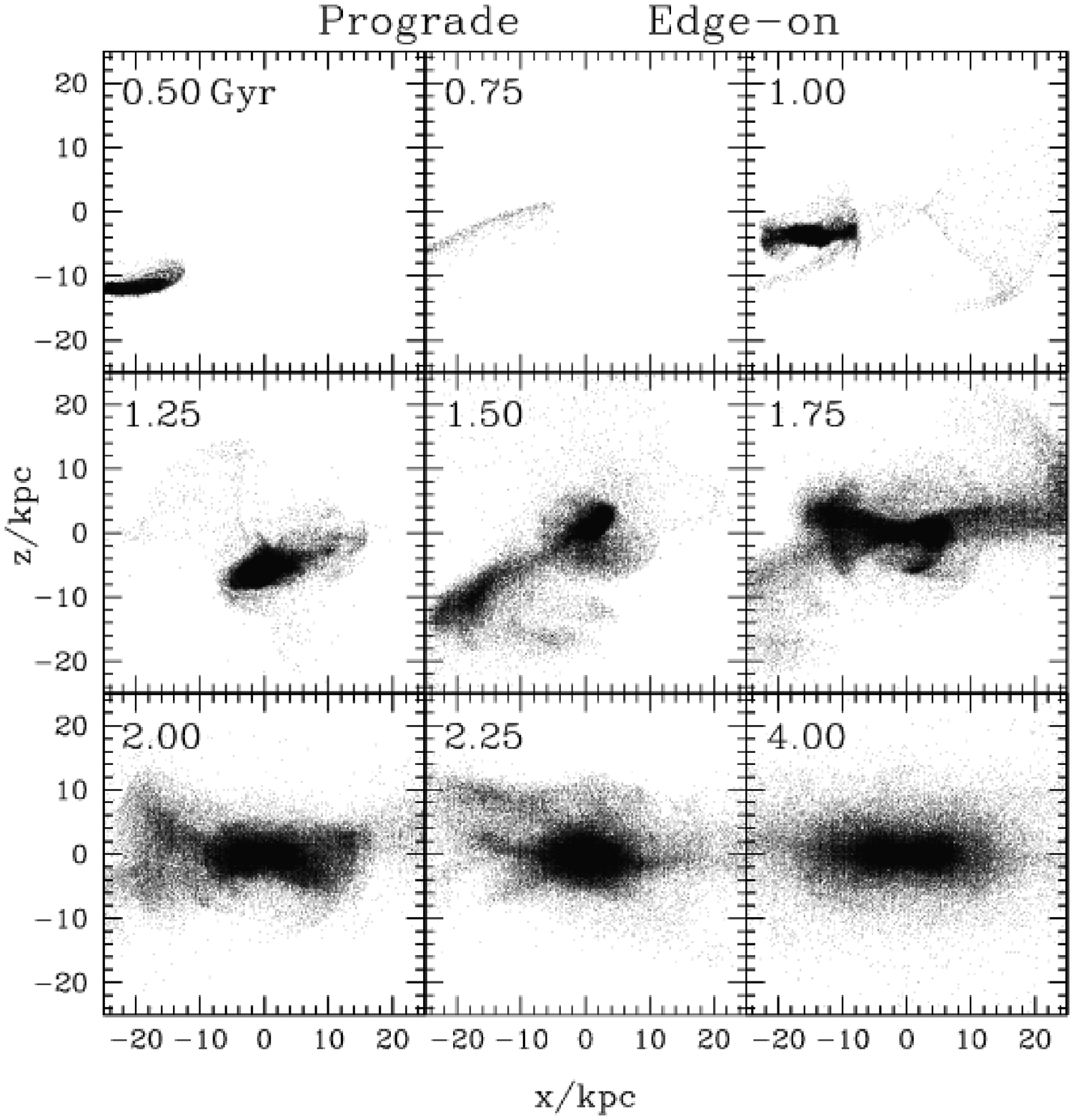}
\caption{ Same as Fig. \ref{snaps.faceedgeon.proret.sati30}, for the
case of the disky satellite.  }
\label{snaps.faceedgeon.proret.diskysati30}
\end{center}
\end{figure*}

\begin{figure*}
\begin{center}
\includegraphics[width=95mm]{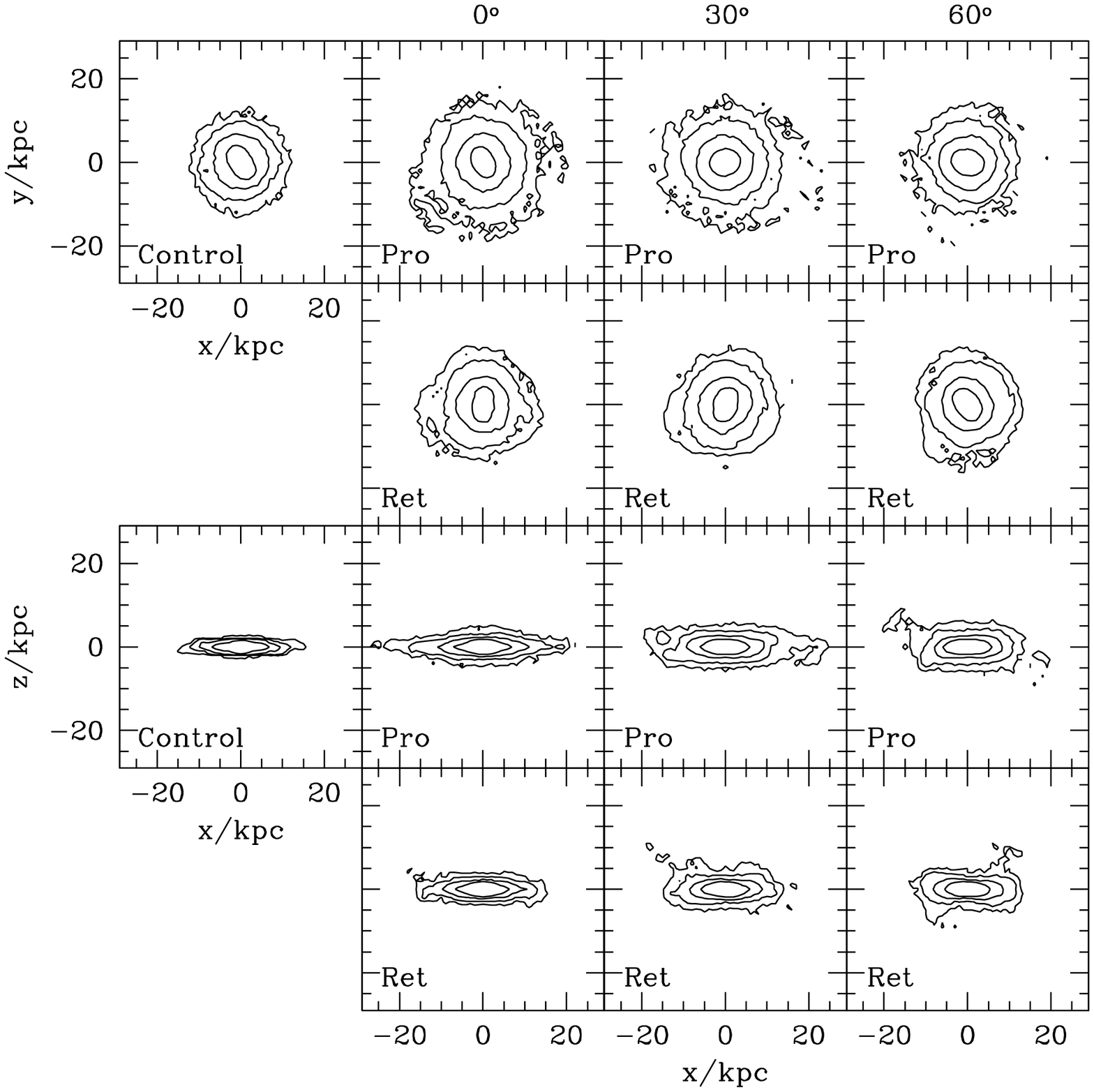}
\includegraphics[width=75mm]{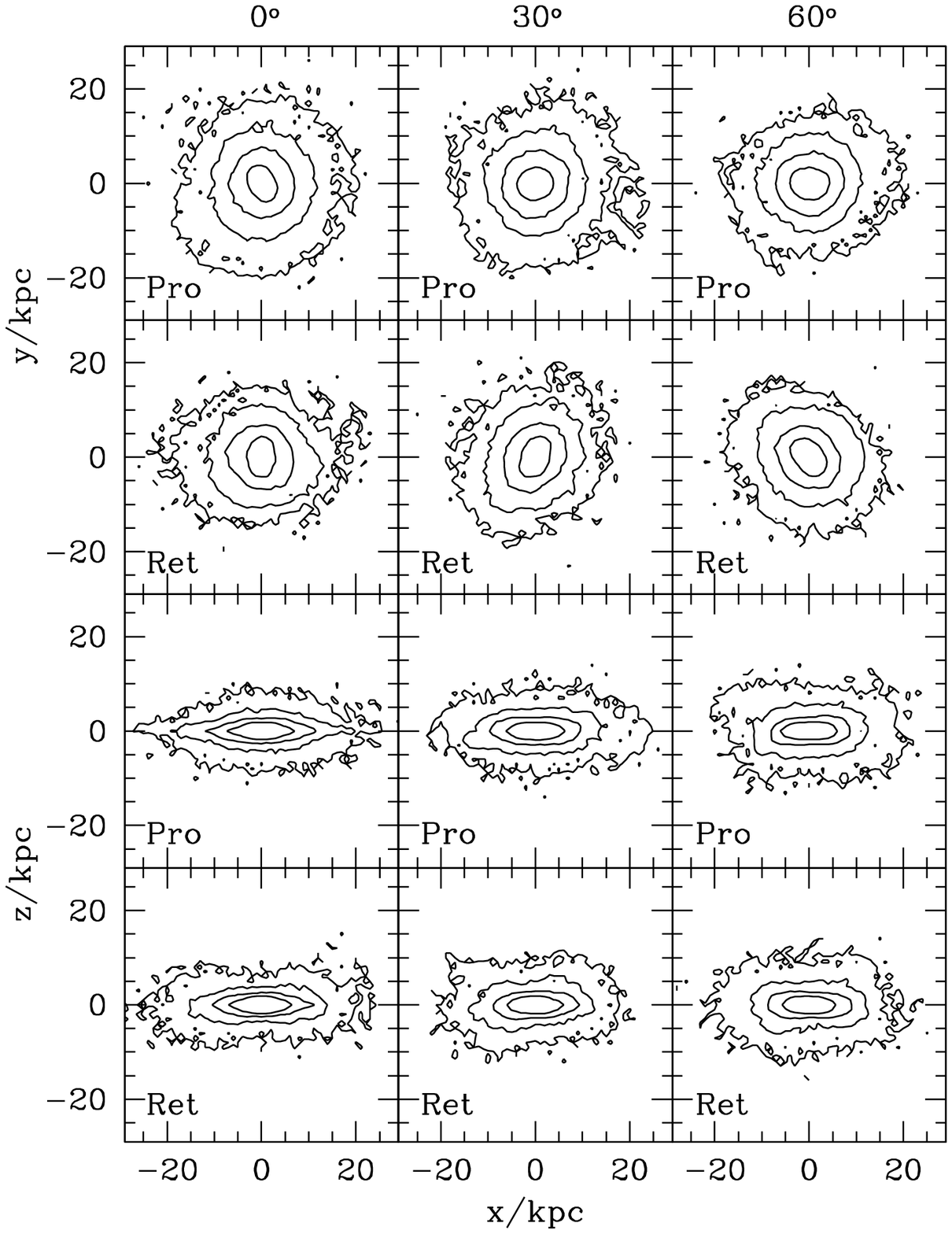}
\caption{ Face-on and edge-on views of the final morphologies of
heated disks (left) and thick disks (right) at the end of the
simulations in ``z=1'' experiments, 4 Gyr after the infall of the
spherical satellite.  In each case the tilting induced by the
satellites has been eliminated to facilitate the comparison with the
coeval control disk. The contours correspond to 4.5, 6.2, 8 and 9.7
magnitudes below the central surface brightness of the remnant system.
For the experiments shown here, and assuming a mass-to-light ratio
$\Upsilon_V = 2 \Upsilon_{\odot,V}$, these would be located at 22.5
mag/arcsec$^2$ (innermost), 24.2 mag/arcsec$^2$, 26 mag/arcsec$^2$ and
27.7 mag/arcsec$^2$ (outermost) in the V-band.}
\label{contours.snapf.faceonedgeon.proret}
\end{center}
\end{figure*}

\subsection{Description of the Mergers}
\label{sec:descr-mergers}

Figure \ref{snaps.edgeon.proret.diski30} illustrates the morphological
changes in the host disk during merger events
configured at ``z=1''. The initial inclination of the satellite is 
30$\degr$ for both prograde and retrograde orbits.

As the satellite decays in its orbit, it induces the formation of
noticeable spiral arms in the host disk, which transport angular
momentum from the central parts towards the outskirts.  Once the
satellite sinks onto the plane of the host disk it transfers kinetic
energy from its orbit to the particles in the disk, increasing their
vertical motions and causing a visible thickening.  At the same time
the disk responds to the decaying satellite, by tilting its plane in
order to conserve the total angular momentum of the system (although a
significant amount of the satellite's initial angular momentum has by
this time, already been transferred to the host halo).

Figs. \ref{snaps.faceedgeon.proret.sati30} and
\ref{snaps.faceedgeon.proret.diskysati30} show the distribution of
stellar particles for both spherical and disky satellites on 
prograde orbits with initial inclination $30\degr$.  In the early stages of the orbital decay,
the satellite is stripped leaving trails of particles on orbits with
inclinations similar to that of the satellite initially. Since the
stars are initially deeply embedded in the satellite's dark matter
halo, only a small fraction of the stellar debris is deposited at
large radii.  Most of the stars from the satellite end up in a
disk-like configuration, with the same orientation as the final disk,
but one that is somewhat thicker and more extended (see \S
3.4). Noticeable shells of debris material are formed as time goes by.
These structures are a consequence of the interaction of a dynamically
cold system with a larger one \citep{hernquist1988}.  In general the
survival of these shells will depend on the mean phase-space density
of the infalling satellite and also on its orbit.  In the case of
spherical satellites shells are visible typically since $t\sim1.7$
Gyr, lasting $\sim$2 Gyr; and for disky satellites much sharper shells
are seen starting at $t\sim1.5$ Gyr, being still noticeable by the end
of the simulation, i.e., $\sim$2.5 Gyr later.  Shells are rather
common features related to merger events, being observed in many
elliptical and spiral galaxies.  An important characteristic of shells
is that they usually survive for a long time in physical space, as
previous numerical studies have shown
\citep[e.g.][]{hernquist1988,hernquist1989}.  The presence of such a
structures in the solar vicinity and the possible signatures imprinted
on them during the formation of the thick disk will be explored in
Paper II \citep[note that such features have already been proposed to
explain the Monoceros ring, see e.g.][]{helmi2003}.
 
\subsection{Properties of the merger products}

By the end of the simulations the morphological, structural and
kinematical properties of the heated disks and satellite debris have
settled down and do not evolve further.  This occurs $\sim$2 Gyr after
either the satellite has been disrupted or its core has reached the
centre of the host disk.  This means that the properties of the final
thick disks do not change after $t$=5 Gyr and $t$=4 Gyr for the
systems configured at ``z=0'' and at ``z=1'' respectively, in the case
of heavy satellites.  This timescale is $\sim$ 6 Gyr for the lighter
satellite in our ``z=1'' experiment.

We now study in more detail the characteristics of the final disks.
To this end we use a reference frame centred on the centre of mass of
the final product, and aligned with its principal axes in such a way
that the rotation axis defines the $z$-direction.

\subsubsection{Morphological properties}

The left panels in Fig.~\ref{contours.snapf.faceonedgeon.proret} show
the morphologies of the heated host disks at the final time of the
``z=1'' experiments ($t$=4 Gyr), for the case of the spherical satellites.
Prograde orbits and lower inclinations induce on the host moderate
arms and more radial expansion in comparison with retrograde orbits
and higher inclinations.  On the other hand, higher prograde
inclinations are more efficient at thickening the disk, especially in
the outskirts.  For instance, a prograde satellite with inclination of
60$\degr$ only causes a slight increment of the radial extension
compared to the coeval control disk, but induces a noticeable
thickening compared to the same control disk and to the other
inclinations.  Satellites on retrograde orbits have a similar
thickening effect on the disk but a considerably milder influence on
the formation of tidal arms and radial expansion compared to
satellites on low inclination prograde orbits.  Notice also that the
satellites during their decay can induce the formation of weak bars
\citep[see also][] {berentzen2004}.  Some warping in the disks is also
visible in the case of mergers with inclinations of $60\degr$ for both
prograde and retrograde orbits \citepalias[see also][]{
quinn1993,velazquez1999}.

The panels on the right side of
Fig.~\ref{contours.snapf.faceonedgeon.proret} show the thick disks
obtained, now including the contribution of the satellite's stellar
particles. Their final structure is dominated by the heated disk
(compare to the left panels), except in the outer regions, where the
contribution of satellite debris is important. The outskirts are
clearly thicker for satellites on higher inclination orbits, although
their debris does not show the warp feature characteristic of the
heated disks.  Also noticeable is the difference in the distribution
of satellite debris between prograde and retrograde orbits for the
case of coplanar infall (in edge-on views).  

The contour levels shown in this figure have been drawn 4.5, 6.2, 8
and 9.7 magnitudes below the central surface brightness of the
remnant system.  If we assume a mass-to-light ratio $\Upsilon_V = 2
\Upsilon_{\odot,V}$ for the host as well as for the satellite stars,
these contours correspond to 22.5, 24.2, 26 and 27.7 mag/arcsec$^2$ in
the V-band, respectively. 

It is useful to compare this to the sample of late-type edge-on 
galaxies observed by \citet{dalcanton2000} (their Fig.~3) and
\citet{dalcanton2002} \citepalias[][their Fig.~1]{dalcanton2002},
who typically probe up to $\sim$~5 mag below the central surface
brightness of their (thin + thick) disks in the R-band.  Their 
faintest contour would be located in between the first and
second brightest contours shown in the right panels of 
Fig. \ref{contours.snapf.faceonedgeon.proret}. At least 
qualitatively, the surface brightness distribution of the remnants 
in our simulations resemble those observed by these authors.

We further quantify the shapes of the isophotal contours of the
remnants by obtaining their photometry with the task ELLIPSE 
\citep{jedrzejewski1987} of the
data reduction package IRAF\footnote{IRAF is distributed by the
National Optical Astronomy Observatories, which are operated by the
Association of Universities for Research in Astronomy, Inc., under
cooperative agreement with the National Science Foundation.}.  This
task draws an ellipse to approximately match an isophote and then
expands the intensity along the ellipse as a Fourier series.
According to \citet{bender1988}, the most significant non-zero
component of this Fourier analysis is the $a_4$ parameter
(corresponding to the $\cos(4\theta)$ term). Isophotes are then
characterised as either disky ($a_4 > 0$) or boxy ($a_4 < 0$). 

To mimic the observations, we have created artificial images out of
the simulated thick disks from an edge-on point of view, by binning a
central area ($\sim$15$\times$15 scale-lengths of the initial primary
disk) into 1024$\times$1024 pixels. When running ELLIPSE on these
images, we have allowed the geometric centre, ellipticity and position
angle of the isophotes to vary freely, taking linear steps of 5 pixels
along the semi-major axis.  We have also made sure that our results
are robust to the initial guesses for the values of the various
parameters required by ELLIPSE. 

\begin{figure}
\begin{center}
\includegraphics[width=41mm]{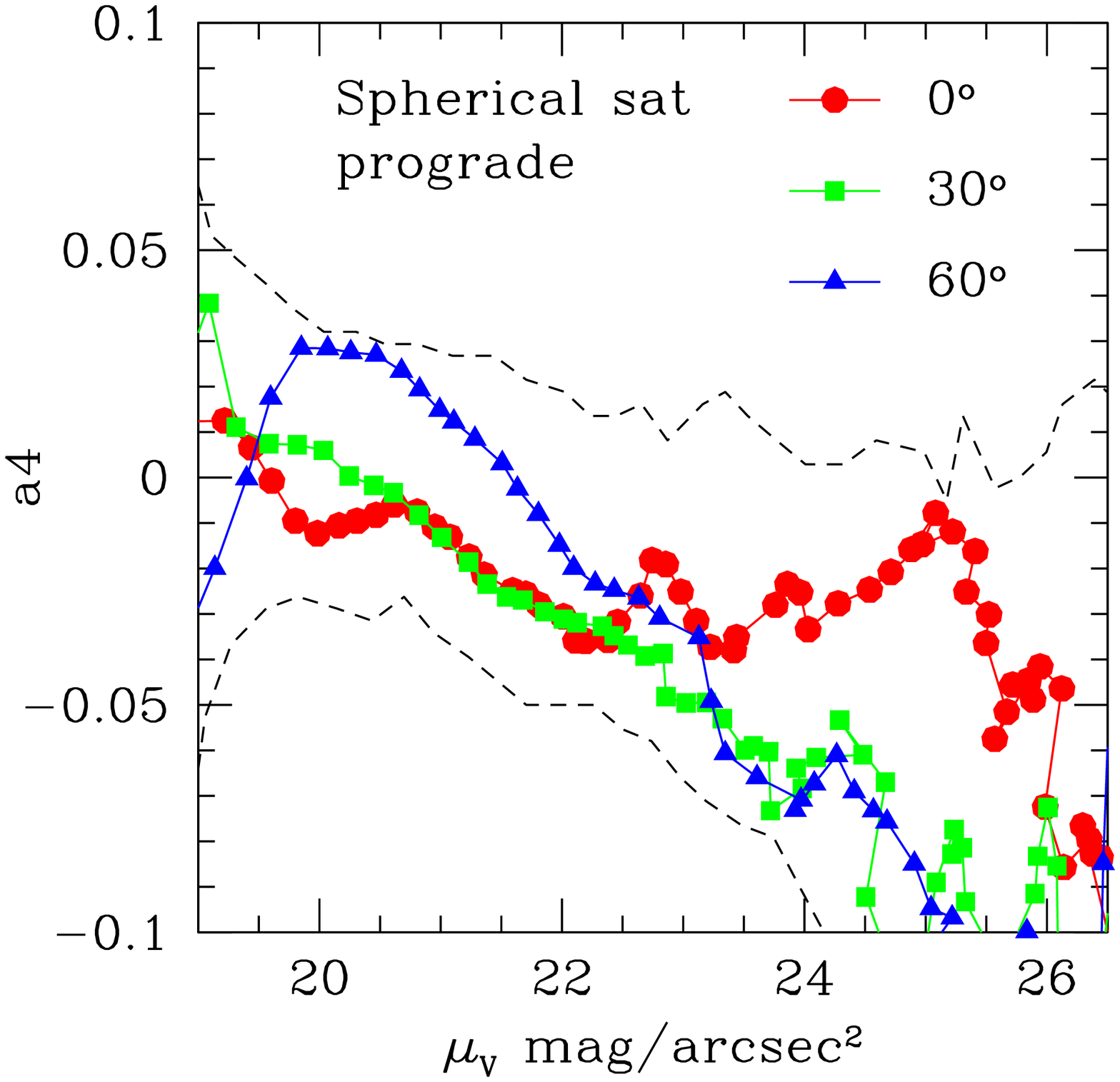}
\includegraphics[width=41mm]{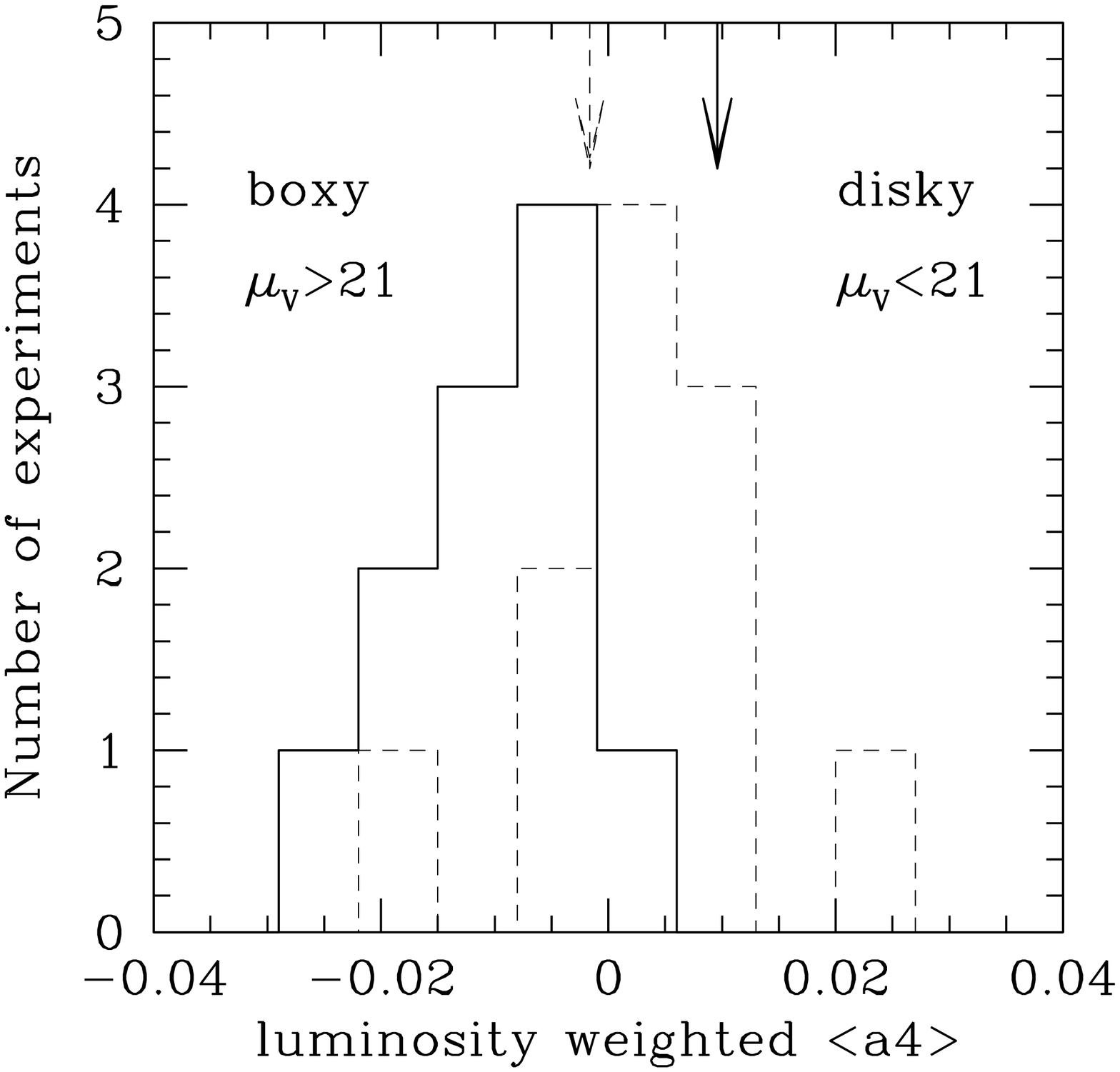}
\caption{{\it Left:} $a_4$ vs surface brightness in the V-band for our
``z=1'' experiments of a spherical satellite on a prograde orbit. The
dashed curves delimit the region where all experiments fall into. Note that
the central surface brightness of our remnant disks are $\mu_{V,0}
\sim 18 - 19$ mag/arcsec$^2$. {\it Right:} Luminosity weighted
distribution of the $a_4$ parameter for all our ``z=1'' experiments. The
dashed histogram corresponds to the $a_4$ obtained by considering the
region enclosed within an isophote with $\mu < \mu_0 + 3$, while the
solid one to $\mu_0 + 8 > \mu > \mu_0 + 3$. The arrows show the values
of $<a_4>$ for the coeval control simulation.}
\label{a4-mu}
\end{center}
\end{figure}

The left panel of Fig. \ref{a4-mu} shows the $a_4$ parameter as a
function of isophotal surface brightness in the V-band for our ``z=1''
experiments. To avoid cluttering only experiments with spherical
satellites on prograde orbits are explicitly shown. The rest fall in
the region delimited by the dashed envelopes.  This figure shows that
the isophotes go from more disky at higher surface
brightness (i.e. in the central regions) to more boxy at
lower surface brightness, presenting a mild trend with initial
inclination of the satellite.

\citetalias{dalcanton2002} performed a similar analysis on their sample
(see their Fig. 11).  They find that inner
isophotes with a surface brightness level of $\sim$3 mag below the
typical peak level for their sample (21 mag/arcsec$^2$) are disky,
while outer isophotes (defined as those $\sim$5 mag below the peak)
are as likely to be boxy as disky\footnote{It is important to keep in
mind that \citetalias{dalcanton2002} use a different procedure to
quantify the isophotal shape, meaning that the values of their shape
parameters are not directly comparable to ours.}. Similarly to
\citetalias{dalcanton2002}, in the right panel of Fig. \ref{a4-mu} we
plot the distribution of $a_4$ (weighted by both errors and
luminosity) for both inner and outer regions including all the ``z=1''
experiments. The inner region is defined to be within 3 mag
from the peak surface brightness, as done by
\citetalias{dalcanton2002}.  The outer region extends down to 8 mag
below the central surface brightness value.  This figure confirms that
inner isophotes are disky while outer ones appear clearly boxy.  This
may suggest that deeper photometry, beyond the limit reached by
\citetalias{dalcanton2002} would be needed to detect the predominantly
boxy shape of the contours in the outskirts of our remnants. 

The boxy nature of the outer isophotes, which is present in all our
experiments, could in principle be used as a discriminant for the
formation of thick disks via mergers such as those studied
here. However, it should be borne in mind that the degree of boxiness
in the remnants also depends on the initial structure of the host
system. For example, studies which have a spherical centrally
concentrated core component (bulge or cored dark matter profile),
produce a remnant which is less boxy
\citep{naab-burkert2003,bournaud05}. This is because such spherical
components act in a stabilising sense for the disk
\citetext{\citetalias{velazquez1999}; \citealt{kazantzidis2007}}, which
therefore retains more closely its original morphology. This would
imply that boxy isophotes are not necessarily direct evidence in
support of the scenario proposed in this paper, but that they should
be more prominent in the thick disks present in bulgeless galaxies.

\subsubsection{Structural Properties}

We now describe in detail the structural properties of the remnant
systems produced in our experiments. We first address their vertical
structure and show explicitly the need to include two components (thin
and thick). We then use the information obtained from the vertical
decomposition to characterise the radial extension of both disk
components.  Finally, the spatial distribution of stars from both the
primary disk and the satellite are compared.

Note that, as mentioned above, before measuring the disks, these have
been properly centred and aligned, so the rotation axis defines the
$z$-direction.  Furthermore, all the properties have been computed
taking into account star particles from both the host disk and the
satellite. Their relative contribution has been appropriately weighted
according to the initial $M_{sat,stars}/M_{host,disk}$ ratio to
account for the fact that the satellite particles have smaller masses.

\paragraph{Vertical structure of the remnants}
\label{vert-decomp}

\begin{figure}
\begin{center}
\includegraphics[width=81mm]{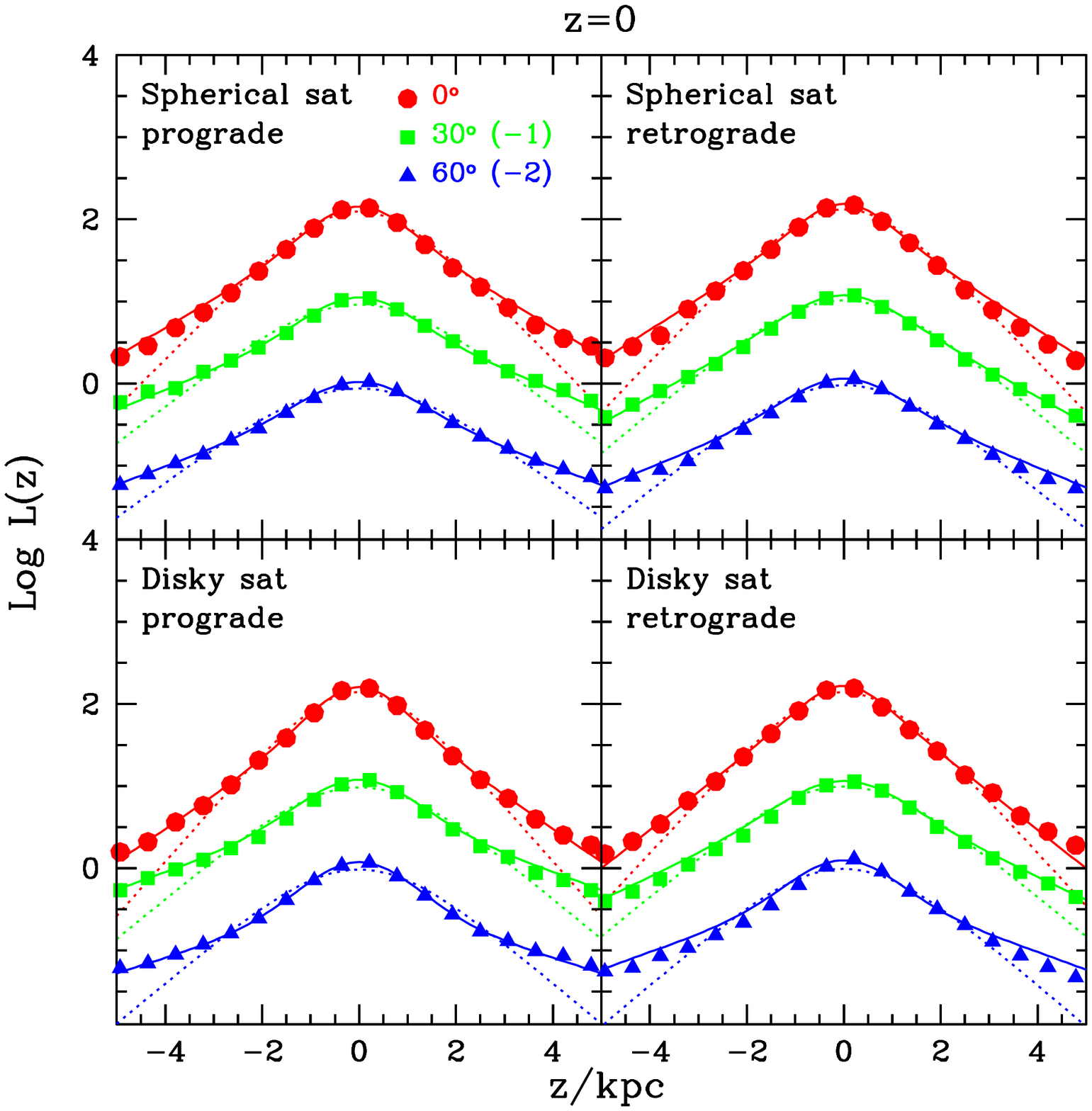}
\includegraphics[width=81mm]{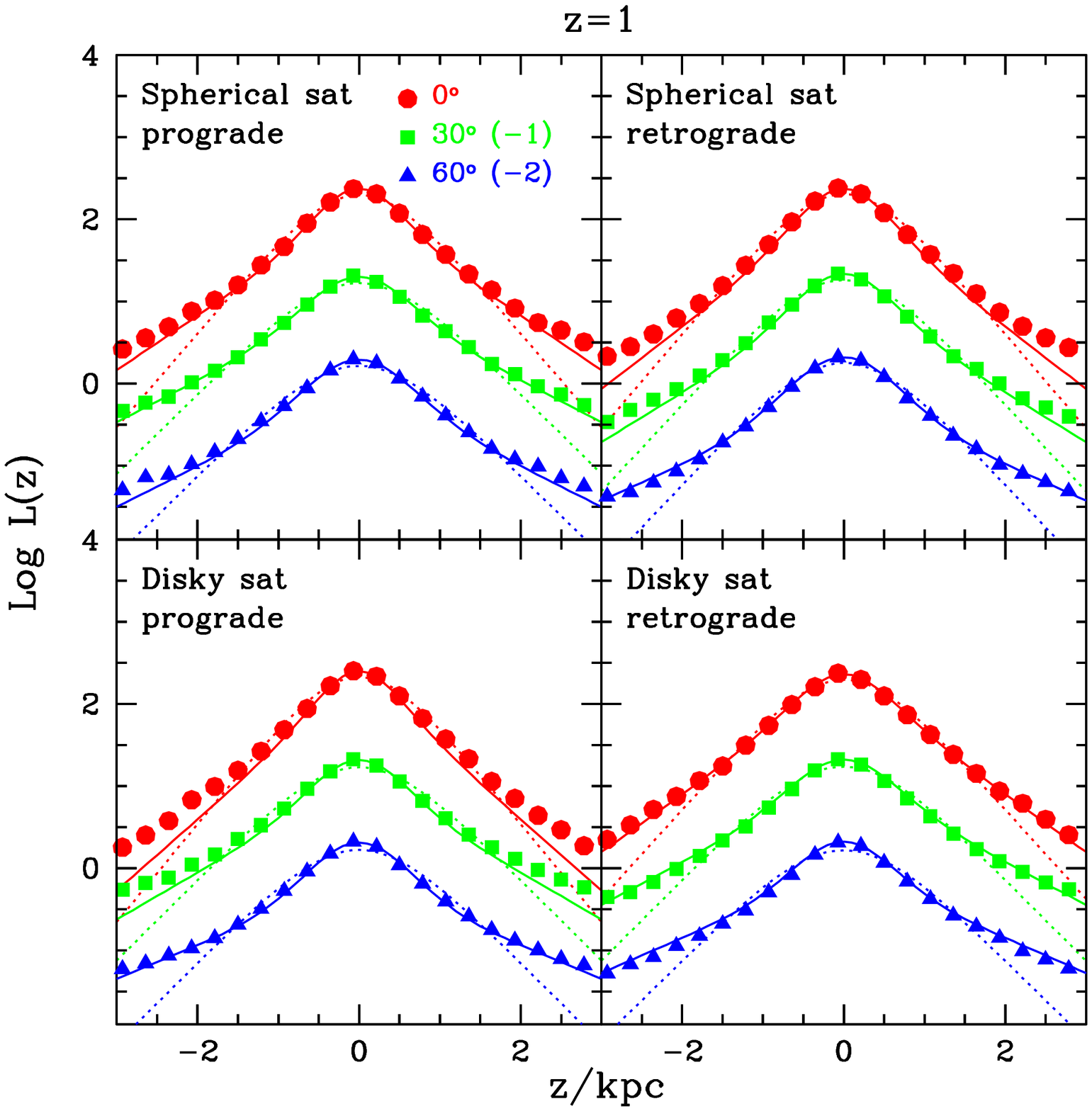}
\caption{Vertical luminosity profiles (integrating over all
radii at each height) for each of our experiments. Fits using
only one component (dotted lines) systematically underestimate the
luminosity far from the midplane.  Two-component fits (solid lines)
are clearly a better representation of the vertical structure of the
our remnant disks.
Note that, for clarity, profiles of inclinations of 30$\degr$ 
and 60$\degr$ include offsets of $\log (L)-1$ and $\log (L)-2$, 
respectively.  
}
\label{1c-vs-2c-allR}
\end{center}
\end{figure}

\begin{figure}
\begin{center}
\includegraphics[width=85mm]{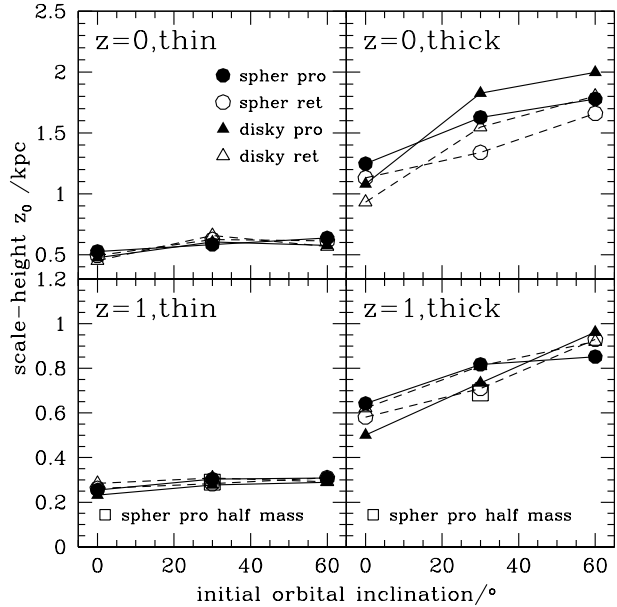}
\caption{Scale-heights of the final systems decomposed into a ``thin''
and a ``thick'' disk.  Solid/dashed lines connect prograde/retrograde
satellites.}
\label{sheight}
\end{center}
\end{figure}

\begin{figure}
\begin{center}
\includegraphics[width=79.8mm]{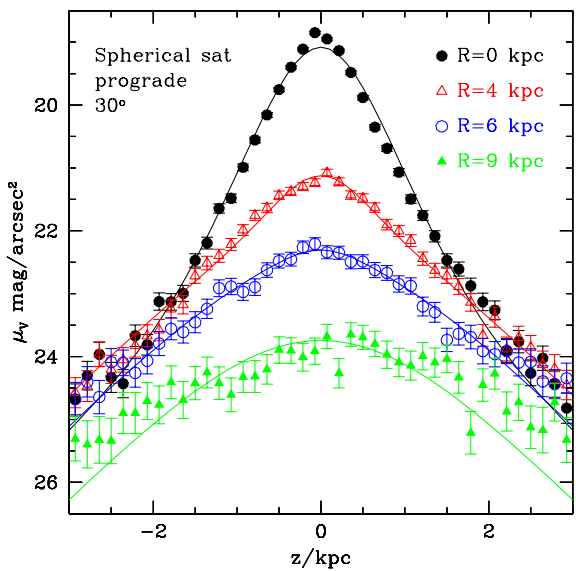}\\
\caption{Surface brightness profiles as a function of height at several projected radii for one of the ``z=1'' experiments.
The thin remnant dominates the surface brightness in the central regions, while  
the thicker component is more dominant at the outskirts.
Solid lines show the two-component fits at each radius.}
\label{surfbright-at-proj-radii}
\end{center}
\end{figure}

Fig. \ref{1c-vs-2c-allR} shows the vertical luminosity
distributions of the remnant systems for all our experiments. These
have been obtained by including all stars at all radii within $|z|<3$~kpc for 
``z=1'', and within $|z|<5$~kpc for ``z=0''. The dotted
lines correspond to a single-component $sech^2$ fit to the vertical
profile. This figure shows that such a model clearly underestimates
the luminosity at large distances from the plane, highlighting the
need for a two-component decomposition.

Therefore, we fit the surface brightness as:
\begin{equation} \label{lumdens}
L(R,z) = \sum^{2}_{i=1} L_{0,i}(R)\
\textrm{sech}^2\left(\frac{z}{2z_{0,i}}\right)
\end{equation} 
where $z_{0,i}$ is a (luminosity weighted) exponential scale-height
and $L_{0,i}$ is the central luminosity (on the midplane) of each
component ($i = 1, 2$).  As usual, $\mu(R,z)=26.4-2.5\log[L(R,z)]$.
To compute the luminosity-weighted scale-heights we proceed as
follows. First we fit independently the vertical brightness profiles
in 1 kpc radial bins (within projected radii $R < 20$ kpc for ``z=0'', and $R <
10 $ kpc for ``z=1'') using two components. For each radial bin we
allow the algorithm to find the best central luminosities and
exponential scale-heights using a Levenberg-Marquardt least-squares
minimisation. The luminosity-weighted scale-height of each component
is then the mean scale-height averaged over all radii and weighted by
luminosity. 

The fits obtained in this way are shown in Fig. \ref{1c-vs-2c-allR} as
solid curves. Clearly the vertical structure of our remnants is
considerably better modeled by considering two disk components with
different scale-heights and central surface brightness. In all cases,
a thin disk is present after the merger with the satellite.

Fig. \ref{sheight} shows the luminosity-weighted scale-heights of each
component of the remnant systems for all our experiments.  Note that
the thinner component has in all cases, a very similar (and only
slightly larger) scale-height to that of the initial host disk. The
scale-height of thicker component is clearly larger for encounters
with higher orbital inclinations.  This is because there is a
significantly larger vertical kinetic energy associated to the
satellites' orbital motion transferred to the disk.  Spherical and
disky satellite do not induce very different vertical heating on the
disks.  Note that less massive satellites produce final thick disks
that are thinner.

Fig. \ref{surfbright-at-proj-radii} shows surface brightness profiles 
as a function of height at several projected radii for one of the experiments.
Two-component fits using the luminosity-weighted scale-heights 
described above are also included.
This figure shows that the remnant thin disk dominates the surface
brightness at small radii. Note that at large radii ($R = 9$ kpc, or
$\mu \sim \mu_0 + 6$) there is an indication that the thick disk is
flared, and no longer follows an exponential distribution with a
constant scale-height at all radii. Such flared disks have already
been observed in previous studies (e.g. \citetalias{quinn1986}),
suggesting that flaring is a rather generic characteristic of disks
heated by mergers \citep[see][for a derivation of how the scale-height
varies with radius due to minor mergers]{kazantzidis2007}.

\paragraph{Radial structure of the remnants}

\begin{figure*}
\hspace*{-0.5cm}
\includegraphics[width=178mm]{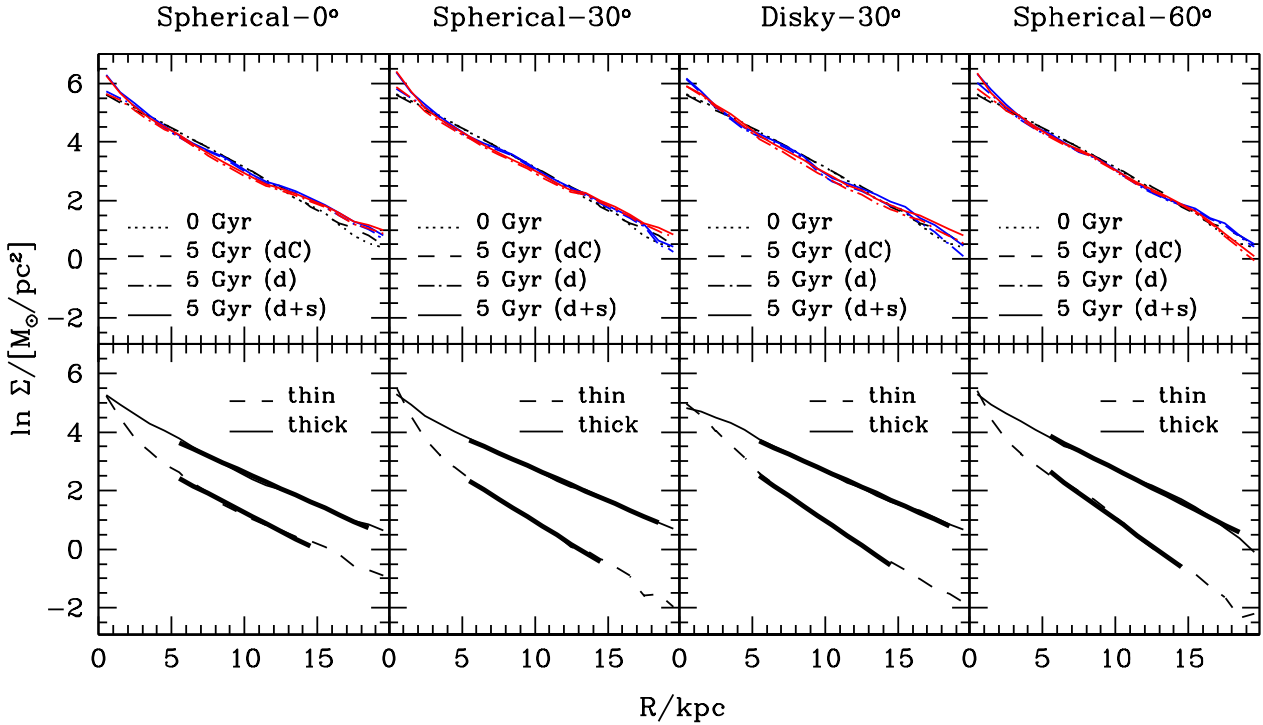}
\caption{ \emph{Upper panels}: surface density profiles of final thick 
disks as a function of cylindrical radius for ``z=0'', including stars
within $|z|<5$~kpc. 
Red and blue lines correspond to prograde
and retrograde orbits, respectively.  The measurements include
stars of both disk and satellite (solid) or only stars from the disk (dashed-dot).
As reference, both initial (dotted) and
final states (dashed) of the disk in the control model are also shown.
\emph{Lower panels}: Surface density profiles of regions dominated by
thin (dashed; defined by $|z| < 0.5 z_{0,thin}$) and thick (solid;
$1 z_{0,thin} < |z| <5$~kpc) remnants, including the section
used to compute the scale-lengths (heavy solid).}
\label{grid.morfo}
\end{figure*}
\begin{figure*}
\hspace*{-0.2cm}
\includegraphics[width=178mm]{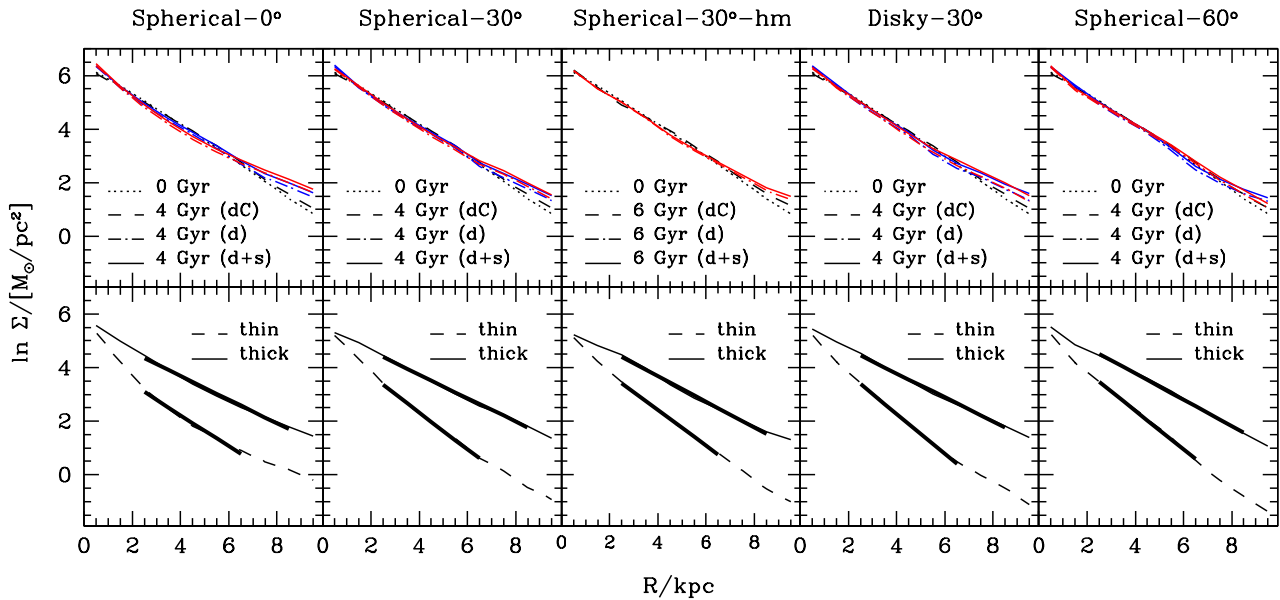}
\caption{ Same as Fig.~\ref{grid.morfo} but now for the ``z=1''
experiments, including stars within $|z|<3$~kpc 
in the upper panels. In this case the region dominated by thick remnants is defined 
as $1 z_{0,thin} < |z| <3$~kpc.
The column ``hm'' corresponds to the lighter 
satellite (with half the mass).
}
\label{grid.morfo_1}
\end{figure*}

Figures \ref{grid.morfo} and \ref{grid.morfo_1} show the mass surface 
densities of the simulated thick disks as a function of radius, for
the ``z=0'' and ``z=1'' configurations, respectively. Each panel is
for a different inclination or different type of satellite (spherical,
disky or half-mass) on prograde (red) and retrograde orbits (blue).
The control disk galaxies are also shown at the
initial (dotted curves) and final times (dashed curves) to
calibrate the effect of the mergers against the intrinsic evolution of
the host (which in all cases is negligible). In order to estimate the
contribution of the satellite stars, the
mass surface densities of only host disks are plotted separately 
(dash-dotted curves).

For both ``z=0'' and ``z=1'' experiments, the surface density
profiles show a mild dependence on orbital inclination (upper panels), having
slightly higher surface brightness in the outskirts 
for lower inclinations. This tendency is mostly due to the host disk
material that is transported radially outwards during the merger by
transfer of energy and angular momentum.  

As expected, the lighter spherical satellite (in the ``z=1''
experiment) produces a smaller variation in the surface density at
larger radius.  In general the contribution of satellite particles to
the spatial structure of final thick disks is very small.  This is
true at all radii except at the centre where the very dense spherical
satellites accreted in the ``z=0'' experiment are not completely
disrupted, retaining a core and giving rise to a small ``bulge-like''
component (see also Fig. \ref{massloss.spher.disky.z0}).  The surface
density profile does not show any strong dependence on orbit
direction.

The lower panels of Figs. \ref{grid.morfo} and \ref{grid.morfo_1} show the surface density
profiles of regions dominated either by the thin or thick remnants
(for prograde experiments only)
according to the decomposition performed in \S\ref{vert-decomp}.  These regions are defined as
$|z|<0.5z_{0,thin}$ for the thinner component, and for the thicker one
from $1 z_{0,thin}$ upto 5~kpc for the ``z=0'' case (and upto 3~kpc for
``z=1'').  The scale-lengths of each component are computed by
applying a linear fit to $\textrm{ln} \Sigma(R)$, avoiding
non-axisymmetries associated to both the central regions and the very
outskirts. Note that the linear fits consider a more extended region for thick remnants.
This is to account for the dominance of thick remnants at larger radii 
(see Fig. \ref{surfbright-at-proj-radii}).

\begin{figure}
\begin{center}
\includegraphics[width=85mm]{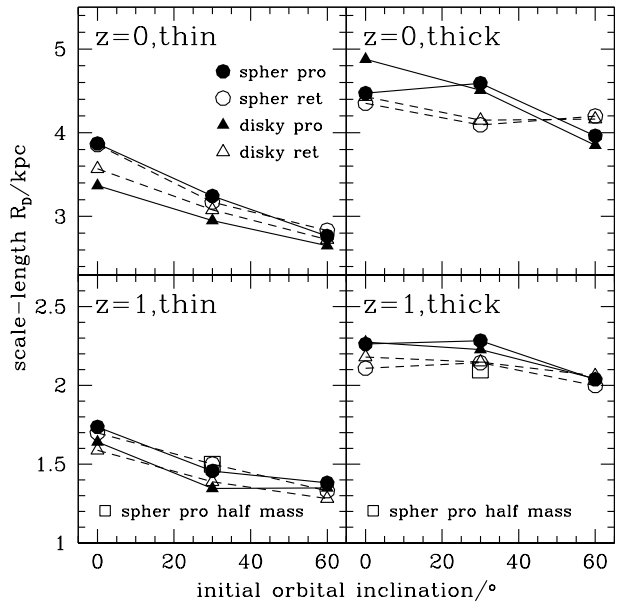}
\caption{ Scale-lengths of the remnant system for all experiments.
They are obtained by decomposing the final disks into thin and 
thick components.
Solid/dashed lines connect prograde/retrograde satellites. }
\label{slength}
\end{center}
\end{figure}

Fig. \ref{slength} shows the final scale-lengths of the thick disks as
a function of the initial orbital inclinations of the satellites.  In
all cases, the remnant thin disks have smaller scale-lengths than
their thicker counterparts, and comparable to the initial values. Low
inclination encounters induce larger thick-disk scale-lengths in
comparison to higher inclinations.  This is because in the former
cases, the orbital energy of the satellite is deposited mostly into
radial motions of the stars in the disk. Similar trends are observed
for the galaxies in \citet{yoachim2006}.  However, this should be
taken with great care because our comparison is to the remnant thin
disk and not to the present-day thin disk of those galaxies (because
we do not model this). Furthermore we have not modeled the response
of the remnant thin or thick disks to the formation of the new thin
disk. 

We compute the total mass associated to each of disk component using
the fits just derived. We find that the total mass associated to the
remnant thin disk is $\sim$15\%--25\% of the total stellar mass of the
system for both ``z=0'' and ``z=1'' experiments.

In general, the presence of a thin remnant after the merger is in
agreement with results by \citet{kazantzidis2007}, although the total
mass associated to this component is significantly smaller in our case
($< 25$\% versus $\sim 80$\%). This is maybe due to the fact that
\citet{kazantzidis2007} do not follow the full merger event, but only
let their satellites have one passage around their host disk, hence
perhaps increasing its chances of remaining relatively cold. Note however,
that in their work this bombardment is repeated in a sequence using
six different satellites.

\begin{figure}
\begin{center}
\includegraphics[width=80mm]{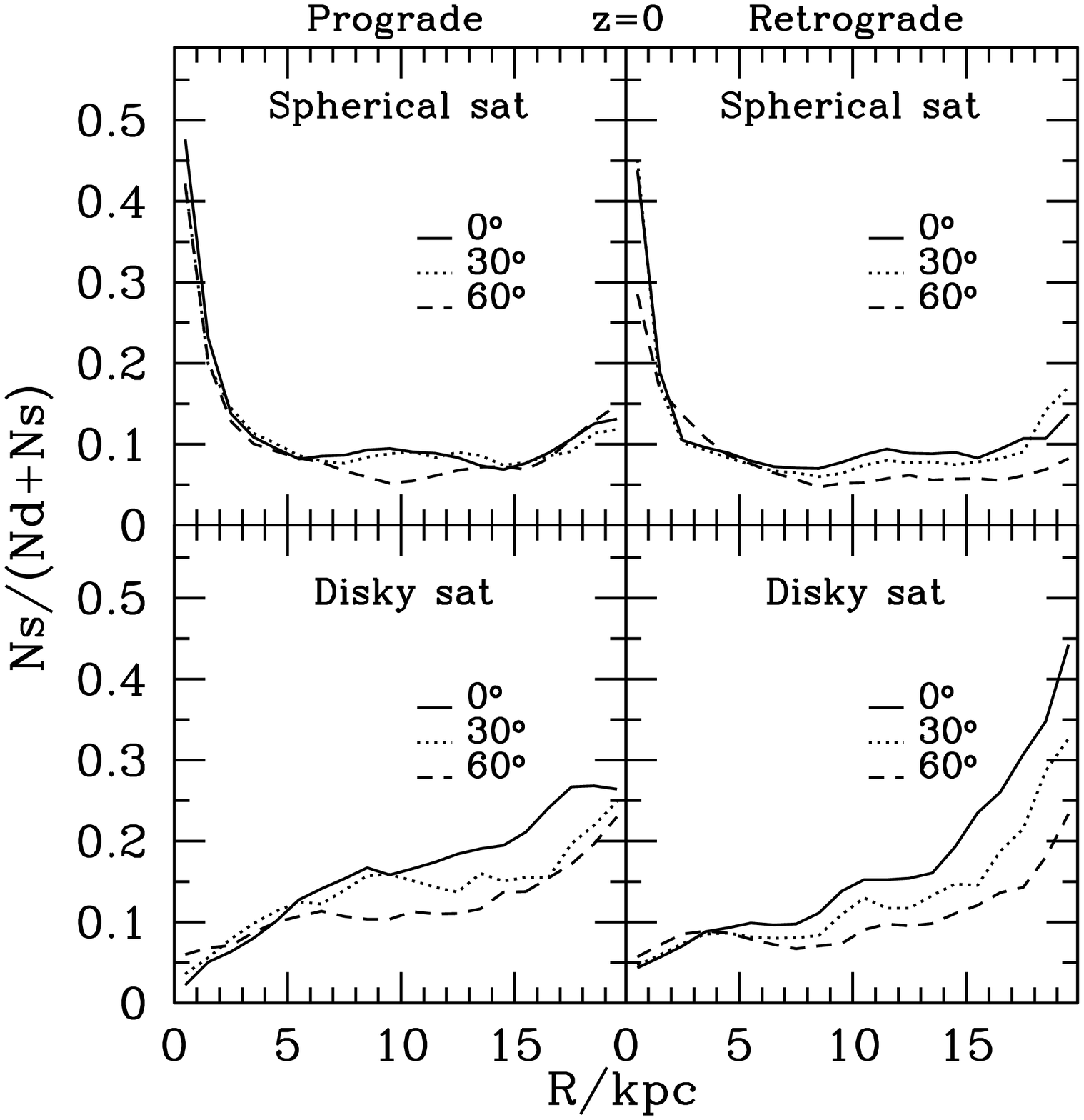}
\includegraphics[width=80mm]{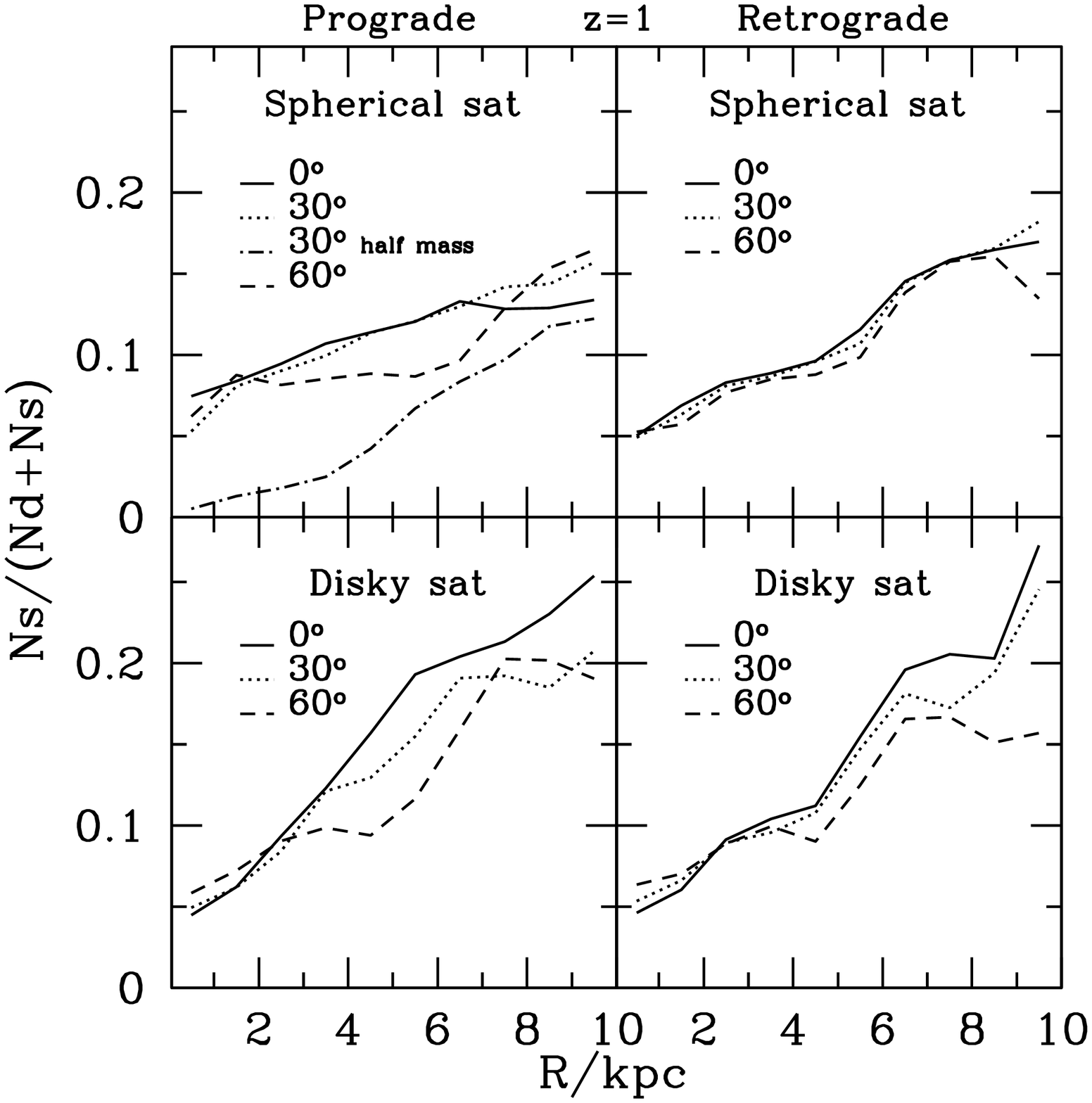}
\caption{ Relative number of satellite particles in the final systems
as a function of cylindrical radius for the ``z=0'' (top) and ``z=1''
(bottom) experiments.}
\label{ns.nt.r}
\end{center}
\end{figure}

\begin{figure}
\begin{center}
\includegraphics[width=80mm]{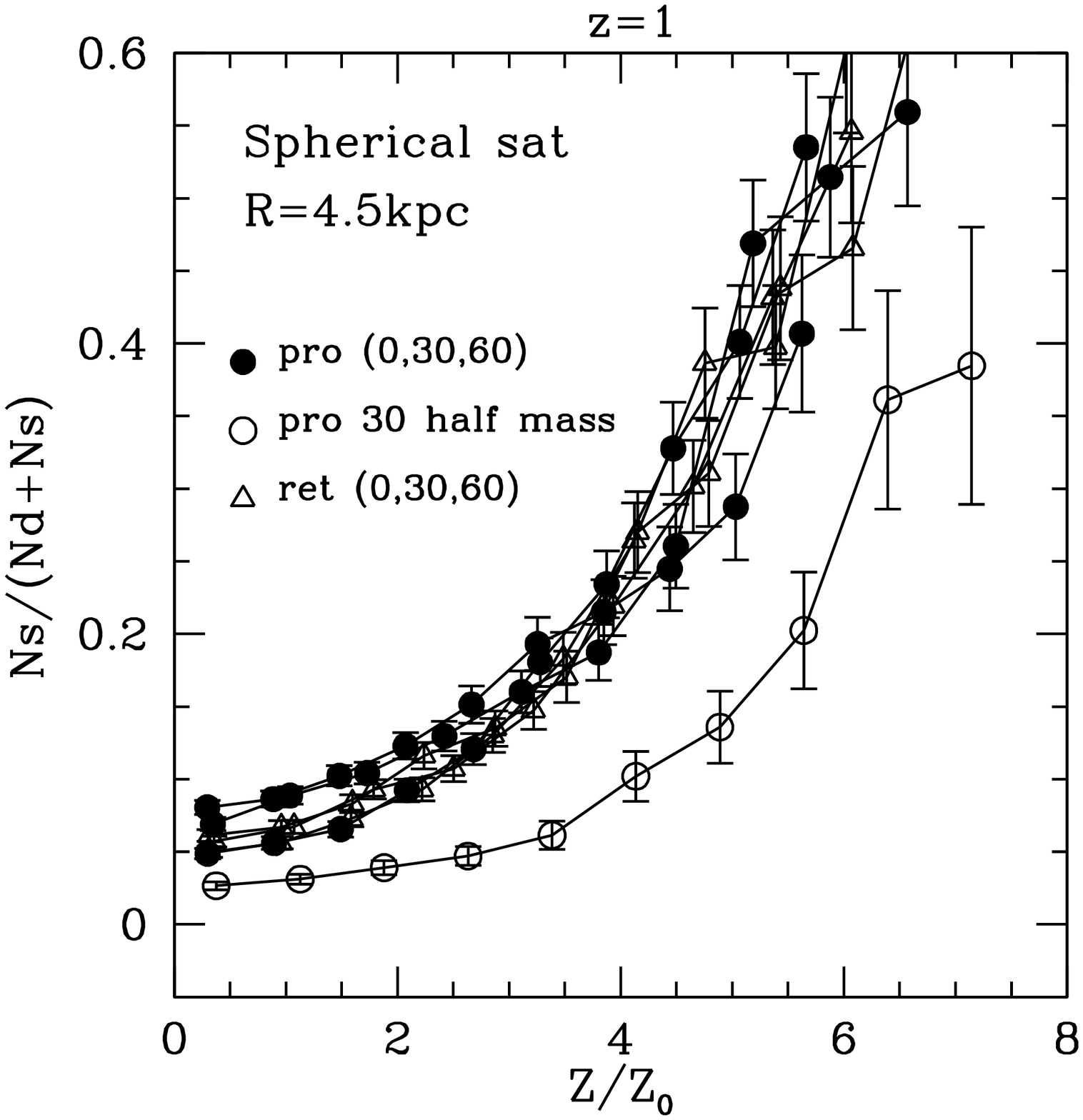}
\includegraphics[width=80mm]{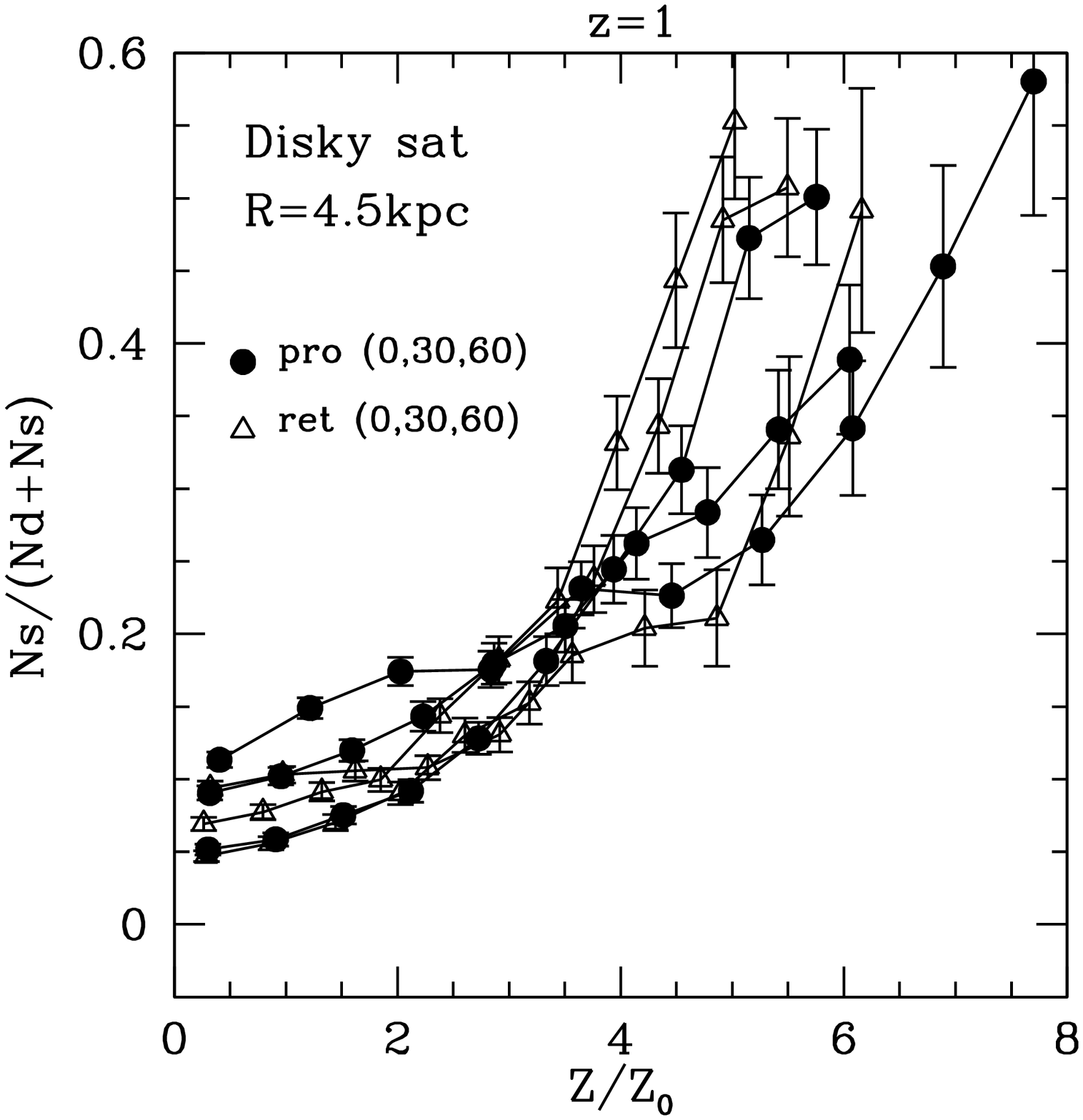}
\caption{ Relative number of satellite particles for our ``z=1''
experiments, as a function of distance from the disk plane.  Distances
from the plane are normalised by the respective scale-height at a
distance of $R=$ 4.5 kpc (i.e. 2.4 scale-lengths of the original
disk). Note that the scale-height here is obtained by fitting locally
a single $sech^2$ law to the vertical density distribution, and hence
it is close to a luminosity weighted average of the scale-heights of
the thin and thick disk shown in Figure \ref{sheight}.}
\label{ns.nt.z}
\end{center}
\end{figure}

\paragraph{Distribution of satellite vs host disk stars}

Fig. \ref{ns.nt.r} shows the final relative number
\mbox{$N_s/(N_s + N_d)$} of stellar satellite particles as a function
of radius in the resulting thick disks for mergers configured at
``z=0'' and ``z=1''.  Recall that the number of satellite particles
has been renormalised according to \mbox{$N_s = N_{sat,stars} \times
M_{sat,stars}/M_{disk}$}.

As mentioned before, in the ``z=0'' experiments, spherical satellites
have a higher mean density that the host disk.  This causes the core
to reach the host disk centre almost intact, representing $\sim$50\%
of the total number of particles near the centre of the final thick
disk.  In comparison, this ratio drops to 5\% for the disky
satellites.  At both intermediate and larger radii the fraction of
particles from the spherical satellites is roughly constant,
independently of either orbital inclination or sense of rotation.  On
the other hand disky satellites are disrupted at large radii, where
their debris is deposited.  Furthermore, the lower the inclination,
the higher the fraction at a given radius, as naturally
expected.

For ``z=1'' both spherical and disky satellites are completely
destroyed.  The relative fraction of satellite stars increases with
radius and, as expected, at the centre the relative number of
satellite particles is smaller for the lighter satellite.  The
observed trend with orbital inclination in spherical and disky
satellites at ``z=0'' is confirmed for the ``z=1'' configuration.

Fig. \ref{ns.nt.z} shows the fraction of satellite particles plotted
now as a function of the vertical direction at a radius $R$=4.5 kpc,
for spherical and disky satellites in the ``z=1'' experiments. The
radius corresponds to 2.4 initial scale-lengths in this experiment.
In this figure the distances from the plane are normalised by the
scale-height obtained by fitting locally a single $sech^2$ law to the
vertical density distribution. Therefore $Z_0$ it is close to a
luminosity weighted average of the scale-heights of the thin and thick
disks given in Fig.~\ref{sheight}. This figure shows that the fraction
of accreted particles as a function of distance from the plane, when
normalised by this scale-height, \emph{only} depends on the mass ratio
between the satellite and host.  E.g., at $z=4Z_0$ the fraction of
particles reflects the mass ratio of the merger. The same behaviour is
observed in the ``z=0'' experiments.

\subsubsection{Kinematical Properties}

\begin{figure}
\begin{center}
\includegraphics[width=80mm]{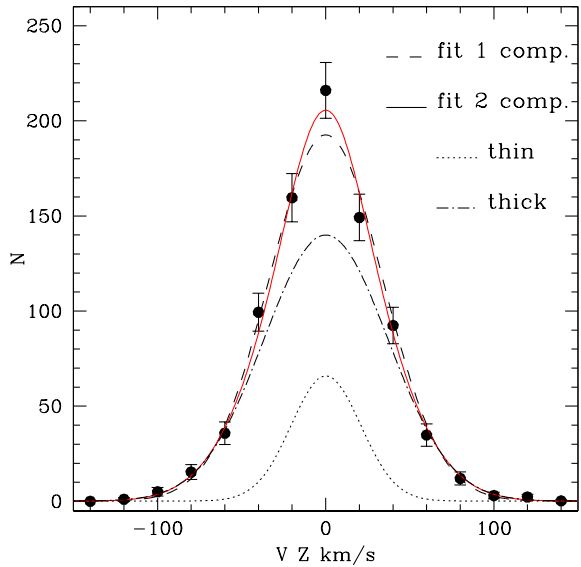}
\caption{Example of decomposition of the vertical velocity
distribution into a cold component (associated to the remnant thin
disk) and a hot one (the thick disk). The normalisation of each
component has been fixed according to the photometric decomposition of
\S\ref{vert-decomp} (see text). The contribution of the two components
is separately shown.  }
\label{vz-decomp}
\end{center}
\end{figure}

The structural decomposition made in \S\ref{vert-decomp} should also
be reflected in the kinematics of the stars in our systems in order to
be physically meaningful.  Fig. \ref{vz-decomp} shows this is indeed
the case.  Here we plot the $z$-velocity distribution within a
spherical volume of 1 kpc radius centred at R $\sim$ 4 kpc on the
midplane for one of our experiments (``z=1'', spherical satellite,
prograde, 30$^o$).  The dashed curve shows that a single Gaussian
(corresponding to a one-component system) misses the peak of the
distribution highlighting the need for a second (colder) component.
We therefore proceed to fit all velocity distributions (also the
radial and azimuthal) with two Gaussians. We constrain the relative
normalisation of these by using the photometric decomposition from
\S\ref{vert-decomp}, which determines the relative number of stars from
each component within a given volume. The solid curve in Fig. \ref{vz-decomp} is an
example of the quality of the fit obtained in this way, whose
reduced-$\chi^2$ ($= 0.47$) is lower than that obtained for a single
Gaussian ($=0.68$).

\begin{figure}
\begin{center}
\includegraphics[width=85mm]{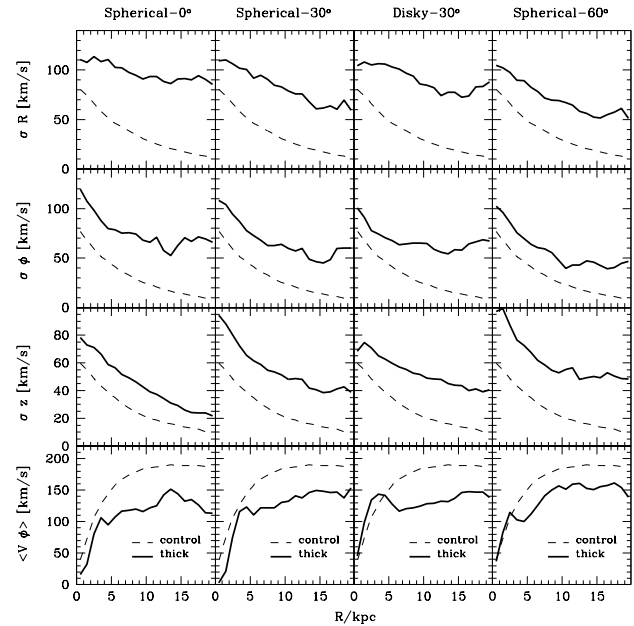}
\caption{Final dynamical properties of the thick disk components for
our prograde ``z=0'' experiments.  These properties have been computed
at $t=5$~Gyr for stars within $|z|<3$~kpc in concentric rings of 1~kpc
width, using the decomposition described in \S\ref{vert-decomp}.  As
reference, the final state of the disk in the control model is also
shown (dashed).}
\label{grid.kine1}
\end{center}
\end{figure}

\begin{figure}
\begin{center}
\includegraphics[width=85mm]{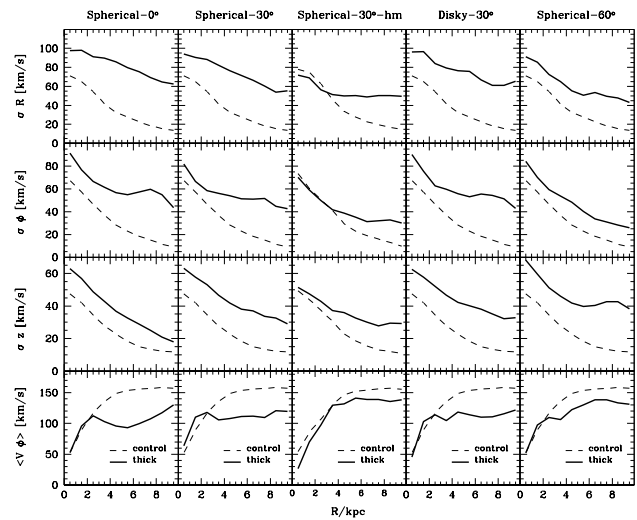}
\caption{Final dynamical properties of thick disk components for our
prograde ``z=1'' experiments.  The column ``hm'' show the satellite
with half the mass.  As in the previous figure, the properties have
been computed within concentric rings of 1~kpc width, now for
particles with $|z|<1.5$~kpc at $t=4$~Gyr (except for the ``hm''
satellite, where $t=6$~Gyr). The final state of the disk in the
control model is shown by the dashed curves.}
\label{grid.kine2}
\end{center}
\end{figure}

The transformation of orbital energy into thermal energy (random
motions) in the disk is evident in Figs. \ref{grid.kine1} and
\ref{grid.kine2}.  Here we show the radial, azimuthal and vertical
velocity dispersions along with the mean rotational velocities of the
thick disks present in our systems as a
function of cylindrical radius.  These quantities have been computed
at each cylindrical radius in concentric rings of 1~kpc width,
including stars between $|z|<3$~kpc, for ``z=0'', and $|z|<1.5$~kpc,
for ``z=1'' experiments.

In all our experiments, the vertical and azimuthal velocity
dispersions of the remnant thin disks are very similar to those of the
initial host disk, and hence are not plotted for clarity. On the other
hand, the radial velocity dispersions are generally larger by $5 - 10$~km/s
at all radii.

Figures \ref{grid.kine1} and \ref{grid.kine2} show that the radial
$\sigma_R$ and azimuthal $\sigma_{\phi}$ velocity dispersions of the
thicker component are larger for lower inclination orbits moving in
the prograde sense.  The opposite trend is observed for the vertical
velocity dispersion $\sigma_Z$. This is as expected given our previous
discussion on the evolution of the scale-heights and scale-lengths and
their dependence on orbital inclination. Spherical and disky
satellites give rise to similar velocity distributions. 

\begin{figure*}
\begin{center}
\includegraphics[width=170mm]{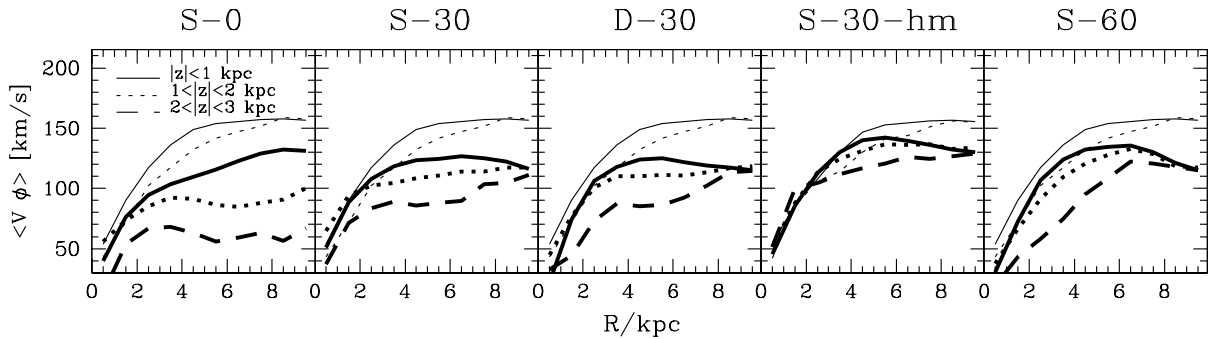}
\caption{ Mean azimuthal velocities of the ``stars'' in our
experiments at several heights above the plane as a function of radius
for our ``z=1'' experiments in prograde orbits.  Thick lines include
particles from both the heated disk and satellite, while the thin
lines correspond to the disk in the control model (the highest
$z$-range is not shown in this case because there are too few disk
particles at that distance from the midplane).  The same behaviour is
observed in our ``z=0'' experiments (not shown).}
\label{grid.vp.fn.z}
\end{center}
\end{figure*}

The resulting velocity dispersions $\sigma_R$ and $\sigma_Z$ for
retrograde orbits are similar to the prograde cases. On the other
hand, the azimuthal velocity distributions of the heated disk stars
generally require an additional component to account for the
contribution of the (accreted) counter-rotating stars (Villalobos \&
Helmi, in prep). The global velocity dispersion (that would be
obtained by imposing a single Gaussian for the thick component) would
be in this case significantly larger at large radii for retrograde
orbits, except for those with high inclination.

In general, the mean rotational velocities $\overline{v_{\phi}}$ of
the thick disks differ noticeably from the coeval control simulation
in all cases, by dropping $\sim$60 km/s although with a mild
dependence on the orbital inclination.  Low inclination encounters
produce thick disks that rotate slower, implying larger asymmetric
drifts.  This is also evidenced by their larger radial and azimuthal
velocity dispersions, as discussed above. The mean rotational velocity
of thick disks which are the result of encounters with satellites on
counter-rotating orbits is lower due to the contribution of the
accreted stars, particularly at large radii.  

The mean rotational velocity also shows noticeable differences with
inclination when it is measured away from the midplane of the thick
disk. In Fig. \ref{grid.vp.fn.z} we plot $\overline{v_{\phi}}$ for the
prograde experiments at different heights above the plane, without
making a distinction between the thin and thick disk components. Note
that for $|z| > 1$ kpc we are really measuring the kinematics of the
thick disk component since the contribution of the remnant thin disk
is negligible. Satellites with lower initial orbital inclinations
induce a rotational lag, whose magnitude increases with height above
the plane.  This is because such satellites are more efficient in
heating the disk radially at every height, leading to a larger
asymmetric drift.

\begin{figure}
\begin{center}
\includegraphics[width=85mm]{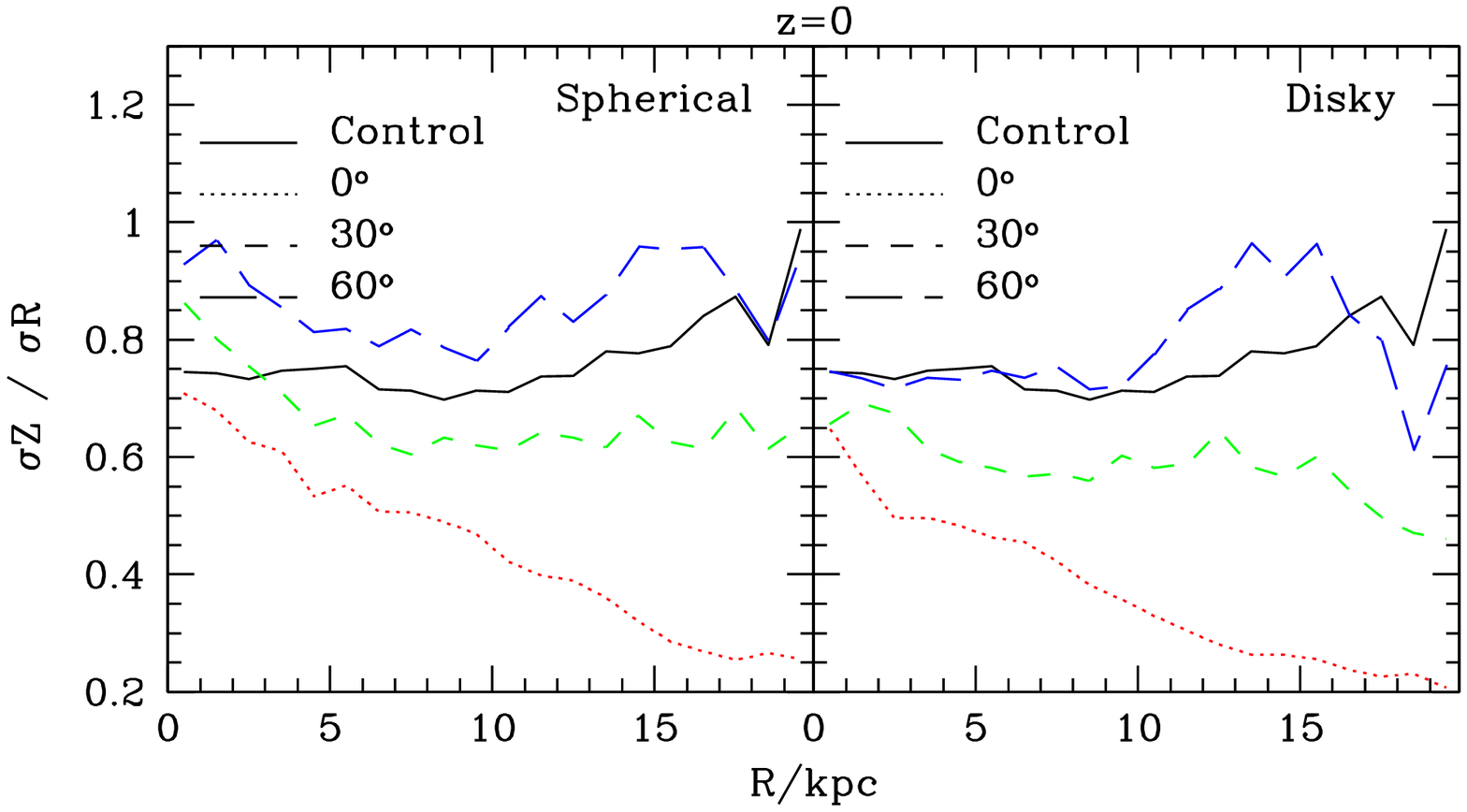}
\includegraphics[width=85mm]{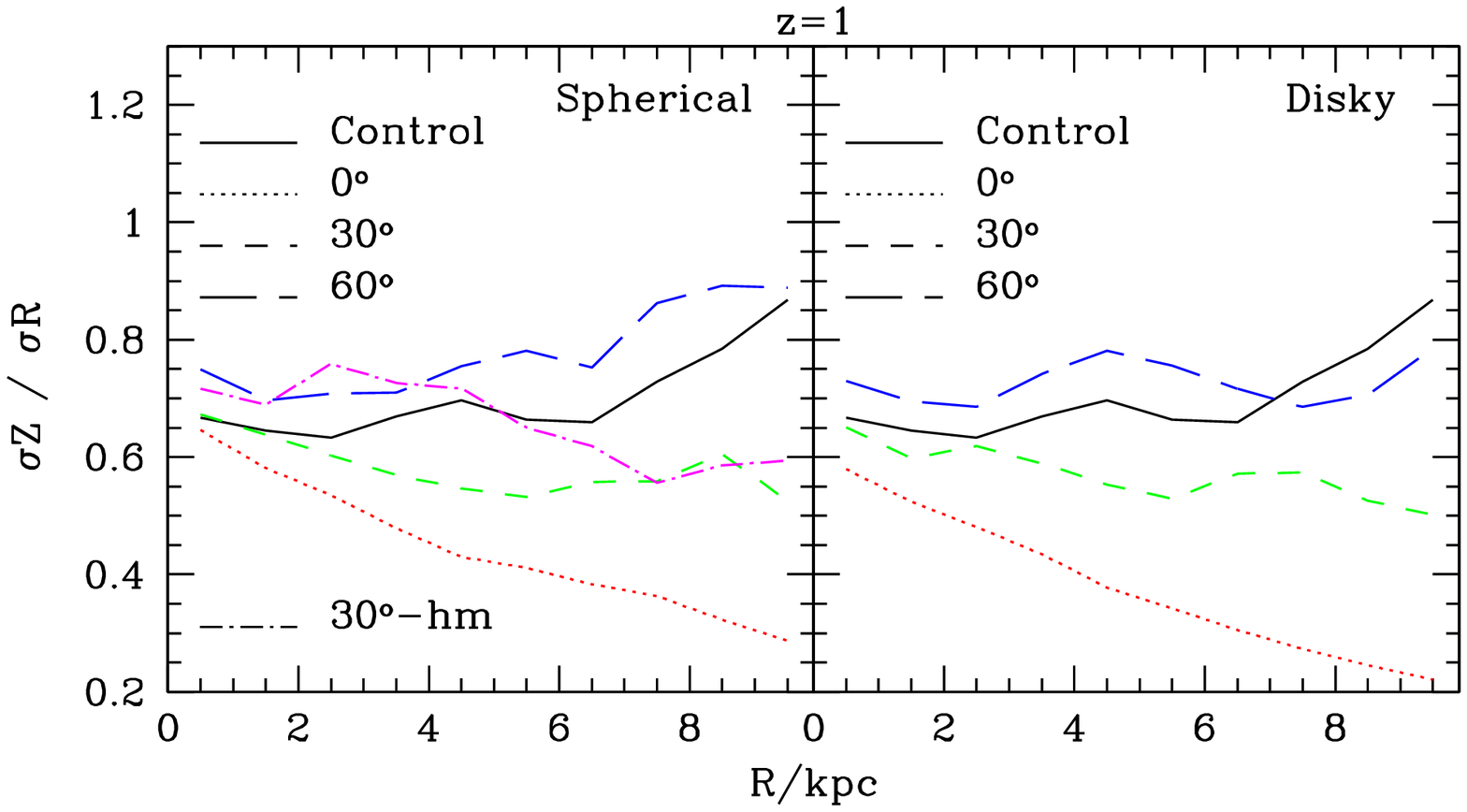}
\caption{Final $\sigma_Z/\sigma_R$ of thick remnants as a function of distance on the
plane for different initial stellar distributions and orbital
inclinations of the satellites. 
The control disks have a steady $\sigma_Z/\sigma_R
\sim 0.7$. Only prograde orbits are shown.}
\label{sigzsigr}
\end{center}
\end{figure}

In Fig. \ref{sigzsigr} we plot the ratio $\sigma_Z/\sigma_R$ of the
thick disk component as a function of radius for different prograde
experiments. Recall that the initial (and the control) disk has a
(nearly) constant $\sigma_Z/\sigma_R \sim 0.7$. This figure clearly
shows that $\sigma_Z/\sigma_R$ can be used as a discriminant of the
initial inclination of the satellite. The reason for this strong
dependence on inclination is essentially due the fact that a satellite
on a highly inclined orbit will induce a much larger change in
$\sigma_Z$ at large radii than one on a co-planar orbit.

\section{Discussion}

Previous works investigating the response of a host disk galaxy to one
or more infalling satellites provide us with a valuable description of
the dynamical effects involved in the process (\citetalias{quinn1986};
\citealt{toth1992}; \citetalias{quinn1993}; \citetalias{walker1996};
\citetalias{huang1997}; \citetalias{velazquez1999};
\citealt{benson2004}).  \citetalias{huang1997} find that a massive
satellite (30\% of $M_{disk}$), as it decays due to dynamical
friction, can tilt the orientation of the disk up to 10$\degr$ and
cause warping. These effects illustrate the strong transfer of angular
momentum from the infalling satellite to the host disk.
\citetalias{velazquez1999} note that satellites in prograde orbits
mostly increase the disk heating. On the other hand retrograde orbits
are more efficient in tilting the disk orientation.
\citet{benson2004} show that more massive and more concentrated
satellites increase the difference between the amount of disk heating
caused by prograde and retrograde orbits.  \citetalias{quinn1986}
indicate that most of the kinetic energy of the infalling satellite is
deposited on the plane of the disk and a small quantity heats up the
vertical motions of disk particles.  This shows that velocity
dispersions in the host disk are not increased isotropically by the
satellite.  \citetalias{quinn1993} note that most of the heating on
the plane of the disk is caused by spiral arms stimulated by the
decaying satellite. These arms also transfer angular momentum radially
outwards, expanding the disk.  \citetalias{quinn1986} and
\citetalias{quinn1993} also show that the vertical structure of the
disk is not uniform across the disk. Instead, the scale-height of the
host disks increases at larger radii.  Our simulations are able to
confirm most of the aforementioned dynamical effects but also show a
significantly larger tilting than previously found, both for the
prograde and retrograde cases. This can be traced back to the fact
that in our experiments, the accreted satellite is initially much more
massive and is launched from a significantly larger distance (see \S
\ref{sec:descr-mergers}).

\subsection{Choice of massive satellites to heat disks up}

Cosmological simulations show that massive mergers like those in our
``z=1'' experiments are likely to have happened during the life-time
of a Milky Way-sized host system.  For instance,
\citet{kazantzidis2007} estimate the number of massive subhalos
accreted since z~$\sim 1$ as at least one object with a mass
$\sim$$M_{disk}$ and five objects more massive than 20\% $M_{disk}$.
This is consistent with simulations by other groups
\citep[e.g.][]{stoehr2006,gdl}, and supported by the detailed study by
\citet{stewart2007}. Even though massive mergers may be less frequent
they are able to reach the centre of the host system thanks to
dynamical friction, causing important changes in the structure and
kinematics of the host disk.  In principle, this could imply that a
Milky Way-like disk would experience only one severe change of
orientation since z $\sim 1$ and also that only one massive satellite
would be needed to heat up a pre-existing disk to give rise to a thick
disk.

\subsection{Observations of thick disks}

\citet{pohlen2004} have carried out photometric thick/thin disk
decompositions and characterised properties such as scale-length and
scale-height for a sample of eight S0 edge-on galaxies by fitting 3D
disk models.  The authors find that the mean scale-height of the thick
disk is between 2.4 and 5.3 times larger than that of the thin disk;
while the mean scale-length is about twice, ranging from 1.6 to 2.6.
In general these values are consistent the results obtained by other
authors for S0, Sab, Sb, Sbc, Scd and Sd galaxies
\citep[e.g.,][]{vdkruit1984,degrijs1996,degrijs1997,pohlen2004,yoachim2006},
including the Milky Way \citep{ojha2001,larsen2003}.  {\it If} we
assume that the structure of our simulated disks are not strongly
affected by the formation of a new thin disk, {\it and} that these new
thin disks will follow the same distribution as the remnant thin disks
in our models, we can take the ratio of scale-heights obtained for our
experiments at the final time at face value. Typically there is an
increase in the scale-height by a factor 3 -- 6, while the
scale-lengths are only slightly larger (with final-to-initial ratios
between 1 and 1.4).  These values appear to be in agreement with the
range of ratios observed in general in spiral galaxies. However, this
should be taken with great care because of the strong assumptions just
made.

Most of our thick disks are significantly flared in the outer regions.
This has also been found in other studies
\citep[e.g.][]{kazantzidis2007,read2008}, with varying strengths,
depending on the initial configuration of the system. For example,
galaxies with a bulge or embedded in a cored dark-matter halo are more
stable, and suffer from less flaring
\citep{naab-burkert2003,bournaud05,kazantzidis2007}. This implies that we
might expect to find in nature thick disks with varying degrees of
flaring. However, quantifying this requires reaching extremely low
surface brightness levels, approximately $6 - 7$ magnitudes below the
peak brightness of the {\it thick} disk. This is very challenging and
has not yet been achieved in studies of surface photometry, which
typically reach about 5 magnitudes below the peak value of the {\it
thin + thick} disks. For example, Neeser et al (2002) detected a thick
disk in a low-surface brightness galaxy. They reach a surface
brightness level in the R-band of 29 mag/arcsec$^2$. This is almost 7
magnitudes below the central thin + thick disk value, and a very
marginal indication of a flare is apparent on the east side of the
thick disk at a distance of $\sim 3$ scale-lengths. Morrison et
al. (1997) in their study of NGC891 (a late type disk with a small
bulge) do not find evidence for flaring in the thick disk, but
``only'' reach down to 26 mag/arcsec$^2$ in R, which is about 6
magnitudes below the central surface brightness of the dominant disk
component.

\citet{dalcanton2002} have found a vertical colour-gradient in their
sample of galaxies, in the sense that the thick disks are redder, and
consistent with a relatively old ($>6$ Gyr) stellar population. In the
scenario we envisage for the formation of thick disks we would expect
two sources of color-gradients.  The first would be due to the
transition from the population of young thin disk stars to the thick
disk stars formed in situ, i.e. originating from a pre-existing thin
disk. Unfortunately, because we do not model the formation of a new
disk from freshly accreted gas, we cannot make a quantitative
comparison.  The second source of a change in stellar populations
would be due to the difference between accreted satellite stars and
those formed in situ in the (heated) disk. Our Figures \ref{ns.nt.r}
and \ref{ns.nt.z} show that the fraction of satellite particles is
generally quite low. Only at distances above the plane of
approximately 6 thick disk scale-heights, the number of accreted stars
is comparable to that of the heated disk. This implies that this
second gradient should be weaker, and is unlikely to be detected in
current observations because at these heights, the typical surface
brightness levels are exceedingly low (see Figure
\ref{surfbright-at-proj-radii}).

Kinematically, the values of the velocity ellipsoids of our thick
disks, measured at $R \approx 2.4R_D$, are in good agreement with the
one observed at the solar radius in the Milky Way
($\sigma_R$,$\sigma_{\phi}$,$\sigma_Z$)$\sim$(65,54,38) km/s
\citep[e.g.,
see][]{layden1996,chiba2001,soubiran2003,alcobe2005,vallenari2006,veltz2008}.
The values of $\sigma_Z/\sigma_R$ measured by most of these authors
are typically $\sim0.6$\footnote{Although Veltz et al find
$\sigma_Z/\sigma_R \sim 0.9$, which could be due to the fact that the
assumption of isothermality for the velocity distribution is not
valid.}, suggesting that the thick disk of the Galaxy could have been
produced by a merger of intermediate inclination (see
Fig. \ref{sigzsigr}).  The differences between the mean rotational
velocity of the final thick disk and that of the remnant thin disk are
$\Delta \overline{v_{\phi}}$$\sim$40-50 km/s.  Note that these values
cannot be interpreted directly as the observed rotational lag between
the thin and thick disks in the Milky Way since we do not include gas
in our simulations that later on can collapse and actually form a new
thin disk (see \S \ref{gas-in-sims}).

\subsection{Our simulations in the context of M31}

It is important to notice that in our experiments configured at
``z=0'' the initial properties of the host disk, before the merger
with the satellite, resemble the properties of the current thin disk
of the Galaxy.  This implies that the final thick disks generated in
our experiments are too massive and do not represent direct analogues
of the thick disk of the Milky Way.  However this set of simulations
may be useful for understanding the evolution of M31, which is likely
to have experienced recent merger events as may be inferred from the
rich and complex structures present in its outskirts \citep[][and
references therein]{ibata2007}.

For example, \citet{ibata2005} discovered an extended and clumpy
disk-like structure around M31 with a scale-length similar to that of
the main disk, rotating $\sim$40 km/s slower and with a rather low
velocity dispersion of $\sim$30 km/s.  Its stellar population has
homogeneous kinematics and abundances over the entire region where it
is observed, which suggests that it was formed in a single global
event.  However, it is not straightforward to link the extended disk
of M31 to the thick disk modeled here, mainly because of the much
higher velocity dispersions of the final thick disks in our
simulations.

The giant stream of M31, thought to originate in the disruption of a
satellite with mass $\sim$10$^{9} M_{\sun}$, and the so-called eastern
and western ``shelfs'' \citep{ibata2007}, may be related given the
similarities of their stellar populations.  In the context of our
simulations, this would be quite natural. Analogous structures are
observed as the satellite is disrupted as shown in
Figs. \ref{snaps.faceedgeon.proret.sati30} and
\ref{snaps.faceedgeon.proret.diskysati30} for prograde orbits in
face-on view at 1.5 Gyr \citep[see also][]{fardal2007,mori2008}.  It
is interesting to note that much sharper and longer lasting shells are
generated by disky satellites compared to spherical ones.

\subsection{Caveats}

\subsubsection{Lack of gas physics in our simulations}
\label{gas-in-sims}

In this paper we have focused on the collisionless interactions
between a disk galaxy and a satellite, with both dark matter and
stellar components, without including gas physics nor star formation
which may affect the interactions and the final thick disk's
properties.  This is potentially the most crucial simplifying
assumption in this study since disk galaxies were presumably much more
gas rich at high redshift \citep{robertson2006}.

The lack of gas physics implies that the modeled disks do not grow in
stellar mass or in size during the timespan of the merger (except of
course through the dynamical heating processes described
above). Furthermore, it is likely that such mergers would trigger a
(strong) burst of star formation.  These new stars would be relatively
old at the present day and located in a thinner structure.  On the
other hand, some orbital energy deposited by the infalling satellite
into the gas could be radiated away from the system, reducing the
dynamical damage done to the disk \citepalias{quinn1993}. For example,
in recent work \citet{hopkins2008} based on \citet{younger2008}
suggested that the change in the structural parameters depends on the
fraction of gas $f_g$ available as $\delta H \sim (1 - f_g) \delta
H_*$, where $\delta H_*$ corresponds to the scale change in the purely
dissipationless case.

Any remaining (and presumably heated) gas would eventually cool and
settle down to form a new thin disk.  This slow accumulation of gas on
the midplane should induce a contraction of the thick disk. For
example, \citet{elmegreen2006} estimate that this contraction leads to
a decrease in the scale-height of the contracted thick by $\sim$$40$\%
and an increase of the velocity dispersion by
$\sim$$50$\%. Furthermore, the accretion of fresh gas from the
inter-galactic medium is likely to also be important, and will lead to
further changes to the properties of the merger products studied here.

\subsubsection{Time-dependence of gravitational potential}

In general, the structure of a dark matter halo evolves with time
through mergers and slow accretion.  However, in our simulations we
have neglected any cosmological evolution of the structure of the host
halo during and after the merger with the satellite.  This
simplification may be justified by recent studies
\citep[e.g.,][]{wechsler2002,romano-diaz2006}, which have shown that
the structure of dark halos is very stable within the scale radius
$r_s$ after the phase of active mergers, which for a disk galaxy must
have ended at redshifts $\sim 0.5 - 1$. This is indeed the region that
we follow dynamically in our simulations, after the initial decay of
the satellite due to dynamical friction, which lasts typically less
than 1 Gyr. Therefore, the final thick disks are well within this
scale radius, being $r_s\sim6$ and $\sim 11$ times the final
scale-lengths in the ``z=0'' and ``z=1'' experiments, respectively.

\section{Summary and Conclusions}

We have performed numerical simulations of the heating of a disk
galaxy by a single relatively massive merger.  These mergers lead to
the formation of thick disks whose characteristics are similar, both
in morphology as in kinematics, to those observed in the Milky Way and
other spiral galaxies.

The simulations explore several configurations of the progenitor
systems whose properties have been scaled at two different redshifts
in order to study the formation of thick disks at different epochs.
The satellites have total masses of 10\% and 20\% that of the host
galaxy and have been modeled self-consistently as a stellar 
component immersed in a dark matter halo.  
The stellar components have either a spherical or disky
distribution.  The satellites have been released far away from the
host disk, with initial orbital parameters that are consistent with
cosmological studies.  Additionally, three different initial orbital
inclinations of the satellites have been studied in both prograde and
retrograde directions with respect to the rotation of the host disk.
 
We find that as the satellite galaxies spiral in through dynamical
friction, significant asymmetries are visible, both in the host disk
and in the satellite debris.  Particularly interesting are the low
surface brightness shells, especially visible in the outskirts of the
final thick disks, that last for about 1.5 to 2 Gyr after the merger
has been completed.  These shells acquire relevance in the case of
Andromeda where according to recent studies a couple of these features
are likely associated to the event that also gave rise to the giant
stream \citep[][and references therein]{ibata2007}.

Despite the relatively large mass ratios, the infalling satellites do
not fully destroy the host disk, but merely heat it and tilt it.  The
host disks are found to change their orientation both for the prograde
and retrograde encounters. Furthermore, a remnant thin component
containing between 15 and 25 per cent of the total stellar mass of the
system is present at the final time in all our experiments. 
This prediction of the minor merger model might be potentially
relevant to understand the presence of a very old thin disk in the 
Milky Way.

The scale-lengths of the final thick disks are slightly more extended
than those of the original host disk while the scale-heights are
between three and six times larger, depending on the initial
inclination of the satellite.  The scale-heights have also increased
in proportion to the inclination of the encounter, and the outer disks
are noticeably flared.  If this is the case for the thick disk of the
Milky Way, part of the flared material could be (confused with) the
Monoceros ring.

In our simulations, the outer isophotes of the final thick disks
(measured at surface brightness levels $>$6 mag below the central
value) are consistently more boxy than the inner ones. The eventual
detection of such degree of boxiness, especially for bulgeless
galaxies, would provide support for a formation process as that
modeled here.

Interestingly, the fraction of satellite particles at a given galactic
radius as a function of height above the plane, when normalised by the
(luminosity-weighted) scale-height, \emph{only} depends on the mass
ratio between the satellite and host and not on stellar morphology of
the satellite or type of orbit. For instance, at a distance of 4
scale-heights the fraction of satellite particles reflects the mass
ratio of the merger.

We find that satellite stars do not dominate the luminosity of the
thick disk until rather far above the midplane.  In this sense, the
existence of a counter-rotating thick disk, detected by
\citet{yoachim2005} only $\sim$2 thick-disk scale-heights above the
midplane, can only be explained in the context of our models if the
(young) thin disk formed from freshly accreted counter-rotating gas.
The remaining possibility is, of course, that the thick disk formed
exclusively by direct accretion of stars from an infalling satellite.
Relatively fast rotating thick disks (like the one of the Milky Way)
may be more easily explained by disk heating formation, since a random
distribution of accreted satellites would seem to have less chance of
producing thick disks with strong coherent rotation.

{\it If taken at face value} the velocity ellipsoids of the simulated
thick disks are in good agreement with observations of the Galactic
thick disk at the solar radius. The rotational lag may also be
consistent with observations. These statements are however only valid
if we neglect further significant evolution due to the formation of
a thin disk component from freshly accreted gas.  The observed
trend of the ratio $\sigma_Z/\sigma_R$ with radius in the final thick
disks is found to be a very good discriminant of the initial
inclination of the decaying satellite. In the case of the Milky Way,
the observed $\sigma_Z/\sigma_R$ at the position of the Sun is
$\sim0.6$ \citep[e.g.,][]{chiba2001,vallenari2006}, suggesting that
the thick disk of the Galaxy could have been produced by a merger of
intermediate inclination.  Measurements of the mean rotational
velocity in the final thick disks, at several heights from the
midplane, indicate that satellites with lower initial inclinations are
more efficient in introducing asymmetric drifts dependent on height.
This implies that the possible existence of vertical gradients in the
mean rotational velocity in the thick disk of the Galaxy
\citep{girard2006} would also favour mergers with either low or
intermediate orbital inclination. 
We defer to paper II a more detailed
analysis of the phase-space structure of the merger product. We expect
this will lead to new constrains on the mechanism described here for
the formation of the Galactic thick disk.

\section*{Acknowledgements}
We are grateful to the referee for the extensive and very insightful
comments which have led to a number of significant improvements in our
manuscript.  We thank M. C. Smith and L. Sales for stimulating
discussions and suggestions, and S. C. Trager and R. Sanders for
useful remarks.  We acknowledge financial support from the Netherlands
Organisation for Scientific Research (NWO).  The simulations were run
in the Linux cluster at the Centre for High Performance Computing and
Visualisation (HPC/V) of the University of Groningen in The
Netherlands.

\bibliographystyle{mn}
\bibliography{bibl2n,bibl3} 

\appendix
\section{Setting up the initial conditions for the host and satellite systems}

\subsection{Main disk galaxy}
The main disk galaxy is a self-consistent two-component system,
containing a dark matter halo and a stellar disk.

\subsubsection{Dark Matter Halo}
\label{sec:dark-halo}

The dark matter (DM) halos in our simulations follow a NFW mass
density profile 
\citep*[][hereafter NFW]{navarro1997}:

\begin{equation} \label{nfw}
\rho_{NFW}(r) = \frac{\rho_s}{(r/r_s)(1+r/r_s)^2}
\end{equation}
where $\rho_s$ is a characteristic scale density and $r_s$ a 
scale radius.  The advantage of using this density profile
is that it is consistent with cosmological simulations, and its
evolution with redshift (or time) is relatively well-known (e.g.,
\citealt{wechsler2002}). This implies that it is easy to re-scale its
properties to study the formation of thick disks at redshifts greater
than zero.

In this paper we adopt a flat cosmology defined by $\Omega_m(\textrm{z}=0)=0.3$
and $\Omega_{\Lambda}=0.7$ with a Hubble constant of $H(\textrm{z}=0)=70$
km/s/Mpc.

The virial radius of the halo $R_{vir}(\textrm{z})$ is defined as the
radius within which the mean density is $\Delta_{vir}(\textrm{z})$
times the critical density $\rho_c(\textrm{z})$ of the universe at a
given redshift:
\begin{equation} 
M_{vir}(\textrm{z}) = \frac{4\pi}{3}\Delta_{vir}(\textrm{z})\rho_c(\textrm{z})R_{vir}^3
\end{equation}
where the virial overdensity $\Delta_{vir}(\textrm{z})$ is taken from
the solution to the dissipationless collapse in the spherical top-hat
model.  Its value is $18\pi^2$ for a critical universe but has a
dependency on cosmology.  In the case of flat cosmologies,
$\Delta_{vir}(\textrm{z}) \approx (18 \pi^2 + 82x + 39x^2)$, where $x
= \Omega(\textrm{z}) - 1$, and $\Omega(\textrm{z})$ is defined as the
ratio between mean matter density and critical density at redshift z.
Another important related quantity is the concentration $c$ defined as
$c=R_{vir}/r_s$.  From \citet{wechsler2002} we take the relation
linking $M_{vir}$ to the concentration parameter $c$ at redshift z=0
as:
\begin{equation} 
c \simeq 20 \left(\frac{M_{vir}}{10^{11}M_{\sun}}\right)^{-0.13}.
\end{equation}

We follow \citet{wechsler2002} to scale both the virial mass of the
halo and its concentration as a function of redshift:
\begin{equation} 
M_{vir}(\textrm{z}) = M_{vir}(\textrm{z}=0) \exp(-2 a_c \textrm{z}),
\end{equation}
\begin{equation} 
c(\textrm{z}) = \frac{c(\textrm{z}=0)}{1+\textrm{z}}
\end{equation}
where $a_c$ is a constant defined as the formation epoch of the halo,
taken as $a_c=0.34$. In practice this means that the structure of the
halo of the main galaxy at any redshift is fully determined by
imposing only a value for the virial mass at redshift z=0.  The values of the
halo parameters used in our simulations are included in the Table
\ref{halo-disk-bulge-param}.

Since the mass of a NFW profile formally diverges with radius, we introduce an
exponential truncation starting at $R_{vir}$ and decaying on a scale
$r_{dec}$ \citep{springel1999}:
\begin{equation} 
\label{trunc}
\rho(r) = \frac{\rho_s}{c(1+c)^2} \left(\frac{r}{R_{vir}}\right)^{\epsilon} \exp \left(- \frac{r-R_{vir}}{r_{dec}} \right) (r>R_{vir})
\end{equation}
where $r_{dec}$ is a free parameter.  By requiring continuity at
$R_{vir}$ between Eqs. (\ref{nfw}) and (\ref{trunc}), and also between
their logarithmic slopes, the exponent $\epsilon$ is computed as:
\begin{equation} 
\label{trunc2}
\epsilon = - \frac{1-3c}{1+c} + \frac{R_{vir}}{r_{dec}}.
\end{equation}
Note that for $r_{dec}=0.1 R_{vir}$ the total mass of the halo
becomes $\sim 10\% $ larger than $M_{vir}$.  We define the maximum
extension of the halo as $R_{max}=R_{vir}+3 r_{dec}$.

We also allow the contraction of the halo in response to the formation
of a stellar disk in its central part \citep{blumenthal1986,mo1998}
The adiabatic contraction first
assumes that the gas (that later forms the disk/bulge) is distributed
in the same way as the dark matter.  Then both the spherical symmetry
of the halo and also the angular momentum of each dark matter orbit
are conserved during the contraction, i.e.,:
\begin{equation} \label{adiab1}
r_i M_i(r_i) = r_f M_f(r_f)
\end{equation}
where $r_i$ and $r_f$ are the initial and final radius of a shell of
dark matter, $M_i$ is the initial total mass (distributed according to
an NFW profile) and $M_f$ is the mass distribution after the disk has
been formed, and also includes the contribution of the disk. Therefore,
\begin{equation} \label{adiab2}
M_f(r_f) = M_d(r_f) + (1 - m_d)M_i(r_i)
\end{equation}
where $m_d=M_{disk}/M_{halo}$.  The final dark matter distribution of
the adiabatically concentrated halo will be:
\begin{equation} \label{adiab3}
M_{halo}(r) = M_f(r) - M_{disk}(r).
\end{equation}

Now that the mass distribution of the halo component has been defined,
it is straightforward to initialise the positions of the particles in
our halos. As the next step, the velocity of each particle is computed
from the distribution function (DF) associated to the adiabatically
contracted mass density profile $\rho_{halo}(r)$.  We follow
\citet{kazantzidis2004} and compute numerically the DF that, in
general, is given by\footnote{For a spherical non-rotating system.}:

\[
f(Q) = \frac{1}{\sqrt{8}\pi^2}\Bigg[\int_0^Q
\frac{d^2\rho_{halo}}{d\psi^2} \frac{d\psi}{\sqrt{Q-\psi}}
\]
\begin{equation} \label{df}
\ \ \ \ \ \ \ \ \ \ \ \ \ \ \ \ \ \ \ \ \ \ \ \ \ \ \ \ \ \ \ \ \ \ \
\ \ \ \ + \frac{1}{\sqrt{Q}}
\left(\frac{d\rho_{halo}}{d\psi}\right)_{\psi=0} \Bigg]
\end{equation}
\citep{binney1987} where $Q = \psi - v^2/2$; $\psi = - \Phi(r)$ is
the effective gravitational potential (including the disk); and $v$ is
the velocity of each particle. Finally, we use the rejection method \citep{press1992}
to generate the velocities for our particles.

\subsubsection{Stellar Disk}
\label{sec:exp-disk}

The disk component is constructed following the procedure outlined by
\citetalias{quinn1993} and \citet{hernquist1993} which consists,
briefly, in initialising particle positions according to a density
profile of the form:
\begin{equation} \label{diskprof2}
\rho_d(R,z) = \frac{M_d}{8\pi R_D^2 z_0}
\exp\left(-\frac{R}{R_D}\right)\ \textrm{sech}^2\left(\frac{z}{2
z_0}\right)
\end{equation}
where $M_d$ is the disk mass, $R_D$ is the exponential scale-length,
and $z_0$ is the exponential scale-height.

The velocity components $v_R$, $v_{\phi}$ and $v_z$ of the disk
particles are calculated from moment equations of the collisionless
Boltzmann equation (CBE) supplemented by observational constraints
\citep{binney1987}. We assume that locally (at
each point in the disk) the velocity distribution can be approximated
by a Maxwellian, whose parameters are set up as follows:
\begin{itemize}
\item The radial velocity dispersion $\overline{v_R^2}(R) \propto
\exp(-R/R_D)$.  This is motivated by observations of external galaxies
\citep{vdkruit1981a,lewis1989}.  The
normalisation constant is set by adopting a certain value of the
stability Q-parameter \citep{toomre1964} at a particular location in the
disk.  In this paper $Q=2$ at $R=2.4 R_D$, which for a Milky Way-like
disk corresponds to the solar radius.
\item The vertical velocity dispersion $\overline{v_z^2}(R) = 2 \pi
\textrm{G} \Sigma(R)z_0$, following the isothermal sheet model.
\item The dispersion in the azimuthal direction is obtained by using
the epicyclic approximation \citep{binney1987} \mbox{$\sigma_{\phi}^2(R) = 
\overline{v_R^2}(R) \kappa^2(R)/ 4 \Omega^2(R)$}, where
$\kappa$ and $\Omega$ are the epicyclic and angular frequencies,
respectively.  The mean values of the azimuthal Gaussian distributions
are calculated from the second moment of the CBE,
\[ 
\overline{v_{\phi}}^2(R) = \overline{v_R^2}(R) \left[1 -
\frac{\kappa^2(R)}{4\Omega^2(R)} - 2\frac{R}{R_D}\right] + v_c^2(R),
\]
where $v_c(R) = R\ \Omega(R)$ is the circular velocity considering all
the components of the system.  
\end{itemize}

Note that velocities derived from the CBE are close but not identical
to the ones derived from the DF of the disk.  Unfortunately, the DF is
unknown for the disk in Eq.~ (\ref{diskprof2}).  Therefore, we can
expect some initial evolution in the disk properties.  As shown in \S
\ref{app:isolation}, this evolution is indeed minimal.

\subsection{Numerical Parameters}
\label{app:numerics}

The $N$-body systems are evolved using \mbox{Gadget-2.0}
\citep{springel2005} a well documented massively parallel TreeSPH
code.  Depending on the system under study, this code has to be
provided with suitable values for the so-called numerical parameters,
being these: the number of particles $N$ to represent a given
component in the system; the softening $\epsilon$ of gravitational
forces to avoid strong artificial accelerations between particles
passing close to each other; and finally the timestep $\Delta t$,
that controls the frequency at which positions and velocities are
computed for each particle. In general, these three parameters set the
mass, spatial and time resolution in a numerical simulation.  At the
moment of defining $N$, $\epsilon$ and $\Delta t$ the usual problem is
that they are interrelated in a complicated way.  For instance, $N$
will depend on the available CPU power to run the simulations;
$\epsilon$ will depend on both $N$ and the mass distribution of the
system to be simulated; and $\Delta t$ will depend on the smallest
spatial resolution that is possible to resolve, $\epsilon$, and again
on the available CPU power.  The optimal choice of these parameters
will establish a compromise between quality and efficiency in a
numerical simulation.

\subsubsection{Number of particles}

Tables \ref{halo-disk-bulge-param} and \ref{halo-disk-sat-param} list
the numbers of particles used for each component in our simulations.
As shown by \citetalias{walker1996}, using self-consistent simulations
of an isolated halo-disk system, large numbers of particles in the
halo are needed to suppress the formation of bar perturbations in the
disk.  This is because large $N_{halo}$ decreases the graininess of
the potential, which bars are seeded from. \citetalias{walker1996}
suggest the use of $\sim$ 500000 particles in the halo in order to
smooth out the potential for time scales comparable to the orbital
decay of satellites in our simulations.

For our purposes, bars are an unwanted effect because they represent
an additional source of disk heating, besides the one of interest
here.  Although the complete elimination of bar formation in a
self-consistent simulation is difficult, its effect on the disk can be
constrained by evolving the main disk galaxy in isolation, for the
same timescale as the merger simulation.
  
The number of particles in the host disk $N_{host,disk} = 10^5$, and
is similar to previous studies on disk heating by mergers with
satellites. The satellites are modeled with a relatively large number
of particles (particularly in comparison to previous works) to study
the distribution of the debris, which is the focus of Paper II. In all
cases we can follow accurately the structure and evolution of each
component during the simulations.

\subsubsection{Softening}

\begin{figure}
\begin{center}
\includegraphics[width=85mm]{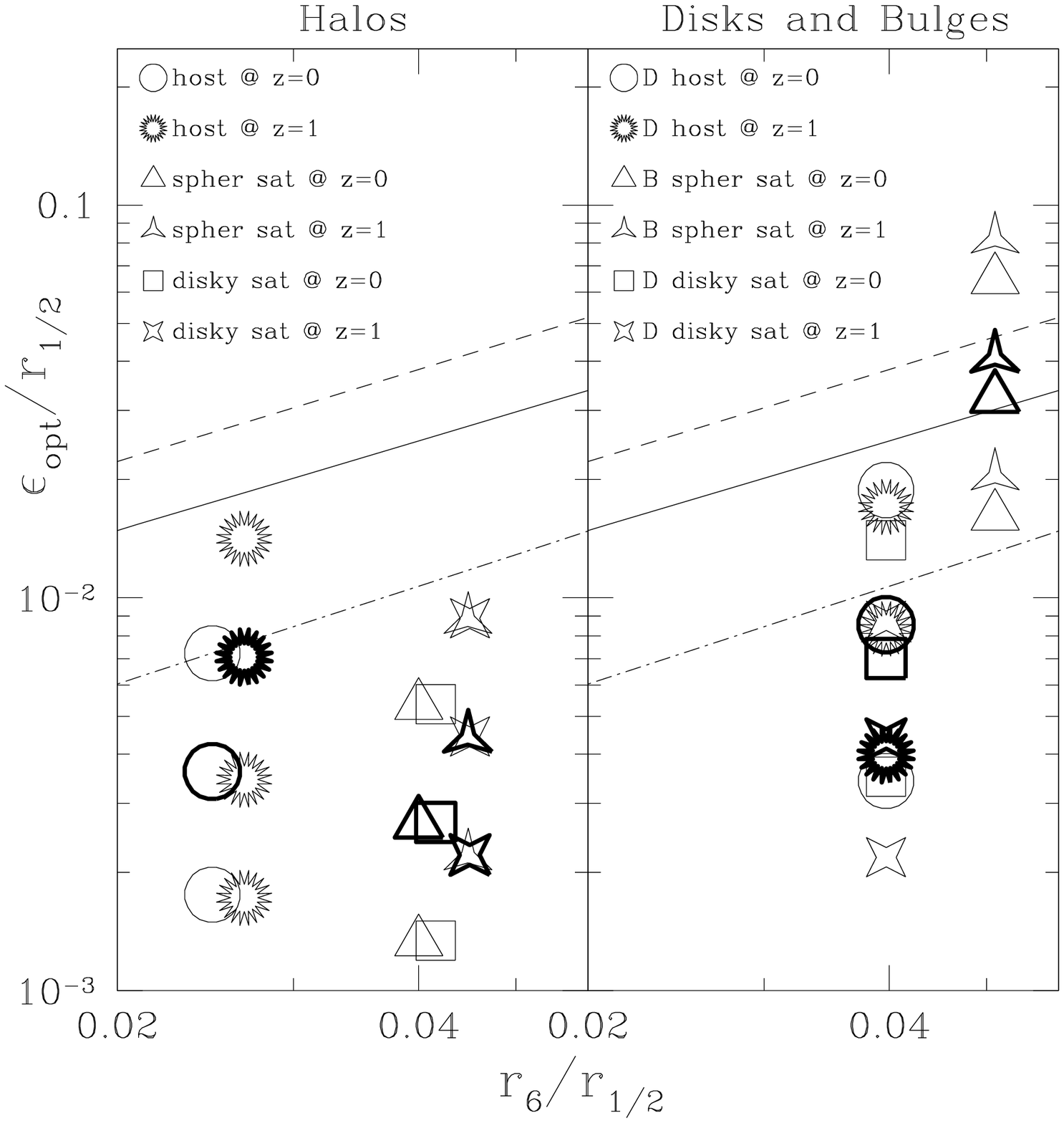}
\caption{Numerical softening as a function of the mean distance from
each particle to the 6th closest neighbour in units of the half-mass
radius $r_{1/2}$.  Straight lines are extrapolations towards smaller
$r_6$ taken from the results of \citet{athanassoula2000} [see Fig. 11
in that work].  Dashed, solid and dot-dashed lines show the values for
a homogeneous, Plummer and Dehnen sphere (with $\gamma=0$),
respectively. The symbols indicate the values of softenings explored
in this paper for each component.  Darker symbols show the optimal
softenings that produced the best stability in each case.}
\label{atha}
\end{center}
\end{figure}

Many studies have been carried out on how to choose the optimal
gravitational force softening $\epsilon$ in order to faithfully
represent the system \citep[see][for an excellent
review]{sellwood1987}.  We use here the prescription by
\citet{athanassoula2000}. These authors present a simple method to
estimate $\epsilon$ for arbitrary mass distributions as a function of
the number of particles.  The optimal softening $\epsilon_{opt}$ is
the one that minimises the error in the forces between particles in a
system with a given mass distribution. They find a correlation between
$\epsilon_{opt}$ and the distance $r_{6,mean}$ from every particle to
its sixth closest neighbour, which is defined as:
\begin{equation}
r_{6,mean} = \left( N^{-1} \sum_{i=1}^{N} r_{6,i}^{-1} \right)^{-1}
\end{equation}
where $r_{6,mean}$ depends on both the number of particles and the
mass distribution.  Fig. \ref{atha} shows $\epsilon_{opt}$ as a
function of $r_{6,mean}$ for three mass distributions discussed by
\citet{athanassoula2000} in order of increasing density: homogeneous
sphere, Plummer profile and Dehnen model (with $\gamma = 0$).  To
estimate the $\epsilon_{opt}$ for the systems of our simulations the
procedure followed is: 1) compute $r_{6,mean}$ for each of our
components; 2) compare the central density of our components to the
ones of the homogeneous, Plummer and Dehnen spheres.  By doing so the
optimal softening for each component can be constrained within a range
on the plane $r_6 - \epsilon_{opt}$; and 3) $\epsilon_{opt}$ is found
by running a few simulations with a set of $\epsilon$ within this
range and choosing the one that offers the best stability.
Specifically, a softening is considered optimal when each component of
both the host and satellite systems present the least evolution in
their structure and kinematics (in the case of the host minimising the
effect of non-axisymmetries) during the amount of time required for
the satellite to sink and become fully disrupted.

In Fig. \ref{atha} darker symbols show the adopted values of $\epsilon_{opt}$ for
each component at every redshift.  Note that for each component the
values of $\epsilon_{opt}$ are well constrained on the $r_6 -
\epsilon_{opt}$ plane, facilitating the extrapolation of these optimal values
to similar systems with a different number of particles. For each
component, the adopted values of $\epsilon_{opt}$ are listed in Table
\ref{halo-disk-bulge-param} and Table \ref{halo-disk-sat-param}.

The distinctive location of the optimal softenings on the
$r_6-\epsilon_{opt}$ diagram basically depends on both the central
concentration of the systems and on the number of particles used to
model them.  For instance, halos of hosts and satellites at ``z=0''
and ``z=1'' always lie below the Dehnen model because they are more
centrally concentrated (even more so when the adiabatic contraction is
taken into account).  On the other hand, the separation along the
$r_6$-coordinate between the optimal softening of the halos of hosts
and satellites reflects the difference in the number of particles used
to model them.  In this sense it is easy to associate systems with a
larger number of particles to smaller mean distances between particles
and viceversa, given that the systems are compared in a normalised
scale.

Also note that the disky satellite requires a halo that is better
resolved at ``z=1'' than for a spherical satellite in order to reach
the best stability.  This can be explained by the fact that a better
resolved centrally concentrated region of the halo is able to inhibit
the formation of non-axisymmetries \citep[e.g.,see][]{athanassoula2005}.

In order to check the robustness of our choices of both number of
particles and softenings, we have followed the suggestion by the
referee and also simulated one of the ``z=1'' experiments adopting a
the same mass and softening for the stars in the satellite and in the
host disk stars. Reassuringly we found practically no difference in
the global properties of the final thick disk in comparison to our
``standard'' choice of numerical parameters.

\subsubsection{Timestep}

The timestep $\Delta t$ has been defined for our simulations according
to the standard criterion of Gadget-2.0.  This means that the timestep
for each particle is calculated as $\Delta t = \sqrt{2 \eta
\epsilon/|\bf{a}|}$, where $\eta$ is a dimensionless parameter
controlling the accuracy of the timestep criterion, $\epsilon$ is the
softening associated to each particle, and $\bf{a}$ is the
gravitational acceleration suffered by each particle.  $\Delta t$ is
also limited by a maximum value in order to prevent particles having
too large timesteps.  The maximum timestep is defined as a few percent
of the timescale $t_c = 2 \pi \epsilon/V_c(\epsilon)$ calculated for
the component that has the smallest $\epsilon$ in the system, where
$V_c$ is the circular velocity.  This ensures us that we follow the
evolution of even the smallest components in the system with enough
time resolution.  We have set $\eta = 0.025$ and the maximum timestep
to $0.25$ Myr, which give us typical conservations of energy and
angular momentum that are better than 0.1\% over 9 Gyr of evolution
for our main disk galaxy configured at ``z=0''.

\subsection{Evolution of Isolated Host}
\label{app:isolation}

\begin{figure}
\begin{center}
\includegraphics[width=40mm]{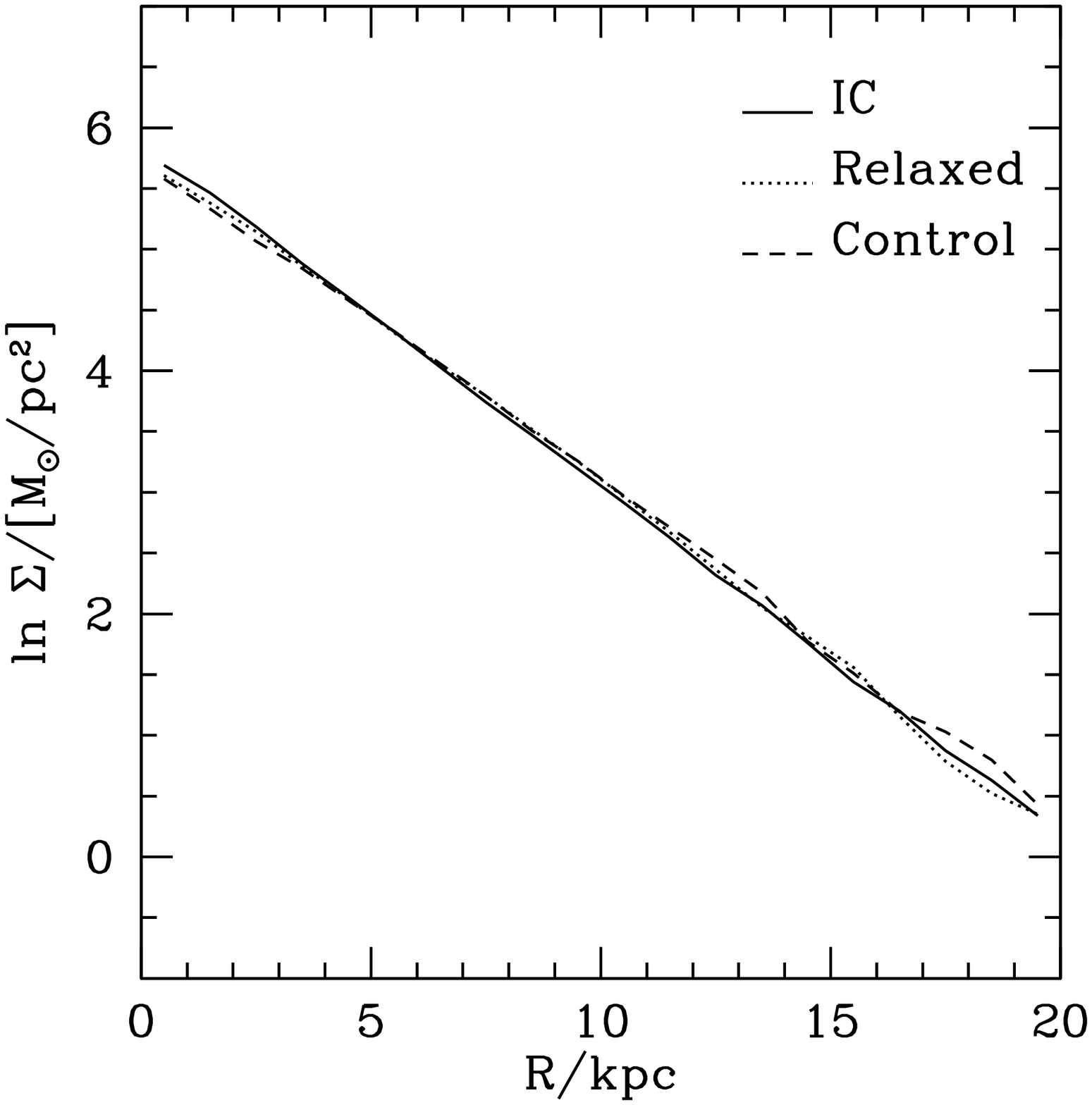}
\includegraphics[width=40mm]{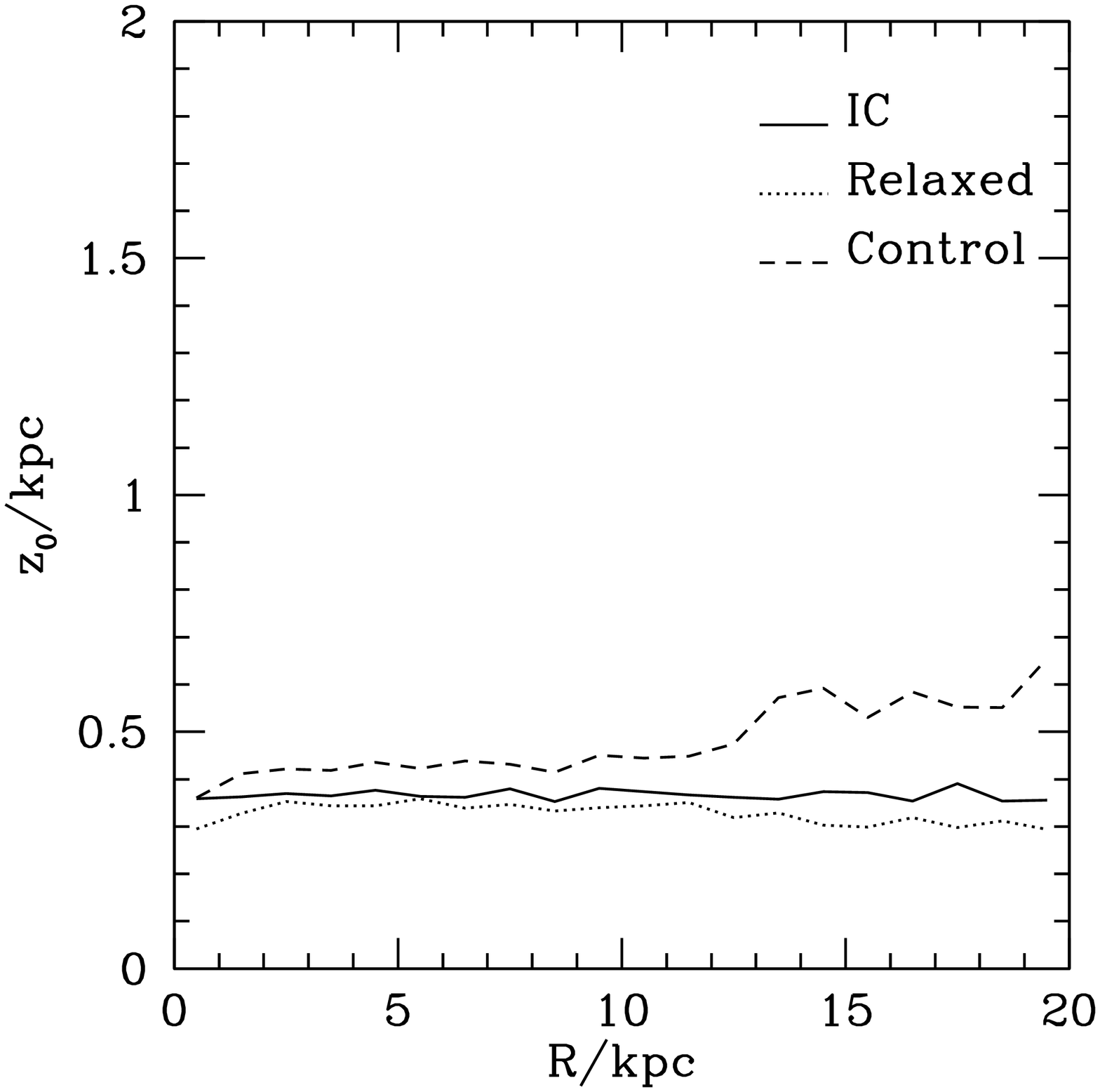}
\includegraphics[width=40mm]{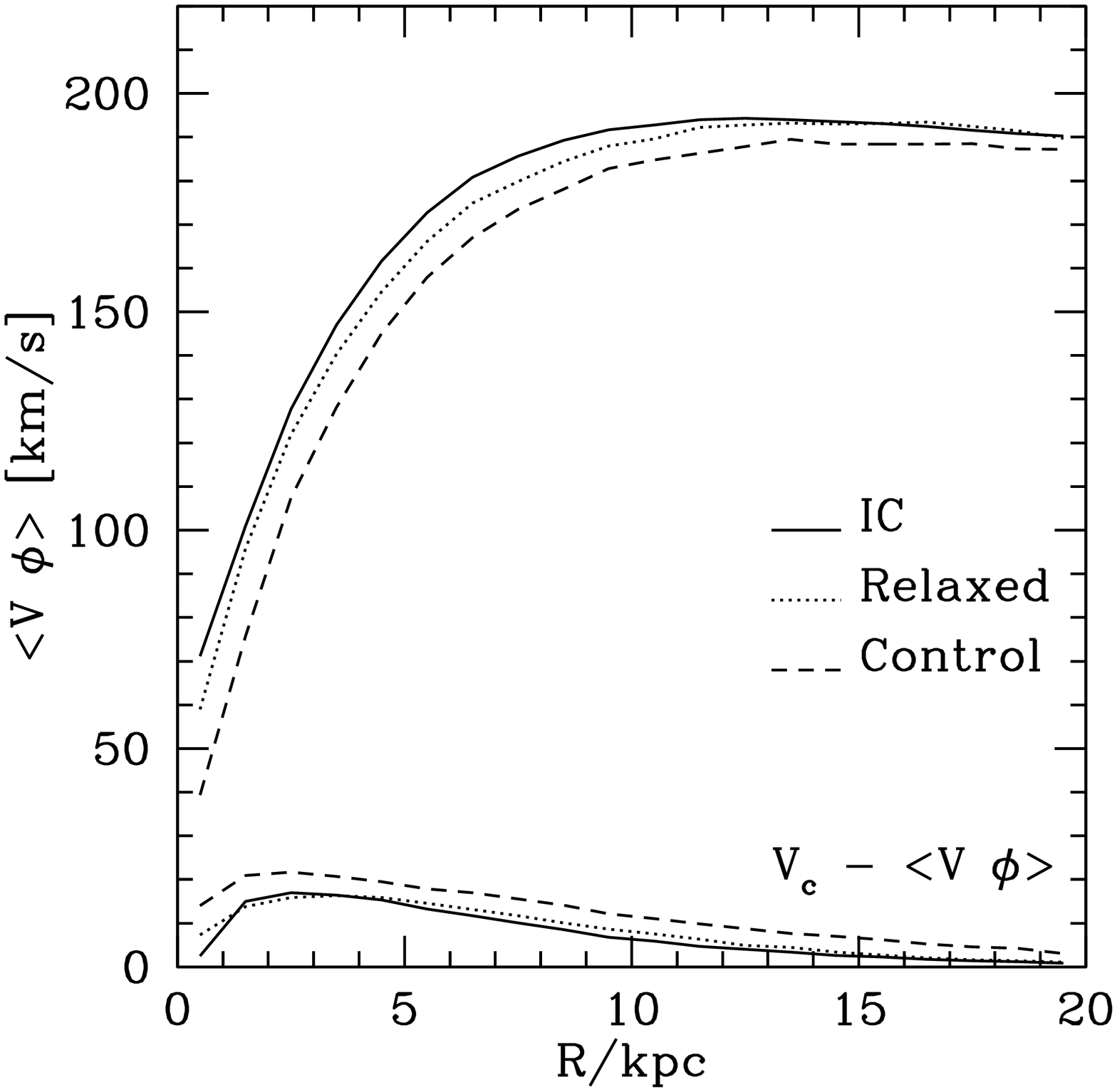}
\includegraphics[width=40mm]{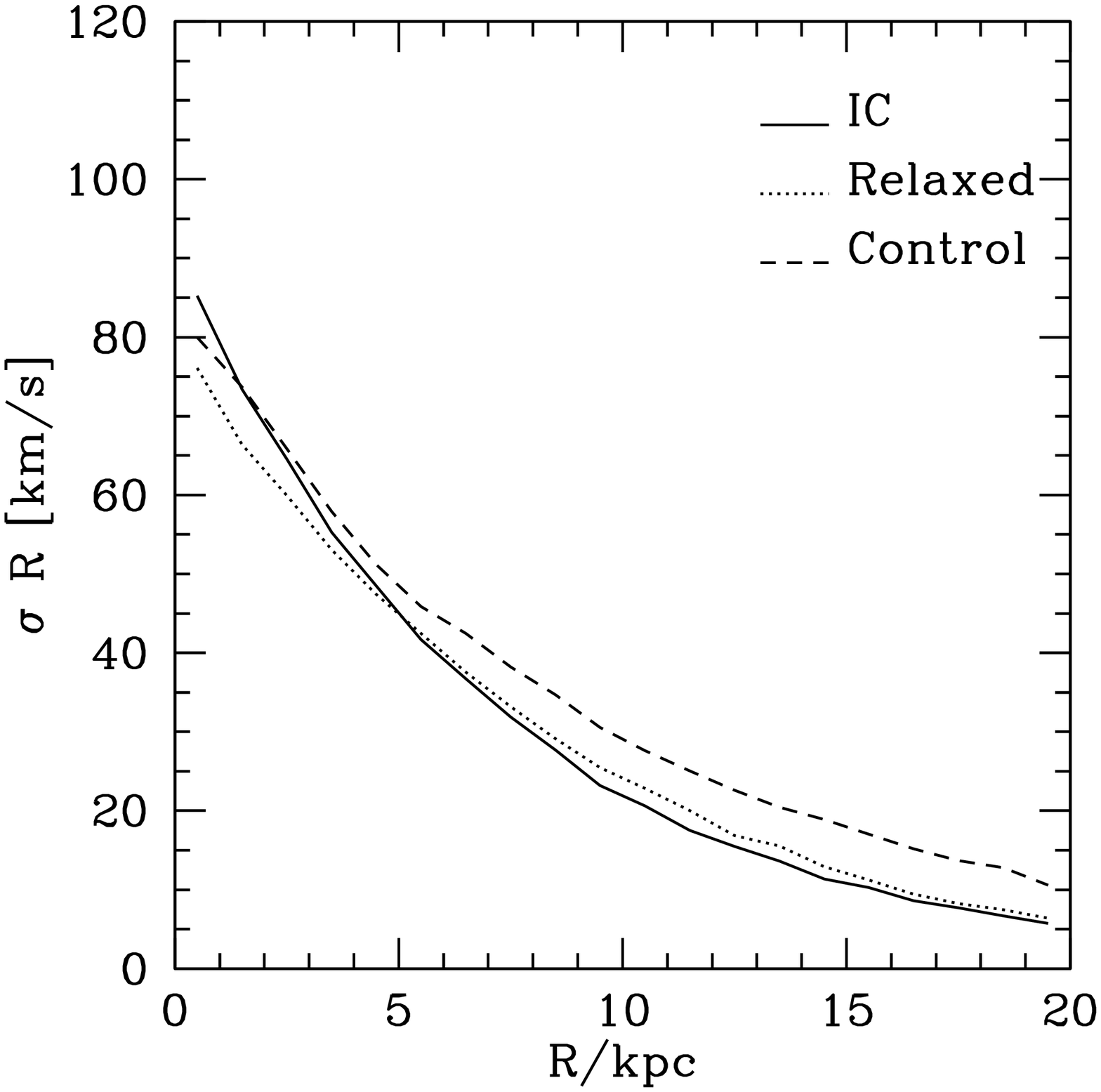}
\includegraphics[width=40mm]{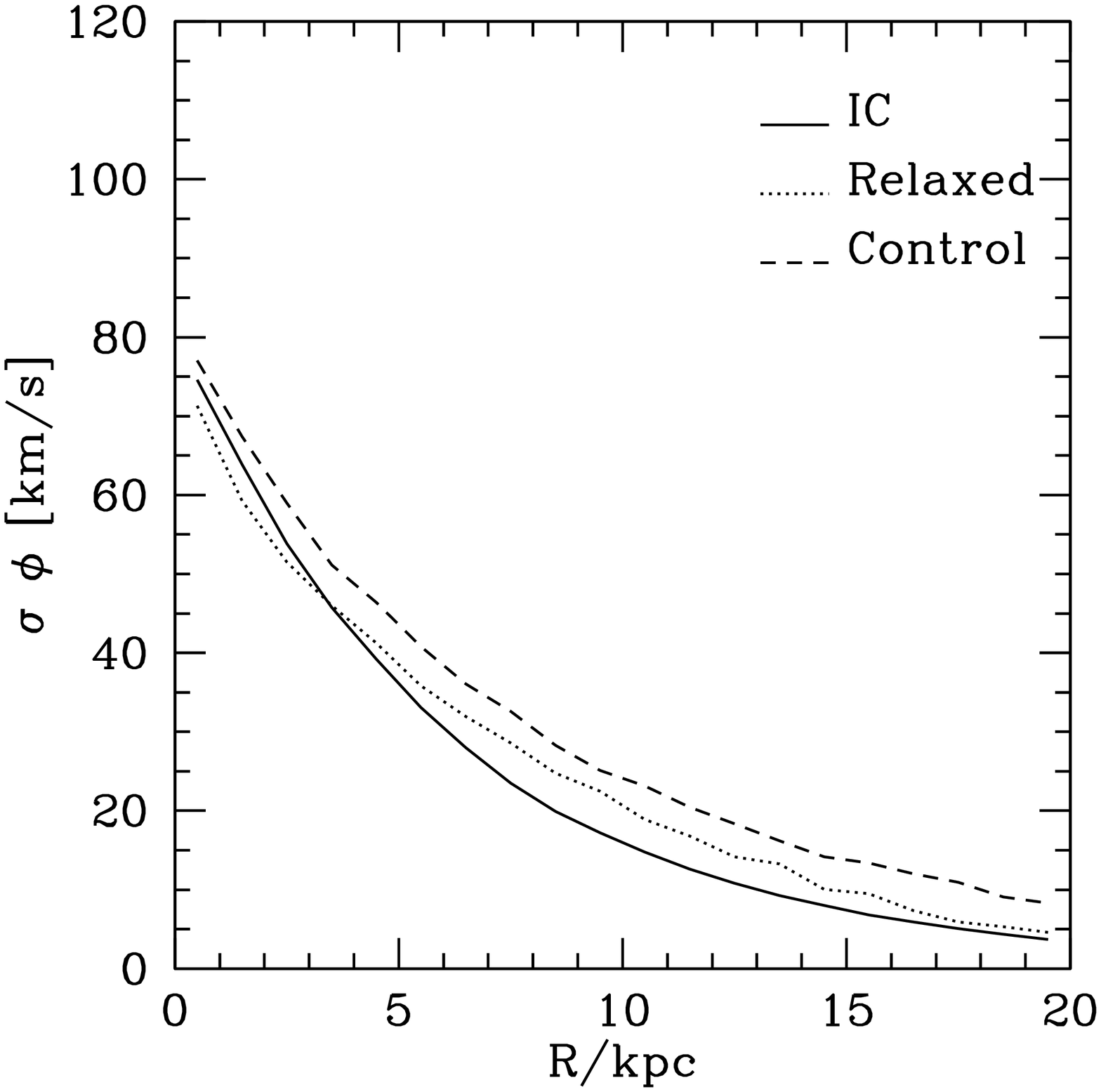}
\includegraphics[width=40mm]{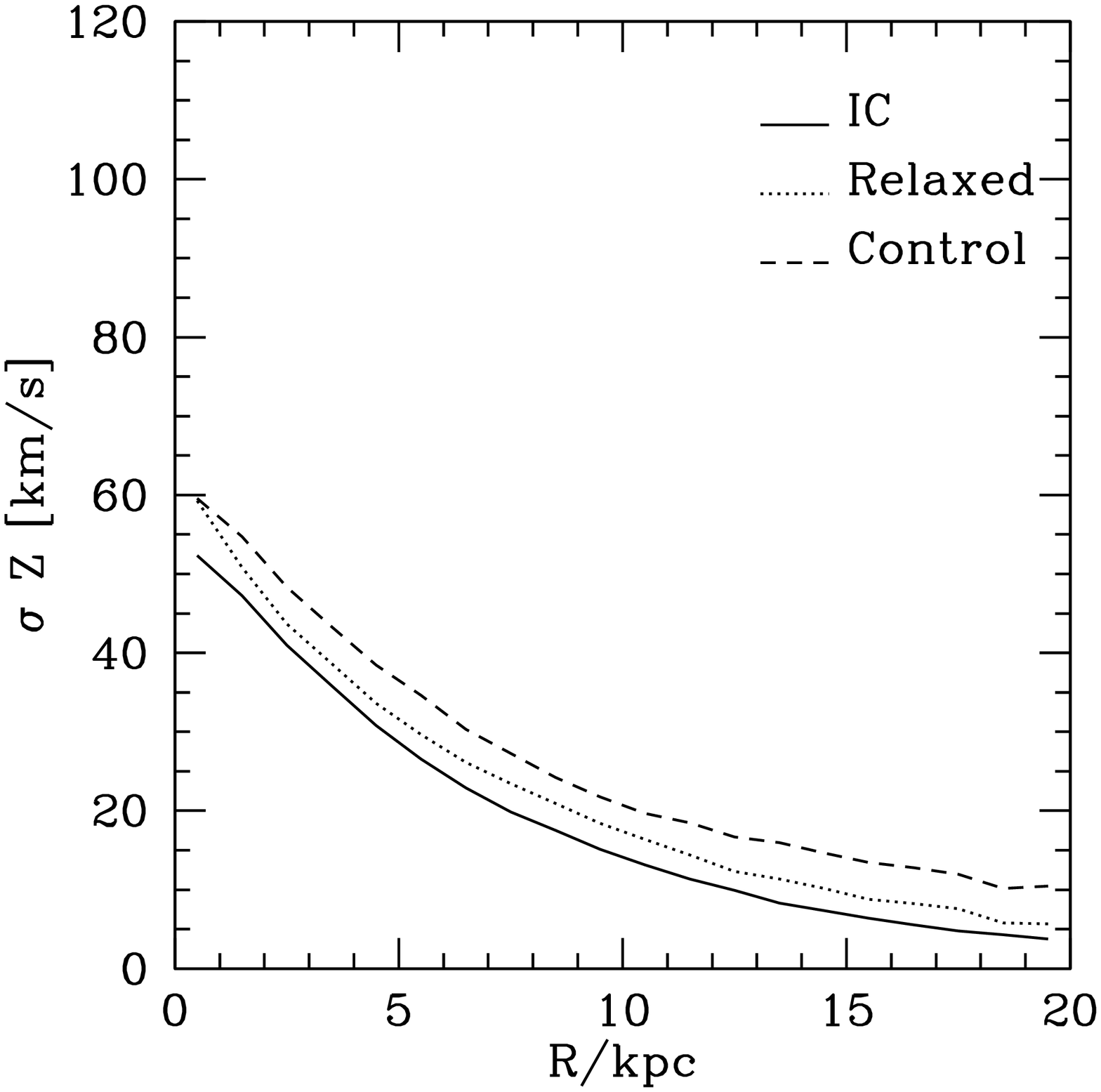}
\caption{ Evolution of the structural and kinematic properties of the
isolated host disk, for the ``z=0'' configuration.  The solid lines
show the initial conditions.  The dotted lines show the disk after 1
Gyr of evolution within the fixed halo potential.  The dashed lines
show the disk's evolution after 5 Gyr within the $N$-body halo.  The
latter model is used as the control case in the comparison to the
minor merger simulations. A similar behaviour is obtained for the disk
in the ``z=1'' configuration.}
\label{maindiskrelaxcontrolz0}
\end{center}
\end{figure}

To test the stability of the host galaxy, this is simulated in
isolation, i.e.  in the absence of any external perturbation.

We first relax the disk component within a ``rigid version'' of the
halo potential (which mimics the N-body halo described in \S \ref{sec:dark-halo})
for a few rotational periods (normally 1 Gyr). Once the disk component
is relaxed (i.e., there are no further changes in either its
morphological or kinematical properties) the ``rigid'' halo is simply
replaced by its $N$-body (``live'') counterpart, and evolved for additional
5 Gyr in isolation for the configuration at ``z=0'', and during
4 Gyr for that at ``z=1''.  As described in \S \ref{sec:orbit-evol}, these
time windows are enough to study the merger events that are of
interest to us. 
Strong bar instabilities appear in the host galaxies evolved in isolation
only after 9 Gyr and 7
Gyr for the configurations at ``z=0'' and ``z=1'', respectively.

Fig. \ref{maindiskrelaxcontrolz0} shows how the initial properties of
the host disk at ``z=0'' change after being relaxed within a fixed
halo for 1 Gyr, and after 5 Gyr in the live halo. Its properties are
measured in concentric rings of 1 kpc of width, including
particles out to $\sim$15 initial scale-heights. The surface density
profiles $\Sigma(R)$ (top left panel) indicate that the scale-lengths
of the disks (given by the inverse of the slope in log-linear scale)
stay practically unchanged.  Similarly, the vertical structure of the
disk does not show significant evolution (top right panel), except a
moderate amount of flaring in the outer disk, which are due to spiral
instabilities induced by swing-amplified Poisson noise in the disk.
The scale-heights, measured at $R=2.4R_D$, change from $z_0$= 0.35 kpc
to 0.41 kpc in the ``z=0'' configuration and from 0.17 kpc to 0.24 kpc
in the ``z=1'' one. The disks are also slightly slowed down (middle left panel)
, while the velocity dispersions show an increase of $\sim 5$
km/s in the first Gyr in the fixed halo, and a total of $< 10$ km/s
after 5 Gyr of evolution in the live potential (middle-right and bottom panels). The
velocity ellipsoid of the disk (also measured at $R=2.4R_D$) increases
from ($\sigma_R$,$\sigma_{\phi}$,$\sigma_Z$)=(28,20,17.5) km/s to
(35,28,24) km/s in the ``z=0'' configuration and from (25,18,14) km/s
to (32,28,22) km/s in the ``z=1'' one.

\label{lastpage}

\end{document}